\documentstyle{amsart}
\newtheorem{thm}{Theorem}[subsection]
\newtheorem{cor}[thm]{Corollary}
\newtheorem{lemma}{Lemma}[thm]

\newtheorem{defn}[thm]{Definition}

\theoremstyle{remark}

\newtheorem{rem}[thm]{Remark}
\newtheorem{hist}[thm]{Historical remark}
\newtheorem{ex}[thm]{Example}

\newenvironment{pr}{{\em Proof}\newline }{\epr}

\numberwithin{equation}{section}

\begin{document}
\input amssym.def
\input amssym
%
%
\newcommand{\bC}{{\Bbb C}}
\newcommand{\bN}{{\Bbb N}}
\newcommand{\bQ}{{\Bbb Q}}
\newcommand{\bR}{{\Bbb R}}
\newcommand{\bZ}{{\Bbb Z}}
%
%
\newcommand{\cA}{{\cal A}}
\newcommand{\cB}{{\cal B}}
\newcommand{\cC}{{\cal C}}
\newcommand{\cD}{{\cal D}}
\newcommand{\cE}{{\cal E}}
\newcommand{\cF}{{\cal F}}
\newcommand{\cG}{{\cal G}}
\newcommand{\cH}{{\cal H}}
\newcommand{\cI}{{\cal I}}
\newcommand{\cJ}{{\cal J}}
\newcommand{\cL}{{\cal L}}
\newcommand{\cM}{{\cal M}}
\newcommand{\cN}{{\cal N}}
\newcommand{\cO}{{\cal O}}
\newcommand{\cP}{{\cal P}}
\newcommand{\cR}{{\cal R}}
\newcommand{\cS}{{\cal S}}
\newcommand{\cT}{{\cal T}}
\newcommand{\cX}{{\cal X}}
%
%
\newcommand{\ub}{\underline{b}}
\newcommand{\uc}{\underline{c}}
\newcommand{\ud}{\underline{d}}
\newcommand{\ue}{\underline{e}}
\newcommand{\uG}{\underline{G}}
\newcommand{\ui}{\underline{i}}
\newcommand{\uI}{\underline{I}}
\newcommand{\uJ}{\underline{J}}
\newcommand{\uL}{\underline{L}}
\newcommand{\uQ}{\underline{Q}}
\newcommand{\uU}{\underline{U}}
%
%
\newcommand{\tc}{\tilde{c}}
\newcommand{\te}{\tilde{e}}
%
%
\newcommand{\gog}{{\goth{g}}}
\newcommand{\goh}{{\goth{h}}}
\newcommand{\gon}{{\goth{n}}}
%
%
\newcommand{\trr}{\triangleright}
\newcommand{\trl}{\triangleleft}
\newcommand{\btrr}{\blacktriangleright}
\newcommand{\btrl}{\blacktriangleleft}
\newcommand{\utrr}{{\underline \triangleright}}
\newcommand{\utrl}{{\underline\triangleleft}}
\newcommand{\epr}{\begin{flushright} $\Box$ \end{flushright}}
%
%
%
%
\begin{center}
{\bf Non-Gaussian generalizations of Wick's theorems,\\
related to the Schwinger-Dyson equation.}\\
Olivier de Mirleau.\;\;\;mirleau@@fwi.uva.nl\\
\end{center}
\setcounter{tocdepth}{2}
\tableofcontents
%
%
%
%
\newpage
\section{Abstract.}
In this work we present a number of generalizations of Wick's theorems on
integrals with Gaussian weight to a larger class of weights which we call
subgaussian.
Examples of subgaussian contractions are that of Kac-Moody or Virasoro type,
although the concept of a subgaussian weight does not refer a priori
to two-dimensional field theory. The generalization was chosen in such a way
that the contraction rules become a combinatorical way of solving
the Schwinger-Dyson equation. In a still more general setting we prove
a relation between solutions of the Schwinger-Dyson equation and a map $N$,
which in the Gaussian case reduces to normal ordering. Furthermore, we
give a number of results concerning contractions of composite insertions,
which do not suffer from the Johnson-Low problem of ``commutation''
relations that do not satisfy the Jacobi identity.
%
%
%
%
\newpage
\section{Introduction.}
\subsection{Motivation.}
\subsubsection{The construction of integration theories.}
The motivation of this work comes mainly from the need of defining
functional integration. In general one may say that the construction of
integration theories proceeds in two steps:
\begin{enumerate}
\item First fix the integral of some ``elementary'' functions.
For example the integration
theories over $\bR^n$ are based on the
requirement that
$$Vol([a_1,b_1]\times ... \times [a_n,b_n])=(b_1-a_1)..(b_n-a_n).$$
In other words the integral of the characteristic function of
$[a_1,b_1]\times .. \times [a_n,b_n]$ is prescribed before having
an integration theory.
\item Then, try to extend the notion of integration to more general functions.
In the example of $\bR^n$, this leads e.g. to the definition of the
Lebesgue measure, but also to the original definition of integration as the
inverse of differentiation.
\end{enumerate}
Note however that it is not strictly necessary to take characteristic
functions as ``elementary''. For example, one of the typical starting points
of functional integration theory over the set of differentiable
functions $\phi:\bR^n\rightarrow \bR$ is the following:
Let $n\geq 3$ and on the set of functions $\phi:\bR^n\rightarrow \bR$
define the weight $S(\phi):=\int \partial_i \phi \partial^i \phi dx^1..dx^n
\in [0,\infty]$. Then whatever the details of the integration theory will be,
we require that:
$$\int_{\{\phi:\bR^n\rightarrow \bR\}} e^{-S(\phi)}
\phi(x)\phi(y)D\phi=\frac{K}{|x-y|^{n-2}}.$$
The motivation for this starting point is the Schwinger-Dyson equation:
\subsubsection{The Schwinger-Dyson equation.} For fixed
$S:\bR^D\rightarrow \bR$,
consider the following linear functional:
$$f\mapsto I(f):=\int_{\bR^D} f e^{-S} dx^1..dx^D,$$
where $f$ and $S$ are restricted such that it is well defined, and such that
upon partial integration boundary terms are zero: In that case the
functional satisfies $\forall_{i,f}\;I(\partial_i(S) f)=I(\partial_i f)$, for:
$$0=\int \frac{\partial}{\partial x^i}(e^{-S} f) dx^1...dx^D
=\int e^{-S} (-\partial_i(S)f +\partial_i f) dx^1...dx^D
=I(\partial_i f)-I(f\partial_i S).$$
This is the Schwinger-Dyson equation
\footnote{\label{diffsd} We will stick to the practice, in the context of
functional integration, of calling
this formula the Schwinger-Dyson equation,
although the above formula was already given for functional integrals by
Feynman in \cite[formula 45]{feynman}(1948).
Another often seen formulation is
$\{J_i-(\partial_i S)(\frac{\partial}{\partial J})\}Z(J)=0$, by setting $Z(J):=
I(e^{J_i x^i})$.}
for the functional $I$.
For Gaussian weights, i.e. where $S$ is quadratic, this equation has
a unique solution up to normalization
for polynomial integrands. Now the point is that even though
the equation is motivated by finite dimensional integration, we may also
try to solve it in infinite dimensions, and take the solution as
a ``starting point'' for the construction of measures.
Appendix \ref{gaussiansection} contains a review of the Gaussian
Schwinger-Dyson equation, together with the proof of the above given
formula for the functional integral.
\begin{rem}Note that the Schwinger-Dyson equation remains valid
if we multiply $I$ by a constant, so if possible we will restrict ouselves to
solutions $I$ such that
$I(1)=1$, and such normalized solutions will be denoted by $f\mapsto \langle
f\rangle $, in
view of the fact that if we set
$$\langle f\rangle :=\frac{\int e^{-S} f d\mu}{\int e^{-S} d\mu},$$
then $\langle .\rangle $ is a normalized solution of the Schwinger-Dyson
equation.
When we say that the solution of the Schwinger-Dyson equation is unique, we
will always mean up to normalization.
\end{rem}
\subsection{Aim of this work.} We have a number of goals:
\begin{enumerate}
\item The first aim of this work is to analyse the conditions for the
weight $S$
under which the Schwinger-Dyson equation has a unique solution.
For example Gaussian weights have a unique solution, but there is more.
\item Next we aim to extend the notion of normal ordering to non-Gaussian
weights in such a way that it is naturally assicated to such weights.
The need for such a normal ordering also comes from functional integration:
In the Gaussian case, normal ordering can be used to regularise certain
types of short distance singularities (see appendix \ref{motsection}),
and it seems desirable to find
the analogue for non-gaussian weights. Furthermore, when using functional
integration for geometric purposes, it is essential to only use
natural constructions. Therefore one is led to look for naturally
constructed normal ordering.
\item We will define the notion of a subgaussian weight for which we
will be able to prove a number of generalizations of Wick's theorems
\cite{wick}.
\item Finally we will look for theorems concerning what are called
composite insertions. This will be explained in a moment.
\end{enumerate}
\subsection{Overview of the article.}
\subsubsection{Change of variables.}
Given a weight $S(x)$, our first step will be to go to a new system of
variables $S_i:=\partial_{x^i} S$. (Thus, for Gaussian weights $S=\frac{1}{2}
g_{ij} x^i x^j$, we just
have $S_i=g_{ij}x^j=x_i$. But the only thing we will assume for now is that
the $S_i$'s form a coordinate system.) Then the Schwinger-Dyson equation reads:
$$\langle \partial_{i_1}(S)...\partial_{i_n}(S)\rangle =
\sum_{k=2}^n \langle \partial_{i_2}(S)..\partial_{i_1}\partial_{i_k}(S)..
\partial_{i_n}(S)\rangle .$$
Now for Gaussian integrals, the second derivatives $\partial^2 S$
are numbers, so that the above equation becomes a recurrence relation.
Our simplest generalization consists in dropping the assumption that
$\partial^2 S$ is a number, and replacing it by the assumption that
$\partial^2 S$ is at most linear in $\partial S$. This is what we
call the subgaussian case, and it obviously leads to a recurrence relation,
so that the subgaussian case has a unique solution too.
The next generalization consists in assuming that $\partial^2 S$ can
be written as a polynomial in $\partial S$, and we will call that
the polynomial case. For the polynomial case there is also
an easy condition which
guarantees uniqueness of solutions for the Schwinger-Dyson equation, which
is invertibility of normal ordering:
\subsubsection{Normal ordering.}
Given a weight $S$, we define normal ordering
\footnote{
Gaussian normal ordering was introduced in \cite{houriet}. A number of
definitions can be found for other cases besides the Gaussian case, see for
example:
\cite{lowenstein},
\cite[formula 7b]{knizhnik},\cite[formula 4]{bershadsky},
\cite[formula 6]{zamolod.multicrit}. However none of these definitions
is directly in terms of the weight $S$.
} inductively using the
new variables $S_i$, as follows: $N(1):=1$, and
$$N(S_{i_0}..S_{i_n}):=S_{i_0} N(S_{i_1}..S_{i_n})-\frac{\partial}{
\partial x^{i_0}}
N(S_{i_1}..S_{i_n}).$$
It is not very difficult to show that if $N$ is invertible, then
$\{I$ satisfies the Schwinger-Dyson equation $\Leftrightarrow
I(f)=ZN^{-1}(f)I(1)\}$,
where $Z$ denotes the projection of polynomial functions of the $S_i$'s
on their constant part, e.g. $Z(3+aS_1+b S_1 S_5)=3$.
This is the main idea of this work.
\subsubsection{The nonabelian case.}
In order to make the link with
two-dimensional field theory, we will have to generalize the above
to the case where instead of using the commuting vectorfields $\partial_i$,
we assume given a not necessarily Abelian Lie algebra $L$ of vectorfields,
and instead of the variables $\partial_i S$, we will (roughly) use
the variables $T_a S$, where $T_a$ is a basis of that Lie algebra.
In section \ref{existsection} we will prove a generalization of the theorem
$\langle . \rangle = ZN^{-1}$ to the non-Abelian case. In this case the
definition of normal ordering needs an explicit symmetrization, which will
make the proof more complicated:
$$N(X_1(S)...X_n(S)):=\frac{1}{ n} \sum_{i=1}^n X_i(S) N(X_{[1,n]\backslash
i}(S))-
X_i(N(X_{[1,n]\backslash i}(S))).$$
Here $X_{[1,n]\backslash i}(S)$ is shorthand for $X_1(S)..X_n(S)$ with
$X_i(S)$ left out.
\subsubsection{Composite insertions and left extensions.}
The fact that in $\langle \partial_i(S) s_1 .. s_n\rangle$
we may eliminate $\partial_i(S)$ in favor of the
sum of terms with $\partial_i s_j$ relies on the special form of $\partial_i
S$.
In general
it will not be possible to find a derivation $D$ such that
$\langle X(S)Y(S) f\rangle =\langle Df\rangle $. However it \underline{may}
happen sometimes, if we choose $X$ and $Y$ in the right way. In that
case we say that with $D$ we have constructed a left extension. (Namely of
the corresponding contraction: See section \ref{examplesection} for the
definition of contractions. In general an ``insertion'' is any factor
in an expression between brackets $\langle . \rangle$.
We will call it composite if it is not a first derivative of $S$).
Section \ref{extendsection} gives examples of such extensions.
\subsubsection{Section by section overview.}
In section \ref{examplesection} we will give precise definitions of what we
mean by the non-Abelian subgaussian case, and prove a number of theorems
concerning them. We will also prove theorems concerning left-extensions
in the subgaussian case. Instead of speaking of weights $S$ we will phrase
everything in terms of contractions $[.\trr.]$, which is a formulation
better suited to study the Schwinger-Dyson equation. In that section
we will also review a relatively well known algebra as an
example
of a non-Abelian contraction: The Kac-Moody algebra.
In section \ref{extendsection} we will give an example of an infinite
dimensional left extension. The example is not new and in fact has a
long history, see remark \ref{hist_foot};
What we want to emphasize is
that the construction is an application of a theorem proved for the general
subgaussian case.
Finally in section \ref{existsection} we will be concerned with the proof
that $\langle .\rangle = ZN^{-1}$ in the non-Abelian case.
\subsubsection{What the article is not about.}
One should be aware that even if the solution of the Schwinger-Dyson equation
is unique, then this does not replace by any means the notion of
integration. Indeed, in the Gaussian case, the solution of the
Schwinger-Dyson equation only determines the integral of
\underline{polynomial} integrands. If one is interested in other integrands,
then it necessary to go on to what we called step two in the
construction of integration theories. Stated precisely:
\begin{defn}
Fix an infinitely differentiable function $S:\bR^D\rightarrow \bR$.
\begin{enumerate}
\item By the algebra $\cS$ of simple functions we
then mean the algebra of functions
generated by $\{\partial_{i_1}..\partial_{i_n} S|i_j\in\{1,..,D\},n\geq 1\}$.
\item We say that $I:\cS\rightarrow \bR$ is a solution of the Schwinger-Dyson
equation iff it is linear and
$\forall_{s\in \cS}\; I(\partial_i s)=I(s\partial_i S ).$
\item $I$ is said to be positive iff $s > 0\Rightarrow I(s)>0$.
\item A measure $d\mu$ on $\bR^D$ is said to be compatible with such a solution
iff $s\in \cS$ is measurable, and $\int_{\bR^D} sd\mu=I(s)$. Such a measure may
allow
one to integrate other functions besides the $s\in \cS$.
\end{enumerate}
\end{defn}

\begin{rem}
\label{asym_foot}
Note that it is not a good idea to try to extend $I$ ``by analyticity'' like
in $I(x\mapsto \sum_i a_i x^i):=\sum_i a_i I(x^i)$, because this equality
need not hold for usual integrals. For example
$$\int e^{-\frac{1}{2}x^2-x^4} dx
\neq \sum_n \frac{1}{ n!}\int (-x^4)^n e^{-\frac{1}{2}x^2}dx=
\sum_n \frac{(-1)^n (4n-1)!!}{ n!} \sqrt{2\pi}=divergent.$$
We will have nothing more to say about measures; This article
will restrict its attention to the algebra $\cS$.
The reader who wishes to know more about the reconstruction and use of
functional measures starting from exact Gaussian results
is referred to the book by Jaffe and
Glimm \cite{jaffe}, noting in particular Minlos' theorem (theorem 3.4.2),
and the
$2D$ non-Gaussian integrals of section 8.6. in that book. Minlos' theorem is
also used in Berezin's book \cite{berezin}.
\end{rem}
%
%
%
%
\newpage
\section{Contraction algebras.}
\label{examplesection}
\begin{rem}
\label{mot_cont_def}
The following definition is motivated as follows:
By a contraction we basically mean the expression of $\partial^2 S$
in terms of $\partial S$. For a Lie algebra of non-Abelian vectorfields
with basis $\{T_a\}$, this roughly means that $T_a T_b S$ is expressed in terms
of $T_a S$, but not exactly:
When using general vectorfields $X$ on some manifold $M$, we can no more
use the fact that the measure $dx^1..dx^n$ on $\bR^n$ is invariant under
the vectorfields $\partial_i$ that we used before. Therefore there is no
point in that case in splitting off the weight $e^{-S}$ from the
volume form $\mu=e^{-S}dx^1..dx^n$, so that we will only talk about the
volume form $\mu$ from now on. By taking Lie derivatives of the integrand,
the Schwinger-Dyson equation then reads: $I(X(f)+f\nabla(X))=0$, where
the divergence is defined by $L_X\mu=\nabla(X)\mu$.
When specializing to $\mu=e^{-S} dx^1..dx^n$ and $X=\partial_i$, we see
that $\nabla(X)=-X(S)$. So instead of expressing $T_a T_b(S)$ in terms of
$T_c(S)$, we will rather be led by the expression of $T_a \nabla(T_b)$
in terms of $\nabla(T_c)$'s.

Recall that the condition that the contraction was polynomial meant
that $\partial^2 S$ could be written as a polynomial
in $\partial S$, and that we could decompose elements of $\cS$
in a unique way as polynomials in $\partial S$.
This condition is now translated into the fact that the map
$X\mapsto X(S)$, or rather $X\mapsto -\nabla(X)$ induces an isomorphism
$Sym(L)\rightarrow \cS$, i.e. the polynomials
in the variables $S_i$ in the abelian case
now get replaced by elements of $Sym(L)$, the
algebra of formal polynomials in elements of the Lie algebra, by the
mapping $X\mapsto -\nabla(X)$, $XY\mapsto \nabla(X)\nabla(Y)$, etc.
The analogue of the assumption that $\partial^2 S$ can be written as a
polynomial in $\partial S$ is that $XY(S)$ is a polynomial in the $X(S)$'s,
or more precisely that $X(-\nabla(Y))$ is a polynomial in the $\nabla(X)$'s,
i.e. it leads to a map: $[.\trr.]:L\otimes L\rightarrow Sym(L);
[X\trr Y]:=-X(\nabla(Y))$, which we extend on the right by derivations to
$Sym(L)$. By the general property of divergences that
$\nabla([X,Y])=X(\nabla(Y))-Y(\nabla(X))$,
see theorem \ref{vol_div_th}, the contraction map $[.\trr.]$
thus obtained will always satisfy the properties
$$[X\trr Y]-[Y\trr X]=[X,Y]\in Sym(L),$$
$$[X\trr[Y\trr Z]]-[Y\trr [X\trr Z]]=[[X,Y]\trr Z].$$
This is what motivates the following definition.
The subgaussian case, i.e. where $\partial^2 S$ was at most linear
in $\partial S$ now corresponds to the contraction being a map
$L\otimes L\rightarrow K\oplus L \leq Sym(L)$.
\end{rem}
\begin{defn} We define a number of special contraction algebras:
\label{pol_contr_def}
\begin{enumerate}
\item
A polynomial contraction algebra is a Lie algebra with a map
$[.\trr.]:L\otimes L\rightarrow Sym(L)$, extended by derivations
on the right to $Sym(L)$ , which satisfies
$$[X\trr Y]-[Y\trr X]=[X,Y]\in L\leq Sym(L),$$
$$[X\trr[Y\trr Z]]-[Y\trr [X\trr Z]]=[[X,Y]\trr Z],$$
\item It is called Gaussian if $[.\trr.]:L\otimes L\rightarrow K=Sym^0(L)$.
\item A subgaussian contraction algebra is one in which
$[.\trr.]:L\otimes L\rightarrow K\oplus L=Sym^{[0,1]}(L)$. In that case, we
extend
$[.\trr.]$ by $[1\trr 1]:=[1\trr X]:=0$, which makes $K\oplus L$ into
a pre Lie algebra.
\footnote{
(I wish to thank C.D.D. Neumann for pointing out the following to me).
A pre-Lie algebra is a vectorspace with a bilinear operation
$[.\trr.]$ satisfying $[a\trr[b\trr c]]-[b\trr[a\trr c]]=
[[a\trr b]-[b\trr a]\trr c]$, see Gerstenhaber
\cite[formula 6]{gerstenhaber}.
In that case $[a,b]:=[a\trr b]-[b\trr a]$ is a Lie composition.}
\item Normal ordering is the map $N:Sym(L)\rightarrow Sym(L)$, defined by
$N(1):=1$, and by
$$N(X_1...X_n):=\frac{1}{n} \sum_{i=1}^n X_i N(X_{[1,n]\backslash i})-
[X_i\trr N(X_{[1,n]\backslash i})].$$
\item The contraction is said to be non-degenerate iff its normal
ordering is invertible.
\item In that case we set $s_1 * s_2:=N^{-1}(N(s_1)N(s_2))$, which in the
Gaussian case corresponds to what is usually called the
operator product: $Sym(L)\otimes Sym(L)\rightarrow Sym(L)$.
\end{enumerate}
\end{defn}

For a review of Gaussian contractions, see appendix \ref{gaussiansection}.
\subsection{Subgaussian contraction algebras.}
\begin{ex}
A simple example of a subgaussian contraction is the one-dimensional Abelian
Lie algebra with basis element $e$, and contraction
$[e\trr e]:=\lambda 1 + \mu e$, for some scalars $\lambda$ and $\mu$.
\end{ex}
\begin{thm}
Set $[X\circ Y]:=\frac{1}{2}[X\trr Y]+\frac{1}{2}[Y\trr X]$.
In subgaussian algebras the following holds:
\label{subgaussthm}
\begin{enumerate}
\item
$$[[X\trr Y]\circ Z]+[Y\circ [X\trr Z]]-[X\trr[Y\circ Z]]=
\frac{1}{2}([[Y\trr X]\trr Z]]+[[Z\trr X]\trr Y]]).$$
\item
$$N([X\trr Y])=[X\trr Y]$$
\item $$N([X\trr Y]Z)=[X\trr Y]Z-[[X\trr Y]\circ Z].$$
\item \label{wickgeneralize} $$[X\trr N(YZ)]=N([X\trr Y] Z + Y[X\trr Z]
+\frac{1}{2}[[Y\trr X]\trr Z]
+\frac{1}{2}[[Z\trr X]\trr Y]).$$
\end{enumerate}
\end{thm}
\begin{pr}
\begin{enumerate}
\item This is an identity which holds in any pre Lie algebra:
$$2.LHS=[[X\trr Y]\trr Z]+[Z\trr [X\trr Y]]+[Y\trr [X\trr Z]]+[[X\trr Z]\trr
Y]$$
$$-[X\trr [Y\trr Z]]\;\;\;-[X\trr [Z\trr Y]]$$
$$=[[Y\trr X]\trr Z]+[Z\trr [X\trr Y]]+[Y\trr [X\trr Z]]+[[Z\trr X]\trr Y]$$
$$-[Y\trr [X\trr Z]]-[Z\trr [X\trr Y]]$$
$$=[[Y\trr X]\trr Z]+[[Z\trr X]\trr Y].$$
\item $[X\trr Y]\in Sym^{[0,1]}(L).$
\item Idem, together with the symmetrized definition of $N$.
\item
$$LHS=[X\trr YZ-[Y\circ Z]]=[X\trr Y]Z+Y[X\trr Z]-[X\trr [Y\circ Z]]$$
$$=N([X\trr Y]Z)+N(Y[X\trr Z])$$
$$+[[X\trr Y]\circ Z]+[Y\circ [X\trr Z]]-[X\trr [Y\circ Z]]=RHS.$$
\end{enumerate}
\end{pr}
\begin{rem}
\label{wickgenrem} I would like to draw particular attention to point
\ref{wickgeneralize} above: It is a generalization of the
Gaussian fact that $[X\trr N(YZ)]=N([X\trr YZ])$.
In the next section we will make ample use of
this formula to handle expressions like $N(\psi(z)\psi(z))$
A more familiar use of the formula is in Gaussian form:
It is then the essential statement for proving that in the Gaussian case
 $\langle N(X_1X_2)N(Y_1 Y_2)\rangle $, say, can be expanded in terms of
contractions between $X$'s and $Y$'s only; Indeed,
$$\langle N(X_1X_2)N(Y_1 Y_2)\rangle =\langle (X_1N(X_2)-[X_1\trr N(X_2)])N(Y_1
Y_2)\rangle $$
$$=\langle N(X_2)[X_1\trr N(Y_1 Y_2)]\rangle =\langle N(X_2)N([X_1\trr Y_1
Y_2])\rangle .$$
\end{rem}
\begin{thm} (Subgaussian reconstruction.) Let $L$ be finite dimensional
Lie algebra.
Let $\langle .\rangle :Sym(L)\rightarrow K$ be the solution of the Schwinger-
Dyson equation determined by a subgaussian contraction on $L$.
Then this contraction can be reconstructed from $\langle .\rangle $ if
$c_{ij}:=\langle T_i T_j\rangle $ is invertible ($T_i$ a basis for $L$), as
follows:
Let $c_{ijk}:=\langle T_i T_j T_k\rangle $, and let $[T_i,T_j]=:f_{ij}^k T_k$.
Then the contraction is given by $[T_i\trr T_j]=g_{ij}+\Gamma_{ij}^k T_k$,
where
$$g_{ij}:=c_{ij},$$
$$\Gamma_{ij}^k:=\frac{1}{2} g^{kl}(c_{ijl}+f_{ij}^m g_{ml}
-f_{il}^m g_{jm}-f_{jl}^m g_{im}).$$
\end{thm}
\begin{pr}
The proof is similar to the uniqueness proof of the Levi-Civit\`a
connection: We know the contraction
is subgaussian, so it is of the form $[T_i\trr T_j]=g_{ij}+\Gamma_{ij}^k T_k$,
and it remains to prove the above relations. Indeed,
$$c_{ij}=\langle T_i T_j\rangle =\langle [T_i\trr T_j]\rangle =\langle
g_{ij}+\Gamma_{ij}^c T_c\rangle =g_{ij}.$$
Further, we have
$$c_{ijk}=\langle [T_i\trr T_j]T_k\rangle +\langle T_j[T_i\trr T_k]\rangle =
\Gamma_{ij}^m c_{mk}+c_{jm}\Gamma_{ik}^m;$$
Using the fact that $\Gamma_{ij}^k-\Gamma_{ji}^k=f_{ij}^k$, since
$\Gamma_{ij}^kT_k-\Gamma_{ji}^kT_k=[T_i\trr T_j]-[T_j\trr T_i]=[T_i,T_j]$,
we arrive at
$$c_{ijk}+c_{jik}-c_{kij}
=\Gamma_{ij}^m c_{mk} + \Gamma_{ik}^m c_{jm}
+\Gamma_{ji}^m c_{mk} + \Gamma_{jk}^m c_{im}
-\Gamma_{ki}^m c_{mj} - \Gamma_{kj}^m c_{im}$$
$$=(\Gamma_{ij}^m + \Gamma_{ji}^m)c_{mk} + f_{ik}^m c_{jm} + f_{jk}^m c_{im}$$
$$=(2\Gamma_{ij}^m-f_{ij}^m)c_{mk}+ f_{ik}^m c_{jm} + f_{jk}^m c_{im}$$
So that
$$2\Gamma_{ij}^l g_{lk} = c_{ijk}+c_{jik}-c_{kij}+f_{ij}^m c_{mk}
-f_{ik}^m c_{jm} -f_{jk}^m c_{im}.$$
\end{pr}
\subsubsection{Left extensions of a subgaussian contraction.}
\begin{defn}
\label{left_ext_def}
Let $r\in Sym(L)$.
Then we say that a map $[r\trr .]:L\rightarrow Sym(L)$
is a left extension of the contraction to $r$, if it satisfies:
$$\forall_{X\in L}\;[X\trr r]-[r\trr X]\in L,$$
$$\forall_{X,Y\in L}\;[X\trr[r\trr Y]]-[r\trr[X\trr Y]]=
[[X\trr r]-[r\trr X]\trr Y].$$
In that case, we make we make $[r\trr.]$ act on $Sym(L)$ by
derivations.
The extension is said to be subgaussian iff $[r\trr X]\in K\oplus L$, and
$[X\trr r]\in K\oplus L$.
\end{defn}
\begin{thm}
If $[r\trr.]$ is a subgaussian left extension to $r$ then $r$ behaves in
a way similar to elements of $L$:
\begin{enumerate}
\item $$\forall_{X\in L,s\in Sym(L)}\;[X\trr[r\trr s]]-[r\trr[X\trr s]]=
[[X\trr r]-[r\trr X]\trr s].$$
\item If $\langle .\rangle $ satisfies the Schwinger-Dyson equation then:
$$\forall_{s\in Sym(L)}\;\langle rs\rangle =\langle r\rangle \langle s\rangle
+\langle [r\trr s]\rangle .$$
\item $$[r\trr N(YZ)]=N([r\trr YZ]+\frac{1}{2}[[Y\trr r]\trr Z]
+\frac{1}{2}[[Z\trr r]\trr Y]).$$
\end{enumerate}
\end{thm}
\begin{pr}
\begin{enumerate}
\item
Since both sides of the defining equation of left extensions concern
derivations acting on $Y$, we can replace $Y$ by $s$.
\item
We prove this by induction on $deg(s)$. If $s=1$ then
$$\langle rs\rangle =\langle r\rangle =\langle r\rangle \langle s\rangle
=\langle r\rangle \langle s\rangle +\langle [r\trr s]\rangle .$$
Next, assume the identity holds up to $deg(s)$. We will prove it for $Xs$:
$$\langle rXs\rangle =\langle [X\trr rs]\rangle =\langle [X\trr r]s\rangle
+\langle r[X\trr s]\rangle $$
Using that $[X\trr r]$ is in $K\oplus L$, and that $[X\trr s]$ has degree
smaller or equal to that of $s$:
$$=\langle [[X\trr r]\trr s]\rangle +\langle [X\trr r]\rangle \langle s\rangle
+\langle [r\trr [X\trr s]]\rangle +\langle r\rangle \langle [X\trr s]\rangle $$
$$=\langle [[X\trr r]\trr s]+[r\trr [X\trr s]]\rangle +\langle [X\trr r]\rangle
\langle s\rangle +\langle r\rangle \langle Xs\rangle .$$
Thus, using $\langle [X\trr r]\rangle =\langle [r\trr X]\rangle $, since
$[X\trr r]-[r\trr X]\in L$:
$$\langle rXs\rangle -\langle r\rangle \langle Xs\rangle =\langle [r\trr
X]\rangle \langle s\rangle +\langle [[X\trr r]\trr s]+[r\trr [X\trr s]]\rangle
$$
$$=\langle [r\trr X]\rangle \langle s\rangle +\langle [[r\trr X]\trr s]\rangle
+\langle [X\trr[r\trr s]]\rangle $$
$$=\langle [r\trr X]s+X[r\trr s]\rangle =\langle [r\trr Xs]\rangle .$$
\item The proof is identical to that of theorem \ref{subgaussthm} by replacing
$X$ by $r$.
\end{enumerate}
\end{pr}
\subsection{An example from conformal field theory: The Kac-Moody
algebra.}
\begin{thm} The following almost everywhere defined contraction
satisfies the pre-Lie property and
therefore defines a subgaussian contraction algebra:
The algebra is associated to a Lie algebra
$L$ with invariant symmetric bilinear form $g$, namely it is defined by
generating symbols
$J(X,z)$, linear in $X\in L$
\footnote{When we say that a symbol is linear in some argument, we mean
we impose an equivalence relation on the linear span of these symbols.
Thus for example $J(2X,z)\sim 2J(X,z)$, etc.}
, and where $z\in \bC$, with contraction:
$$[J(X,z)\trr J(Y,\zeta )]:=\frac{g(X,Y)}{(z-\zeta)^2}1 + \frac{J([X,Y],\zeta
)}{z-\zeta }.$$
\end{thm}
\begin{pr}
The pre-Lie property is equivalent to the statement that
$[[a\trr b]\trr c]+[b\trr[a\trr c]]$ is symmetric in $a$ and $b$. This
way of proving this property has an advantage over proving that
$[[a\trr b]\trr c]-[a\trr [b\trr c]]$ is symmetric in $a$ and $b$, since
the former way will automatically give an expression for the three-point
function:
$$\langle abc\rangle =\langle [a\trr b]c\rangle +\langle b[a\trr c]\rangle
=\langle [[a\trr b]\trr c]+[b\trr [a\trr c]]\rangle .$$
Following this remark, we first prove the following
\begin{lemma}
$$[[J(X_1,z_1)\trr J(X_2,z_2)]\trr J(X_3,z_3)]
+[J(X_2,z_2)\trr [J(X_1,z_1)\trr J(X_3,z_3)]]$$
$$=\frac{g([X_1,X_2],X_3)+J((z_1-z_3)[[X_1,X_2],X_3]+(z_1-z_2)[X_2,[X_1,X_3]],z_3)
}{(z_1-z_2)(z_1-z_3)(z_2-z_3)}.$$
\end{lemma}
\begin{pr}$LHS=$
$$[\frac{J([X_1,X_2],z_2)}{(z_1-z_2)}\trr J(X_3,z_3)]
+[J(X_2,z_2)\trr \frac{J([X_1,X_3],z_3)}{(z_1-z_3)}]=t(g)+t(J),$$
where $t(g)$ denotes the terms involving $g$, and $t(J)$ those with $J$.
$$t(g)=\frac{g([X_1,X_2],X_3)}{(z_1-z_2)(z_2-z_3)^2}$$
$$+\frac{g(X_2,[X_1,X_3])}{(z_1-z_3)(z_2-z_3)^2}
=\frac{g([X_1,X_2],X_3)}{(z_2-z_3)^2}(\frac{1}{z_1-z_2}-\frac{1}{z_1-z_3})$$
$$=\frac{g([X_1,X_2],X_3)}{(z_2-z_3)^2}
\frac{z_2-z_3}{(z_1-z_2)(z_1-z_3)}
=\frac{g([X_1,X_2],X_3)}{(z_1-z_2)(z_1-z_3)(z_2-z_3)},$$
and
$$t(J)=\frac{J([[X_1,X_2],X_3],z_3)}{(z_1-z_2)(z_2-z_3)}
+\frac{J([X_2,[X_1,X_3]],z_3)}{(z_1-z_3)(z_2-z_3)}$$
$$=\frac{J((z_1-z_3)[[X_1,X_2],X_3]+(z_1-z_2)[X_2,[X_1,X_3]],z_3)
}{(z_1-z_2)(z_1-z_3)(z_2-z_3)},$$
which proves the lemma.
\end{pr}
It remains to prove that the result of the lemma is symmetric under
the exchange of $1$ and $2$. This is clear for the term with $g$,
and for the $J$ term, it suffices to prove that
$$(z_1-z_3)[[X_1,X_2],X_3]
+(z_1-z_2)[X_2,[X_1,X_3]]+(1\leftrightarrow 2)=0.$$
Indeed, using the Jacobi identity:
$$LHS=\{(z_1-z_3)-(z_2-z_3)\}[[X_1,X_2],X_3]$$
$$+(z_1-z_2)\{[X_2,[X_1,X_3]]-[X_1,[X_2,X_3]]\}$$
$$=(z_1-z_2)[[X_1,X_2],X_3]
+(z_1-z_2)[[X_2,X_1],X_3]=0.$$
\end{pr}
\begin{rem}
In the same way, the reader may check that the Virasoro algebra also
defines a subgaussian contraction;
For $c\in \bR$, this is an algebra generated by
symbols $\partial^k T(z)$, where $k\in \bN$ and $z\in \bC$.
The contraction reads
$$[T(z)\trr T(\zeta)]:= \frac{c/2}{(z-\zeta)^4}1
+\frac{2 T(\zeta)}{(z-\zeta)^2}+\frac{\partial T(\zeta)}{(z-\zeta)},$$
together with $[\partial^k T(z)\trr \partial^l T(w)]:=\partial_z^k\partial_w^l
[T(z)\trr T(w)]$.
\end{rem}
\begin{defn}
\label{primary_def}
A module for a pre-Lie algebra is defined to be a module for the induced
Lie-algebra.
The reader may check that the following operations define modules for
the Virasoro and Kac-Moody algebras:
$$[T(z)\trr \phi(\zeta)]:=\frac{h.\phi(\zeta)}{(z-\zeta)^2}+\frac{\partial
\phi(\zeta)}{(z-\zeta)}.$$
$$[J(X,z)\trr\phi(v,\zeta)]:=\frac{J(Xv,\zeta)}{z-\zeta}.$$
Here, $h$ is some fixed number, $\zeta\in \bC$ and $v$ runs linearly over a
representation space
of the Lie algebra $L$.
In that case the symbol $\phi$ is called a primary field for $T$ or $J$,
and the number $h$ is called its weight.
\end{defn}
%
%
%
%
\newpage
\section{Examples of left extensions and contraction morphisms.}
\label{extendsection}
\label{bilinearsection}
%
%
%
%
%
%
%
%
\subsection{The construction of left extensions.}
It is in general difficult to find left exensions $[r\trr .]$ for a
contraction.

Here we will describe a method to guess $[r\trr X]$ from $[X\trr r]$
in special cases. The cases we are thinking about are those where the
Lie algebra has a basis of symbols $\phi_i(x)$ and their derivatives,
where $x$ runs over $\bR^D$, and a contraction of the form
$$[\phi_i(x)\trr \phi_j(y)]=c_{ij}^k (x-y) \phi_k(y).$$ (Or possibly
also involving derivatives in the right hand side).  From the fact
that $\langle \phi_i(x)\phi_j(y) ..\rangle =\langle
\phi_j(y)\phi_i(x)..\rangle $ one concludes that
$[\phi_i(x)\trr\phi_j(y)]-[\phi_j(y)\trr\phi_i(x)]$ is regular in
$(x-y)$, since the only singularities in $(x-y)$ in the expectation
value $\langle \phi_i(x)\phi_j(y) ..\rangle $ come from
$[\phi_i(x)\trr\phi_j(y)]$.  If on top of that we know that
$c_{ij}^k(x-y)$'s are singular algebraic functions of $(x-y)$, then
using symbolic Taylor expansions like
$\phi(x+h)=\phi(x)+h^i(\partial_i\phi)(x)+..$, one can determine
$[\phi_i(x)\trr\phi_j(y)]$ from $[\phi_j(y)\trr \phi_i(x)]$: Indeed,
modulo regular terms, we then have $$[\phi_i(x)\trr
\phi_j(y)]=[\phi_j(y)\trr \phi_i(x)] =c_{ji}^k(y-x) \phi_k(x)$$
$$=c_{ji}^k(y-x) \sum_{n=0}^\infty \frac{1}{n!}[(x^i-y^i)\partial_i]^n
\phi_k(y).$$ Thus from this we can read off $[\phi_i(x)\trr
\phi_j(y)]$ modulo regular terms. But since we know that this
contraction has only singular coefficients, the contraction is
determined. We can use the same rules to produce a candidate for
$[r\trr X]$ from $[X\trr r]$ if $r$ is some higher order element in
$Sym(L)$. There is however no guarantee that this will actually give a
left extension in the sense of definition \ref{left_ext_def}.  We will
give examples from 2D holomorphic field theory, where the procedure
actually gives an extension. It leads e.g. to an embedding of the
Kac-Moody algebra in the $Sym(L)$ of free fermion fields.
\begin{rem} A number of remarks seem to be in place to avoid loss of time
for the reader:
\begin{enumerate}
\item
It seems that the above procedure does not have an analogue in an
arbitrary subgaussian contraction algebra, since even if we start out
with a contraction which does not involve the sign $\partial$, then
still the left-extended contraction may involve that sign. (See the
Sugawara construction below.)
\item
Unfortunately, the procedure that we use to produce left extensions in
certain 2D contraction algebras, does not seem useful in higher
dimension: Although we can still apply the same method, I haven't
found an example that satisfies the properties of \ref{left_ext_def}.
This is very reminiscent of the work by Johnson and Low, who in the
operator language found that anomalous current ``commutation
relations'' do not satisfy the Jacobi identity in higher dimensons.
Here one can see the advantage of the contraction language over the
operator language, since there is no inconsistency in the sole fact of
not being able to find left extensions of a contraction.  (It does
however remain a \underline{challenge} to actually find one.  Also it
would be interesting to see any subgaussian contraction in higher
dimensions, apart from the Gaussian ones.)
\end{enumerate}
\end{rem}
\subsection{The Jordan and Sugawara constructions.}
We will here repeat some relatively old constructions,
using on the one hand the
two-dimensionality which provides a way to construct the left
extensions, and some general formulae for subgaussian contraction
algebras.  There is nothing original about the constructions
themselves: We only wish to comunicate the use of subgaussian formulae.
\subsubsection{Jordan-Tomonaga-Coleman-Gross-Jackiw: $J\mapsto N(\psi\psi)$ in
$2D$.}
\begin{thm}
\label{jordan}
Let $(V,g)$ be a finite-dimensional orthogonal representation space
for a Lie algebra $L$. Set $g_V(X,Y):=Tr_V(XY)/2$. Let $F_V$ be the
contraction algebra on odd generators $\psi(v,z)$ and Gaussian
contraction $[\psi(v,z)\trr \psi(w,\zeta)]:=\frac{g(v,w)}{z-\zeta}$.
Then we have a map $Kac-Moody(L,g_V)\rightarrow F_V$, as follows
\footnote{This needs some explanation since $[\psi(z)\trr \psi(z)]$ is
undefined. We will not use any specific value for this contraction
however: We will only use formula \ref{wickgeneralize} of theorem
\ref{subgaussthm}.  This is really a theorem about pre Lie algebras
$P$ if we define normal ordering to be $Sym(P)\rightarrow Sym(P)$.
Thus, to define the above calculus with undefined contractions, we go
to the universal pre Lie algebra on symbols $\psi^\alpha(z), 1$, and
impose the relation $[\psi(z)\trr\psi(w)]=1/(z-w)$ \underline{only}
for $z\neq w$. Thus, $[\psi(z)\trr\psi(z)]$ remains a symbol.  }:
$$J(X,z)\mapsto\frac{1}{2} X_{\alpha\beta}
N(\psi(e^\alpha,z)\psi(e^\beta,z)).$$ It satisfies:
\begin{enumerate}
\item $$[\psi(v,z)\trr J(X,\zeta )]=-\frac{\psi(Xv,\zeta )}{z-\zeta }.$$
\item $$[J(X,z)\trr \psi(v,\zeta )]=\frac{\psi(Xv,\zeta )}{z-\zeta }.$$
\item $$[J(X,z)\trr J(Y,\zeta )]=\frac{Tr(X Y)}{2(z-\zeta)^2} +
\frac{J([X,Y],\zeta )}{z-\zeta }.$$
\end{enumerate}
\end{thm}
\begin{pr}
\begin{enumerate}
\item We use the notation $Xe_\alpha={X^\beta}_\alpha e_\beta$.
Then $X_{\alpha\beta}=-X_{\beta\alpha}$ by orthogonality. Further,
since $\psi$ is odd, we get a number of extra minus signs that we did
not include in our previous discussion. See remark \ref{sup_lazy_rem}.
Using the formulas for Gaussian contractions: $$LHS=\frac{1}{2}
X_{\alpha\beta} N([\psi(v,z)\trr
\psi(e^\alpha,\zeta)]\psi(e^\beta,\zeta))$$ $$-\frac{1}{2}
X_{\alpha\beta} N(\psi(e^\alpha,\zeta) [\psi(v,z)\trr
\psi(e^\beta,\zeta)])$$ $$=\frac{1}{2} X_{\alpha\beta}\{\frac{v^\alpha}{
z-\zeta}\psi(e^\beta,\zeta)-\psi(e^\alpha,\zeta)\frac{v^\beta}{
z-\zeta}\}$$ $$=\frac{1}{2(z-\zeta)}\{-\psi(v^\alpha X_{\beta\alpha}
e^\beta,\zeta)-\psi(v^\beta X_{\alpha\beta} e^\alpha,\zeta)\}$$
$$=\frac{-1}{z-\zeta}\psi(v^\alpha {X^\beta}_\alpha
e_\beta,\zeta)=\frac{-1}{z-\zeta} \psi(v^\alpha X(e_\alpha),\zeta)
=\frac{-\psi(Xv,\zeta)}{z-\zeta}.$$
\item This is where we construct a left extension, using the rules that
we explained before, i.e. we do the calculation modulo regular terms,
and require commutators to be regular.  $$LHS=[\psi(v,\zeta)\trr
J(X,z)]=\frac{-\psi(Xv,z)}{\zeta-z} =\frac{-\psi(Xv,\zeta)}{\zeta -
z}=RHS.$$ We ask the reader to check for himself that this really
gives a left extension in the sense we defined it.
\item Here we will use (the super version of) theorem \ref{subgaussthm}:
$$LHS=[J(X,z)\trr N(\frac{1}{2}Y_{\alpha\beta}
\psi^\alpha(\zeta)\psi^\beta(\zeta))]$$ $$=N(\frac{1}{2}
Y_{\alpha\beta} [J(X,z)\trr \psi^\alpha(\zeta)]\psi^\beta(\zeta))
+N(\frac{1}{2} Y_{\alpha\beta} \psi^\alpha(\zeta)[J(X,z)\trr
\psi^\beta(\zeta)])$$ $$-N(\frac{1}{4} Y_{\alpha\beta}
[[\psi^\beta(\zeta)\trr J(X,z)]\trr \psi^\alpha(\zeta)]) +N(\frac{1}{4}
Y_{\alpha\beta} [[\psi^\alpha(\zeta)\trr J(X,z)]\trr
\psi^\beta(\zeta)])$$ $$=N(\frac{1}{2} Y_{\alpha\beta}
\frac{\psi(Xe^\alpha,\zeta)}{z-\zeta} \psi^\beta(\zeta)) +N(\frac{1}{2}
Y_{\alpha\beta} \psi^\alpha(\zeta) \frac{\psi(X e^\beta,\zeta)}{
z-\zeta})$$ $$-N(\frac{1}{4} Y_{\alpha\beta} [-\frac{\psi(Xe^\beta,z)}{
\zeta-z}\trr \psi(e^\alpha,\zeta)]) +N(\frac{1}{4} Y_{\alpha\beta}
[-\frac{\psi(Xe^\alpha,z)}{\zeta-z}\trr\psi(e^\beta,\zeta)])$$
$$=\frac{Y_{\alpha\beta}{X_\gamma}^\alpha}{2(z-\zeta)}
N(\psi^\gamma(\zeta)\psi^\beta(\zeta)) +\frac{Y_{\alpha\beta}
{X_\gamma}^\beta}{2(z-\zeta)} N(\psi^\alpha(\zeta)
\psi^\gamma(\zeta))$$ $$+\frac{Y_{\alpha\beta}}{
4(\zeta-z)}\frac{(Xe^\beta)^\alpha}{z-\zeta} -\frac{Y_{\alpha\beta}}{
4(\zeta-z)} \frac{(Xe^\alpha)^\beta}{z-\zeta}$$ Now $Y_{\alpha\beta}
{X_\gamma}^\alpha=(XY)_{\gamma\beta}$, and $Y_{\alpha\beta}
{X_\gamma}^\beta={Y_\alpha}^\beta X_{\gamma\beta}=-{Y_\alpha}^\beta
X_{\beta\gamma}=-(YX)_{\alpha\gamma}$, so we get
$$=\frac{(XY)_{\gamma\beta}}{2(z-\zeta)}
N(\psi^\gamma(\zeta)\psi^\beta(\zeta)) -\frac{(YX)_{\alpha\gamma}}{
2(z-\zeta)}N(\psi^\alpha(\zeta)\psi^\gamma(\zeta))$$
$$-\frac{Y_{\alpha\beta}}{
4(z-\zeta)^2}(X^{\alpha\beta}-X^{\beta\alpha})=RHS.$$
\end{enumerate}
\end{pr}
\subsubsection{Sugawara-Coleman-Gross-Jackiw: Nonabelian case of $T\mapsto
N(JJ)$.}
\begin{thm} Let $(L,g)$ be a finite dimensional reductive Lie algebra, with
invariant metric, such that $ad(T_a T^a)=2 c^\vee id_L$.  Consider the
following map from the Virasoro algebra to the Kac-Moody algebra of
$(L,kg)$: $$T(z):=\frac{1}{2k+2c^\vee} N(J(T^a,z)J(T_a,z)).$$ It
satisfies:
\begin{enumerate}
\item
$$[J(X,z)\trr T(\zeta)]= \frac{J(X,\zeta)}{(z - \zeta)^2}.$$
\item
$$[T(z)\trr J(X,\zeta)]=\frac{J(X,\zeta)}{(z-\zeta)^2} +\frac{\partial
J(X,\zeta)}{z-\zeta}.$$
\item
$$[T(z)\trr T(\zeta)]= \frac{k |L|}{k+c^\vee }\frac{1/2}{(z-\zeta)^4}
+\frac{2 T(\zeta)}{(z-\zeta)^2}+\frac{\partial T(\zeta)}{(z-\zeta)}.$$
\end{enumerate}
\end{thm}
\begin{pr} Making use of $ad(T_a T^a)=2c^\vee  id_L$ and
$$[T_a,T^b]^c T_b=g(T^c,[T_a,T^b])T_b=g([T^c,T_a],T^b)T_b=[T^c,T_a],$$
gives:
\begin{enumerate}
\item
$$(2k+2c^\vee )[J_a(z)\trr T(\zeta)]=[J_a(z)\trr N(J^b(\zeta) J_b(\zeta) )]
=N([J_a(z)\trr J^b(\zeta)]J_b(\zeta)$$
$$+J^b(\zeta)[J_a(z)\trr J_b(\zeta)])
+\frac{1}{2} N([[J^b(\zeta)\trr J_a(z)]\trr J_b(\zeta)]+[[J_b(\zeta)\trr
J_a(z)]\trr J^b(\zeta)])$$
$$=N(2\{\frac{kg_{ab}}{(z-\zeta)^2} +
\frac{J([T_a,T_b],\zeta)}{z-\zeta}\}J^b(\zeta)
+[\frac{J([T_b,T_a],z)}{\zeta-z}\trr J^b (\zeta )])$$
$=\frac{2 k J_a(\zeta)}{(z-\zeta)^2} + A + B,$
where
$$(z-\zeta)A=2N(J([T_a,T_b],\zeta)J^b(\zeta))=2N(J(T_c,\zeta)J([T_a,T_b]^c
T^b,\zeta))$$
$$=2N(J_c(\zeta)J([T^c,T_a],\zeta))=-A(z-\zeta)=0,$$
and
$$B=\frac{1}{z-\zeta} \{\frac{g([T_a,T_b],T^b)}{(z-\zeta)^2}
+\frac{J([[T_a,T_b],T^b],\zeta)}{z-\zeta}\}=0+\frac{2c^\vee
J(T_a,\zeta)}{(z-\zeta)^2}.$$
This gives the right hand side.
\item Calculus modulo regular terms, with $X=T_a$:
$$LHS=[J_a(\zeta)\trr T(z)]=\frac{J_a(z)}{(z-\zeta)^2}
=\frac{J_a(\zeta)+(z-\zeta)\partial J_a(\zeta)}{(z-\zeta)^2}=RHS.$$
\item
$$[T(z)\trr N(J_a(\zeta)J^a(\zeta))]=
N(2[T(z)\trr J_a(\zeta)]J^a(\zeta)+[[J_a(\zeta)\trr T(z)]\trr J^a(\zeta)])$$
$$=N(\frac{2J_a(\zeta)J^a(\zeta)}{(z-\zeta)^2}
+\frac{2\partial J_a(\zeta)J^a(\zeta)}{z-\zeta}
+[\frac{J_a(z)}{(\zeta-z)^2}\trr J^a(\zeta)])$$
$$=\frac{2}{(z-\zeta)^2} N(J_a(\zeta) J^a(\zeta))
+\frac{1}{z-\zeta} \partial N(J_a(\zeta)J^a(\zeta))$$
$$+\frac{1}{(\zeta - z)^2}\{\frac{k g_{ab}g^{ab}}{(z-\zeta)^2}
+\frac{J([T^a,T_a],\zeta)}{z-\zeta}\}$$
$$=(2k+2c^\vee )\{\frac{2T(\zeta)}{(z-\zeta)^2} + \frac{\partial
T(\zeta)}{z-\zeta}\}
+\frac{k|L|}{(z-\zeta)^4.}$$
\end{enumerate}
\end{pr}
\begin{hist}
\label{hist_foot}
This type of extensions has a number of original motivations: First,
in four-dimensional field theory of the early 1930's
scientists following de Broglie \cite[formula 2]{broglie} were trying
to see if the photon field could be described as bilinear in the field
of the newly discovered neutrino. And second, scientists tried to
describe many-electron systems in terms of a formal ``soundwave''
bilinear in the electron field \cite[formula 16]{bloch}.
P.Jordan gave a number of derivations that bilinear combinations of the
neutrino field in two dimensions give operators which satisfy
canonical bosonic commutation relations
\cite[formulas 6,15]{jordan}.
However at that time the normal ordering procedure had not been
invented, and Jordan's claim that the bilinear construction was
possible in two dimensions was not universally accepted, see
e.g. the articles by V. Fock \cite{fock} on the subject,
who advocated the a priori impossiblilty of the construction. Gaussian
normal ordering was introduced in $1949$, in \cite{houriet}.  In
$1950$, Tomonaga \cite{tomonaga} gave a new derivation of the bilinear
construction without normal ordering, but trying to account for
a number of steps by considering them as good approximations.  New
problems with bilinear constructions were found by Johnson and Low
\cite[section 6]{johnson} in 1966, namely that in higher dimensions,
the equal-time ``commutation relations'' of composite operators do not
satisfy the Jacobi identity. (Which by the way means that they can't
be commutation relations of operators. We will see that in the context
of contraction algebras, something similar happens which however is
not logically fatal like the problems with the operator language.)
In the 1960's it became fashionable try
to describe elementary particles only by the currents $J$ of their
conserved quantities: Sugawara \cite[formula 8]{sugawara}
suggested in 1968 to generalize the Euler relation $T_{\mu\nu}:=J_\mu
J_\nu-\frac{1}{2}g_{\mu \nu} J_\rho J^\rho$ to the non-Abelian and
quantum mechanical case, in four
dimensions, and without including normal ordering or any regularization
in his discussion. Coleman, Gross and Jackiw \cite{coleman}
subsequently discovered some paradoxes, but showed that
they disappear when the construction is regularized.  They also
discovered that Sugawara's construction was better behaved in two
dimensions than in four. In 1984, an analysis of the Sugawara
construction for any reductive Lie algebra was given by Knizhnik and
Zamolodchikov \cite{knizhnik.zam}.
\end{hist}
Here are some more contraction morphisms which the reader may check as an
exercise:
\begin{enumerate}
\item Starting from the free fermions of theorem \ref{jordan},
set
$$T(z):=\frac{1}{2}g_{\alpha\beta}
N(\partial \psi^\alpha(z)\psi^\beta(z)).$$
 This defines a morphism from
Virasoro with $c=|V|/2$.
\item Starting with the semidirect product of Kac-Moody
with Virasoro and for $Q\in L$, set
$$\tilde T(z):=T(z)+\partial J(Q,z).$$
Find conditions under which $\tilde T$ is itself of Virasoro type.
\item Starting with $[b(z)\trr c(w)]:=-[c(w)\trr b(z)]:=1/(z-w)$,
and $[b\trr b]:=[c\trr c]:=0$, set
$$T(z):=N\{p(\partial b)(z)c(z)+qb(z)\partial c(z)\}.$$
Check that this gives a Virasoro algebra iff $p-q=1$, that in that
case we have $c=-2(p^2 + q^2 + 4pq)$, $h_b=-q$ and $h_c=p$, where
$h$ is the number occuring in the definition \ref{primary_def}.
\end{enumerate}
\subsection{Remarks on higher dimensional left extensions.}
\begin{defn} We define a Gaussian contraction on odd symbols
$\{\psi_{A\alpha}(x)\}$, where $x\in \bR^D$,  $\{e_A\}$ and $\{e_\alpha\}$ are
bases of vectorspaces, as follows\footnote{Beware that for $D\neq 2$ this is
not the contraction induced by the Dirac
action in $D$ dimensions since that one is proportional to
$(\gamma_\mu)_{AB} \frac{x^\mu-y^\mu}{|x-y|^D}.$
We use this contraction only as an illustration because it is easier
to handle.}:
$$[\psi_{A\alpha}(x)\trr \psi_{B\beta}(y)]:=
(\gamma_\mu)_{AB} k_{\alpha\beta} f^\mu(x,y):=
(\gamma_\mu)_{AB} k_{\alpha\beta}
\frac{x^\mu-y^\mu}{|x-y|^2}
.$$
Next, for $\Gamma$ in the Clifford algebra of $\bR^D$, and matrix $X$, we set:
$$J(\Gamma\otimes X,x):={1\over 2} X^{\alpha\beta} \Gamma^{AB}
N(\psi_{A\alpha}(x)\psi_{B\beta}(x)).$$
\end{defn}
\begin{thm}
$J$ satisfies the following properties, where $A^T$ denotes the
transpose of $A$.
\begin{enumerate}
\item
$$J(\Gamma^T\otimes X^T,x)=-J(\Gamma\otimes X,x).$$
\item
$$[\psi(s\otimes v,y)\trr J(\Gamma\otimes X,s)]
=\frac{1}{2} f_\mu(y,x)
(\psi(\Gamma^T\gamma^\mu s\otimes X^T v,x)-\psi(\Gamma\gamma^\mu s \otimes
Xv,x)).$$
\item
$$[J(\Gamma\otimes X,x)\trr\psi(s\otimes v,y)]
=\frac{1}{2} f_\mu(x,y)
(\psi(\Gamma\gamma^\mu s\otimes X v,y)-\psi(\Gamma^T\gamma^\mu s \otimes
X^Tv,y)).$$
\item
$$[J(\Gamma\otimes X,x)\trr J(\Delta\otimes Y,y)]
=\langle J(\Gamma\otimes X,x) J(\Delta\otimes Y,y)\rangle
+\frac{1}{2} f_\mu(x,y)\times$$
$$J(\Gamma \gamma^\mu \Delta\otimes XY-\Gamma^T \gamma^\mu \Delta\otimes X^T Y
-\Gamma\gamma^\mu \Delta^T\otimes XY^T+\Gamma^T\gamma^\mu\Delta^T\otimes
X^TY^T,y).$$
\end{enumerate}
\end{thm}
\begin{rem}
The above is is an illustration of what a subgaussian contraction in
higher dimensions might have looked like if it really satisfied the
pre Lie property, but it doesn't. Note that $J$ here is not really the
Noether current corresponding to the given contraction, but it is
my experience that even if one restricts to Noether currents, the Pre-Lie
property is not satisfied.
\end{rem}
%
%
%
%
\newpage
\section{An existence theorem.}
\label{existsection}
\label{exist}
In this section we will be concerned with the solution of the
Schwinger-Dyson equation in the polynomial nonabelian case. Our first aim will
be to prove that just like in the abelian case, the solution is
necessarily given by $ZN^{-1}$. I.e. we start with uniqueness, which is easy:
\begin{thm}(Uniqueness). Let $I$ satisfy the Schwinger-Dyson equation.
Then $\forall_{s\in Sym(L)} IN(s)=Z(s)I(1)$. So if $N$ is invertible, then
$I=I(1).ZN^{-1}$.
\end{thm}
\begin{pr}
For $deg(s)=0$, this equation reads $sI(1)=sI(1)$, and for higher degree, we
have:
$$IN(X_1..X_n)=\frac{1}{n} \sum_{i=1}^n I(X_i N(X_{[1,n]\backslash i})-
[X_i\trr N(X_{[1,n]\backslash i})])=0=Z(X_1..X_n).$$
\end{pr}
\begin{rem}
Thus, to prove existence it suffices to prove that $ZN^{-1}$ satisfies the
Schwinger-Dyson equation, i.e. we have to prove that
$ZN^{-1}(Xs-[X\trr s])=0$. Let us start with the easiest case in order
to see what exactly the difficulties are: The case $s=Y\in L$.
Then we have to prove that $ZN^{-1}(XY-[X\trr Y])=0$. Now
$$XY-[X\trr Y]=N(XY)+\frac{1}{2}[X\trr Y]+\frac{1}{2}[Y\trr X]
-[X\trr Y]=N(XY+\frac{1}{2}[Y,X]),$$
So that indeed $Z N^{-1}(XY-[X\trr Y])=0$. Now we have to try to generalize
this procedure to arbirary $s$. Our first step will be to prove the
formula for $N(s)$ instead of $s$, which is the same if $N$ is invertible:
I.e. we will prove
$ZN^{-1}(XN(s)-[X\trr N(s)])=0$. This version is better suited for
proof by induction, in view of the definition of normal ordering.
So the question is: Given $X$ and $s$, can
we construct $R(X,s)$ such that $XN(s)-[X\trr N(s)]=N(R)$ and $Z(R)=0$?
Now in the
Abelian case this was easy for we could take $R(X,s):=Xs$, however we are in
a more complicated situation now because of the symmetrization in the
definition of $N$: What we saw above is that
$R(X,Y)=XY+[Y,X]/2$. It turns out that it is possible to find such an $R$ for
all $s$ in
the nonabelian case, which is the main result of theorem \ref{uec_ident}.
We will also be proving a number of extra identities that are useful
for a slightly generalized case of the Schwinger-Dyson equation, namely
where boundary terms are not assumed to be zero, but instead are assumed
to be given by some presecribed (possibly nonzero) map $J$.
\end{rem}

\begin{rem}
We will be doing calculations in $Sym(L)$ throughout, but note that
the calculations involving contractions apply in a weakened form
to the nonpolynomial contractions of appendix \ref{nonpolsection}
if we allow $N$ and $\nu$ to take values in the universal
contraction algebra $UEC(L)$ defined in that appendix.
\end{rem}
\subsection{Preliminaries on the symmetric algebra of a Lie algebra $L$.}
\begin{rem} We will be defining maps on $Sym(L)$ without going through
the explicit symmetrization every time. To that end we include the
following theorem. It is, say, the statement that in symmetric algebras
every element of $Sym^n(V)$ can be written as a sum of $n$-th powers, for
example $2XY=(X+Y)^2-X^2-Y^2.$
This will simplify matters when proving that
$ZN^{-1}(XN(s)-[X\trr N(s)])=0$, because we will only prove that
$ZN^{-1}(XN(Y^n)-[X\trr N(Y^n)])=0$, which as we will see is easier.
\end{rem}
\begin{thm} (Polarization.)
Let $V,W$ be vectorspaces, $G:V^{\otimes n}\rightarrow W$ linear,
then there is a
unique linear map $G_s:Sym^n(V)\rightarrow W$ such that
$\forall_{v\in V} G_s(v^n)=G(v^n)$.
\end{thm}
\begin{pr}
Existence is evident from the following example:
$$G_s(X_1..X_n):=\frac{1}{n!} \sum_{\sigma\in S_n}
G(X_{\sigma(1)},..,X_{\sigma(n)}).$$
Next, we have the following formula in $Sym(V)$:
$$n! X_1...X_n = \sum_{S\subset \{1,...,n\}} (-1)^{n-|S|}
(\sum_{s\in S} X_s)^n,$$
which is proved by noting that both sides are symmetric and homogeneous
polynomials which can be divided by $X_1$, so that both sides are
equal up to scalar multiplication. To determine this factor, we take
$X_1=X_2=..=X_n$, and use that $\sum_{k=0}^n (-1)^{n-k} k^n
{n\choose r}=n!$
This gives uniqueness, since
$$G_s(X_1..X_n)=
\frac{1}{n!} \sum_{S\subset \{1,...,n\}} (-1)^{n-|S|}
G_s(\sum_{s\in S} X_s)^n$$
$$=\frac{1}{n!} \sum_{S\subset \{1,...,n\}} (-1)^{n-|S|}
G(\sum_{s\in S} X_s)^n$$.
\end{pr}
\begin{defn}
\label{sym_def}
Let $L$ be a Lie algebra.
Define the following maps:
\begin{enumerate}
\item $Z:Sym(L)\rightarrow K\leq Sym(L)$, ``Zero degree projection''.
\item  $M:L\otimes Sym(L)\rightarrow Sym(L);$ ``Multiply'':
$$ M(X\otimes s):=Xs.$$
\item $S: Sym(L)\rightarrow L\otimes Sym(L);$ ``Split'':
$$S(1):=0;\;S(X_1...X_n):=\frac{1}{n}\sum_{i=1}^n X_i\otimes X_{[1,n]
\backslash i}.$$
\item  $\Sigma:L\otimes Sym(L)\rightarrow L\otimes Sym(L);$ ``Symmetrize'':
$$\Sigma(X_0\otimes X_1..X_n):=\frac{1}{(n+1)!}\sum_{\sigma\in S_{n+1}}
X_{\sigma(0)}\otimes X_{\sigma(1)}..X_{\sigma(n)}.$$
\item  $r:=\oplus_n r_n: L\otimes Sym^n(L)\rightarrow Sym^{[1,n]}(L);$ ``Rest
term'',
and  $C:=\oplus_n C_n:L\otimes Sym^n(L)\rightarrow L\otimes Sym^{[0,n-1]}(L)$
``Commutator term'',
inductively as follows:
\footnote{The motivation for the definition of $r$ comes from the proof
theorem \ref{uec_ident}: It is chosen in such a way the equality
referred to in footnote \ref{rdefmot} holds. The definiton of $C$ is
useful since we then have $r=(M+r)C$, as proved in the next theorem.}
$$r(X,1):=0;\;C(X,1):=0,$$
$$C(X,Y^{n+1}):=\frac{n+1}{n+2}\{ [Y,X]\otimes Y^n + Y\otimes r(X,Y^n)\}.$$
$$r(X,Y^{n+1}):=
\frac{n+1}{n+2}\{ [Y,X]Y^n + Yr(X,Y^n)+r([Y,X],Y^n)+r(Y,r(X,Y^n))\}.$$
\item  $M_\lambda:=M+\lambda r$;``Modified multiplication''.
\item  $\pi_\lambda:=SM_\lambda$, ``Projection''.
\footnote{This map
with $\lambda=1$ is used in theorem \ref{solve_boundary}.
In that theorem a condition of the form $J=J\pi$ appears, which motivates
us to try to prove that $\pi^2=\pi$. This is indeed the case, as is
demonstrated
in the next theorem.}
\end{enumerate}
\end{defn}
\begin{thm} These maps satisfy the following properties:
\label{sym_alg_ident}
\begin{enumerate}
\item $M_\lambda=M+\lambda r$, $\pi_\lambda=SM_\lambda$.
\item $r=(M+r)C$.
\item $\Sigma^2=\Sigma$.
\item $\Sigma=SM=\pi_0$.
\item $Zr=ZM=0$.
\item $CS=0$, $rS=0$.
\item $\pi_\lambda \Sigma=\Sigma$.
\item $\pi_\lambda=\lambda \Sigma+\pi_\lambda C$.
\item $\pi_\lambda^2=\pi_\lambda$.
\end{enumerate}
\end{thm}
\begin{pr}
\begin{enumerate}
\item By definition.
\item
$$r(X,Y^{n+1})=
\frac{n+1}{n+2}\{ [Y,X]Y^n + Yr(X,Y^n)+r([Y,X],Y^n)+r(Y,r(X,Y^n))\}$$
$$=\frac{n+1}{n+2} (M+r)\{ [Y,X]\otimes Y^n + Y\otimes r(X,Y^n)\}
=(M+r)C(X,Y^{n+1}).$$
\item Left to the reader.
\item By polarization it suffices to prove that
$\Sigma(X\otimes Y^n)=SM(X\otimes Y^n)$. Indeed:
$$LHS=\frac{1}{(n+1)} (X\otimes Y^n +n Y\otimes X Y^{n-1})
=S(XY^n)=RHS.$$
\item $ZM(X\otimes s)=Z(Xs)=0$. Next, we prove by induction on $|s|$ that
$Zr(X,s)=0$. Indeed, $Zr(X,1)=Z(0)=0$; Suppose that the identity holds up to
degree $n$. Then we have:
$$Zr(X,Y^{n+1})=Z(M+r)C(X,Y^{n+1})=(Zr)C(X,Y^{n+1})=0,$$
by induction since $C$ lowers degree.
\item We will prove by induction on $|s|$ that $CS(s)=0$ and $rS(s)=0$.
Indeed $CS(1)=C(0)=0$, $rS(1)=r(0)=0$, $CS(X)=C(X\otimes 1)=0$, and
$rS(X)=r(X\otimes 1)=0$. So assume these identities
hold up to degree $n+1$. Then:
$$CS(X^{n+2})=C(X\otimes X^{n+1})=\frac{n+1}{n+2} X\otimes r(X,X^n)
=\frac{n+1}{n+2} X\otimes rS(X^{n+1})=0,$$
and
$$rS(X^{n+2})=(M+r)CS(X^{n+2})=0.$$
\item $\pi_\lambda \Sigma = S(M+\lambda r)\Sigma=\Sigma^2 +\lambda SrSM
=\Sigma^2=\Sigma$.
\item $\pi_\lambda=S(M+\lambda r)=SM+\lambda S(M+r)C=\lambda SM+S(M+\lambda r)C
=\lambda \Sigma +\pi_\lambda C$.
\item We will prove by induction on $|s|$ that $\pi_\lambda^2(X\otimes s)
=\pi_\lambda(X\otimes s)$.
Indeed, $\pi_\lambda(X\otimes 1)=X\otimes 1=\pi_\lambda^2(X\otimes 1)$.
Next, assume that the identity holds up to degree $n$, and that $|s|=n+1$.
Then since $C$ lowers degree, we have
$\pi_\lambda^2 C(X\otimes s)=\pi_\lambda C(X\otimes s)$, so that:
$$\pi_\lambda^2(X\otimes s)=\pi_\lambda(\lambda \Sigma +\pi_\lambda C)
(X\otimes s)
=\lambda \pi_\lambda \Sigma (X\otimes s)+\pi_\lambda^2 C(X\otimes s)$$
$$=\lambda \Sigma(X\otimes s)+\pi_\lambda C(X\otimes s)
=(\lambda \Sigma +\pi_\lambda C)(X\otimes s)=\pi_\lambda (X\otimes s).$$
\end{enumerate}
\end{pr}
\subsection{Theorems involving contractions.}
\begin{defn} Given a polynomial contraction on $L$,
we define the following maps:
$N:Sym(L) \rightarrow Sym(L)$, ``Normal ordering'', and
$\nu: L\otimes Sym(L)\rightarrow Sym(L)$,
``Greek normal ordering'', inductively as follows:
$$N(1):=1;\;N(X_1...X_n):=\frac{1}{n} \sum_{i=1}^n X_i N(X_{[1,n]\backslash
i})-
[X_i\trr N(X_{[1,n]\backslash i})].$$
$$\nu(X\otimes s):= XN(s)-[X\trr N(s)].$$
\end{defn}
%
%
%
%
%
%
%
\begin{rem} Let us explain the idea underlying the following theorem.
Recall that we want to prove that there is an $R$ such that
$\nu(X\otimes s)=N(R(X\otimes s))$ and $ZR=0$. In other words, we have
to rewrite $\nu(X\otimes Y^n)$ as the $N$ of something, which we are going
to try by induction on $n$. So given the fact that there is an $r(X,Y^n)$
such that
$$\nu(X,Y^n)=N(XY^n+r(X,Y^n)),$$
we want to construct $r(X,Y^{n+1})$ such that it satisfies this
equation with $n$ replaced by $n+1$. In the course lemma \ref{uec_lemma} we
arrive
at
$$\nu(X,Y^{n+1})=
\nu(Y,XY^n)+N(Yr(X,Y^n)+r(Y,r(X,Y^n))+[Y,X]Y^n+r([Y,X],Y^n)).$$
So at that stage we have expressed $\nu(X,Y^{n+1})$ in terms
of the $N$ of something and $\nu(Y,XY^n)$. So it remains to express
$\nu(Y,XY^n)$ in terms of $N(..)$ and $\nu(X,Y^{n+1})$ in an independent
way, so that we get two equations
$$a\nu(X,Y^{n+1})+b\nu(Y,XY^n)=N(..),$$
$$b\nu(X,Y^{n+1})+d\nu(Y,XY^n)=N(..).$$
Which we may solve for $\nu(X,Y^{n+1})$. This second equation is
fournished by point \ref{uec_sec_eqn} of the theorem,
which makes the proof possible.
The map $R=M+r$ thus defined indeed satisfies $ZR=0$, by definition.
\end{rem}
\begin{thm}These maps satisfy the following properties:
\label{uec_ident}
\begin{enumerate}
\item $N=\nu S + Z.$
\item $\nu(Y,XY^n)=\frac{-1}{n+1}\nu(X,Y^{n+1})+N(\frac{n+2}{n+1}XY^{n+1}).$
\label{uec_sec_eqn}
\item $\nu(X,Y^{n+1})=Y.\nu(X,Y^n)-[Y\trr\nu(X,Y^n)]+\nu([Y,X],Y^n)$.
\item $\nu=N(M+r).$ (This is the main result.)
\label{main_uec_ident}
\end{enumerate}
\end{thm}
\begin{pr}
\begin{enumerate}
\item Since $S(1)=0$, we have $N(1)=(\nu S+Z)(1)$, and for higher degree,
we have to check $N(X_1..X_n)=\nu S(X_1 .. X_n)$, which is true
by definition of $S$, $\nu$ and $N$.
\item
$$N(XY^{n+1})=\nu S(XY^{n+1})=\frac{1}{n+2}(\nu(X,Y^{n+1})+(n+1)\nu(Y,XY^n)),$$
$$\Rightarrow (n+2)N(XY^{n+1})-\nu(X,Y^{n+1})=(n+1)\nu(Y,XY^n).$$
\item
$$LHS=X N(Y^{n+1})-[X\trr N(Y^{n+1})]$$
$$=XYN(Y^n)-X[Y\trr N(Y^n)] - [X\trr YN(Y^n)] + [X\trr [Y\trr N(Y^n)]]$$
$$=Y\nu(X,Y^n)-[Y\trr XN(Y^n)]-[X\trr Y]N(Y^n)+[[X,Y]\trr N(Y^n)]$$
$$+Y[X\trr N(Y^n)]+[Y\trr X]N(Y^n) -Y[X\trr N(Y^n)] + [Y\trr [X\trr N(Y^n)]]$$
$$=Y\nu(X,Y^n)-[Y\trr\nu(X,Y^n)]+[Y,X]N(Y^n) -[[Y,X]\trr N(Y^n)]=RHS.$$
\item We prove that $\nu(X\otimes s)=N(M+r)(X\otimes s)$ by
induction on $n=|s|$. $n=0$: $\nu(X,1)=X=N(X)=N(X.1+r(X,1))=N(M+r)(X,1)$.
Assume true up to $n$. Then by polarization it suffices to prove the following
\begin{lemma}
\label{uec_lemma}
$\nu(X,Y^{n+1})=N(XY^{n+1}+r(X,Y^{n+1}))$.
\end{lemma}
\begin{pr}
$$\nu(X,Y^{n+1})=Y\nu(X,Y^n)-[Y\trr \nu (X,Y^n)]+\nu([Y,X],Y^n)$$
$$=YN(XY^n+r(X,Y^n))-[Y\trr N(XY^n + r(X,Y^n))]$$
$$+ N([Y,X]Y^n+r([Y,X],Y^n))$$
$$=\nu(Y,XY^n)+\nu(Y,r(X,Y^n))+N([Y,X]Y^n+r([Y,X],Y^n))$$
$$=\nu(Y,XY^n)+N(Yr(X,Y^n)+r(Y,r(X,Y^n))+[Y,X]Y^n+r([Y,X],Y^n))$$
$$=
\footnote{\label{rdefmot}This equality motivates the definiton of $r$.}
\nu(Y,XY^n)+N(\frac{n+2}{n+1}r(X,Y^{n+1}))$$
$$=\frac{-1}{n+1}\nu(X,Y^{n+1}) + N(\frac{n+2}{n+1} XY^{n+1}
+\frac{n+2}{n+1}r(X,Y^{n+1}))$$
$$\Rightarrow \nu(X,Y^{n+1})=N(XY^{n+1}+r(X,Y^{n+1})).$$
\end{pr}
\end{enumerate}
\end{pr}
\subsection{Applications to subgaussian algebras and boundary terms.}
In the rest of this section we will apply some of the above formulae
to explicitly construct the inverse of normal ordering in the subgaussian
case, to give a generalization of the subgaussian formula for
$[X\trr N(YZ)]$, proved in theorem \ref{subgaussthm}, and to construct
the solution of the Schwinger-Dyson equation with prescribed boundary term.
\begin{defn} Given a subgaussian contraction algebra, we define
$\bar N:Sym(L)\rightarrow Sym(L)$, and $\rho:(K\oplus L)\otimes Sym(L)
\rightarrow Sym(L)$, inductively by: ($r(1,s):=0$).
$$\bar N(1):=1,$$
$$\bar N(Y^{n+1}):=Y\bar N(Y^n)+r(Y,\bar N(Y^n))+\bar N([Y\trr Y^n]).$$
$$\rho(1,s):=\rho(X,1):=0,$$
$$\rho(X,Y^{n+1}):=[[Y\trr X]\trr Y^n]+\rho([Y\trr X],Y^n)+Y\rho(X,Y^n)$$
$$r(Y,\rho(X,Y^n))+r([X\trr Y],Y^n)+r(Y,[X\trr Y^n]).$$
\end{defn}
\begin{thm} We then have:
\begin{enumerate}
\item
$[X\trr N(Y^n)]=N([X\trr Y^n] +\rho(X,Y^n)).$
\item $N^{-1}=\bar N$.
\end{enumerate}
\end{thm}
\begin{pr}
\begin{enumerate}
\item We will prove this identity by induction on $n$. For $n=0$ it reads
$0=0$, so assume it to be true up to $n$, we will now prove it for $n+1$.
$$[X\trr N(Y^{n+1})]-N([X\trr Y^{n+1}])$$
$$=[X\trr YN(Y^n)]-[X\trr [Y\trr N(Y^{n})]]-N([X\trr Y]Y^n) - N(Y[X\trr Y^n])$$
$$=[X\trr Y] N(Y^n)_1-[[X,Y]\trr N(Y^n)]_2 -[X\trr Y] N(Y^n)_1
-YN([X\trr Y^n])_3$$
$$+Y[X\trr N(Y^n)]_3 -[Y\trr[X\trr N(Y^n)]]_4+[[X\trr Y]\trr N(Y^n)]_2
+[Y\trr N([X\trr Y^n])]_4$$
$$\;\;\;\;\;\;\;\;\;\;\;\;\;\;\;\;\;\;\;\;\;\;\;\;\;\;\;\;\;\;\;\;\;\;\;\;\;\;\;\;\;\;\;\;\;
+N(r([X\trr Y],Y^n))+N(r(Y,[X\trr Y^n]))$$
$$=[[Y\trr X]\trr N(Y^n)]_2+YN(\rho(X,Y^n))_3-[Y\trr N(\rho(X,Y^n))]_4$$
$$+N(r([X\trr Y],Y^n)+r(Y,[X\trr Y^n]))$$
$$=N\{[[Y\trr X]\trr Y^n] + \rho([Y\trr X],Y^n)+Y\rho(X,Y^n)+r(Y,\rho(X,Y^n))$$
$$+r([X\trr Y], Y^n) + r(Y,[X\trr Y^n]))\}=N(\rho(X,Y^{n+1})).$$
\item We will prove by induction on n that $N\bar N(Y^n)=\bar NN(Y^n)=Y^n$.
Indeed, this is true by definition for $n=0$, so assume the identities
hold up to $n$, then:
$$N\bar N(Y^{n+1})=N(Y\bar N(Y^n)+r(Y,\bar N(Y^n)))+[Y\trr Y^n]$$
$$=\nu(Y,\bar N(Y^n))+[Y\trr Y^n]$$
$$=YN\bar N(Y^n)-[Y\trr N\bar N(Y^n)]+[Y\trr Y^n]=Y^{n+1},$$
Next, to prove $\bar NN=id$, we first prove that $\bar N$ is surjective.
This follows from $\bar N(Y^{n+1})=Y^{n+1}$ mod $Sym^{[0,n]}(L)$, which in
turn follows from the definition, by induction on $n$.
Therefore, for every $Y^n$ there is an $s_n$ such that $Y^n=\bar N(s_n)$,
so that:
$$\bar NN(Y^n)=\bar NN\bar N(s_n)=\bar N(s_n)=Y^n,$$
which proves the identity.
\end{enumerate}
\end{pr}
\begin{rem}
\label{boundary_rem}
The following theorem is motivated by integration over manifolds with
boundary: Suppose we already know integration over the boundary $\partial M$
of a manifold. Then in particular, if $\mu$ is a volume form on $M$ we
know $\tilde J:X\otimes f\mapsto \int_{\partial M} f i_X \mu$.
This last association
is related to the integral over $M$ through the Schwinger-Dyson equation
for $I:f\mapsto \int_M f\mu$:
$$I(\nabla(X)f+X(f))=\int_M L_X(f\mu)=\int_M di_X(f\mu)=\int_{\partial M}
f i_X \mu.$$
In algebraic language, this leads us to consider
the equation $I(-Xs+[X\trr s])=\tilde J(X\otimes s)$, i.e. setting
$J(X\otimes s):=-\tilde J(X\otimes N(s))$, we become interested in the
solvability of the equation
$I(XN(s)-[X\trr N(s)])=J(X\otimes s)$, which is what the following theorem
is about:
\end{rem}
\begin{thm} Setting $\pi:=S(M+r)$, the following are equivalent:
\label{solve_boundary}
\begin{enumerate}
\item $I(XN(s)-[X\trr N(s)])=J(X\otimes s)$,
\item $IN=JS+I(1)Z$ and $J=J\pi$.
\end{enumerate}
\end{thm}
\begin{pr} First, using theorems \ref{sym_alg_ident} and \ref{uec_ident},
with $R:=M+r$, we have the following properties:
$ZR=0$, $SZ=0$, $\nu=\nu SR$, $N=\nu S+Z$.
We now prove the theorem.
By definition, $(1)$ is equivalent with $I\nu = J$; Next:\\
$(1)\Rightarrow (2): IN=I\nu S+IZ=JS+I(1)Z;\;J=I\nu=I\nu SR=JSR=J\pi$.\\
$(1)\Leftarrow (2): I\nu=I\nu SR=I(N-Z)R=INR=JSR+I(1)ZR=J\pi+0=J$.
\end{pr}
\begin{rem} If $N$ is invertible, then
what we have done is to solve the inhomogeneous linear equation $I\nu=J$
as $I=JSN^{-1}+E(1)ZN^{-1}$, which as expected is of the form
$I_p+I_0$, where $I_p$ is any particular solution, and $I_0$ is the
general solution of the homogeneous equation. Further, note that
$\pi^2=\pi$, see theorem \ref{sym_alg_ident}.
\end{rem}
%
%
%
%
\newpage
\section{Conclusion and acknowledgements.}
The following is a list of the main results of this work.
\begin{enumerate}
\item A useful generalization of contractions to non-Gaussian weights $S$
is the second derivative of $S$, written as a polynomial in the
first derivatives. In view of the fact that normal ordering can be defined
in terms of contractions, this leads to
a non-Gaussian notion of normal ordering.
\item For the Gaussian case, normal ordering is an invertible operation.
\item Invertibility of normal ordering is interesting for non-Gaussian
integrals too, since in that case the inverse of
normal ordering is directly related to the
solution of the Schwinger-Dyson equation.
\item This statement can be generalized to a nonabelian setting.
\item We defined the notion of a subgaussian weight, for which a number of
generalized Wick rules can be derived. Examples of subgaussian algebras
can be found in two-dimensional conformal field theory.
\item We have avoided the use of operators and Hilbert spaces.
The notion of composite operators was replaced by that of
a left extension of a contraction. This approach avoids the
problems encountered by Johnson and Low, namely that what they call
``commutators'' of ``operators'' do not satisfy the Jacobi-identity.
\item A volume form can be algeberaically characterized up to a constant by
its divergence. This leads to the possibility of defining a calculus
with differential forms
of finite codegree on a possibly infinite dimensional manifold,
starting from a given divergence. These
differential forms are suited to formulate the Schwinger-Dyson equation
for infinite dimensional differential forms.
(See appendix \ref{geometric_section}).
\end{enumerate}
I would like to thank my advisors Prof. Dijkgraaf and Dr. Pijls
for the fruitful discussions I have had the pleasure to entertain with them,
for their critical comments, and the patience needed
before this work was completed.
I also thank C.D.D. Neumann for his particularly
influential remark on Pre-Lie algebras.
This work was financially supported by ``Samenwerkingsverband
Mathematische Fysica van de Stichting FOM en het SMC.''
%
%
%
%
\appendix
%
%
%
%
\newpage
\section{Motivations.}
\label{motsection}
\subsection{Motivation for the need of a functional integral.}
There are a number of reasons to want to have a good definition
of functional integration:
\begin{enumerate}
\item They are useful for the compact description of the
time averaged behaviour of
mechanical systems that consist of a large number of particles, for example
a liter of gas.
\item Idem for the compact description of the behaviour of
elementary particles.
\item They are useful in geometry, and more particularly in
the geometry of low-dimensional manifolds, where other methods seem to fail.
\end{enumerate}
\subsection{Motivation for investigating the Schwinger-Dyson equation.}
\subsubsection{Comparison of different methods.}
A universally applicable definition of functional integration
seems to this date not available.
Various methods are used to compute functional integrals, each method in
itself being a possible definition.
Such definitions have various weak points however:
\begin{enumerate}
\item It is usually hard or impossible to prove existence of the integral.
\item Methods of calculation are often of limited applicability.
\end{enumerate}
This motivates one of our aims, which is to find a definition of functional
integration which does not have these inconvenients.
In order to find out where to start, let us have a look at some methods of
calculation:
\begin{enumerate}
\item Gaussian functional integrals are computed using
Wick's rules.
\item For those that can be deformed into a Gaussian integral, we can
try defining the integral using a perturbation series. However this
often fails because the series does not converge, see remark
\ref{asym_foot}. And even if it does, then we cannot be sure that
it converges to what we are looking for, since when dealing with
singular perturbations the dependence on the perturbations may not
be analytical.
\item In other special cases, one can make use of special symmetry
properties which determine the integral.
\item If the integral has already been defined for certain values of some
parameter, then one may use analytic continuation as the definition
for other values of that parameter.
\item If the base manifold is cylindrical, i.e. of the form $N\times [0,1]$,
then the path integral may
be defined as the exponent of the Hamilton operator,
that is if the operator itself is defined.
\item Only two approaches however do not specify in advance which type of
integral is to be considered:
\begin{enumerate}
\item The Schwinger-Dyson equation is an equation that is
independent of the type of integral.
\item The same holds for the continuum limit of a discretized functional
integral.
\end{enumerate}
\end{enumerate}
Thus if we are to follow any of these approaches, it seems that we have to
choose between the Schwinger-Dyson approach and the discretized approach.
Before making this choice, let us recall some history of usual integration:
\begin{enumerate}
\item First, Newton-Leibnitz defined integration as the inverse of
differentiation. I.e. given a function $f$, the integral over $[a,b]$
of $f$ is $F(b)$ , where $F$ the
unique solution of $F(a)=0$, $F^\prime =f$, if it exists.
\item The next step by Riemann can be seen either as the existence theorem
to the above problem for $f$ Riemann-integrable, or by uniqueness it can
be taken to be the definition of the integral of $f$.
\item When the integral came to be seen as a linear functional, Lebesgue
showed that the Riemann-integral could be extended continuously to
the larger class of Lebesgue-integrable functions.
\end{enumerate}
What we learn from this historical development is that it is easier to
work with defining properties like $F^\prime=f$ rather than with
the actual construction of the integral by a limiting process such as in
Riemann's approach.
This defining property does not in itself refer to any measure.
Furthermore, the limiting approach will not be of any
help to us for the \underline{exact} calculation of explicit integrals
in terms of known functions, for which we
will always use the defining property instead of Riemann sums.
Taking this lesson into account for our search for the definition of the
functional integral, we will at this point give up the limiting approach:
It is probably too ambitious a task to construct the integral before
being accustumed to the defining properties.
So what remains is the Schwinger-Dyson equation,
which as we are suggesting is the analogue of $F^\prime=f$.
\subsubsection{Hints that it is possible.}
There are a number of vague reasons
to beleive that it makes sense to define a linear functional up to a scalar
by the Schwinger-Dyson equation. Here is a list of such reasons:
\begin{enumerate}
\item It works if the weight is Gaussian.
\item Next, when perturbing a Gaussian weight $S_0$ to a nongaussian one
$S_0+\lambda S_1$, the Schwinger-Dyson equation reads:
$$\langle f\partial_i S_0\rangle =\langle \partial_i f\rangle -\lambda\langle
f\partial_i S_1\rangle .$$
One may try solving this iteratively by starting with the unique solution
$\langle .\rangle _0$ at $\lambda=0$, and continuing with
$$\langle f\partial_i S_0\rangle _{n+1}=\langle \partial_i f\rangle
_{n+1}-\lambda\langle f\partial_i S_1\rangle _n.$$
Which leads to a unique formal series in $\lambda$.
\item Finally, let us give a very rough sketch of an argument
independent of the type of weight $S$ used,
but using positivity, that leads to
the Schwinger-Dyson equation being a defining
property apart from the scalar: Assume $I$ is a functional that satisfies
the conditions to Riesz' theorem. Then there is a measure $\nu$ such that
$I(f)=\int f \nu$. "Divide" this measure by $\mu$ to get a positive function,
and take the logarithm, so as to get: $I(f)=\int f e^{-P}\mu$. Since $I$
now satisfies the Schwinger-Dyson equation for both $S$ and $P$, we have
$\forall_f I(\partial_i(S-P) f)=0$, which by positivity gives
$\partial_i(S)=\partial_i(P)$, or $P=S+c$, so that $I(f)=K.\int f e^{-S}\mu$,
so that $I$ is determined up to a positive scalar.
\end{enumerate}
Now each of the above arguments unsatisfactory:
Point $(1)$ is a special case. So is point $(2)$, and it only refers to
formal power series, and finally point $(3)$ is not a correct proof.
Thus we were led to find other ways of proving existence and uniqueness.
\subsection{Motivation for the need of generalized normal ordering.}
\subsubsection{The use of normal ordering to avoid certain divergeces.}
Recall that for $D\geq 3$,
$$\int_{\{\phi:\bR^D\rightarrow \bR\}} e^{-\int_{\bR^D} \phi \Delta \phi}
\phi(x)\phi(y)=\frac{K}{|x-y|^{D-2}}.$$
In particular, we see that this integral is undefined for $x=y$.
The divergent limit $x\rightarrow y$ is known as an ultraviolet
\footnote{Ultraviolet radiation has shorter wavelength than
visible light and is thus associated with the small distance limit
$x\rightarrow y$ in the base manifold $\bR^D$. In the same way one speaks
of infrared divergences, i.e. being associated to large distances.} (UV)
divergence.
UV divergences are not specific to Gaussian integrals, but in the non-Gaussian
cases the exact answer is usually not known, so that these divergences are
more difficult to illustrate.
$(UV)$ divergences prevent one from taking the average of more complicated,
functions of $\phi$ at one point. Examples involving these so-called
composite insertions at the point $x$ are $\langle \phi^2(x)\phi(y)\rangle $
and
$\langle e^{i\phi(x)}\phi(y)\rangle $.

For Gaussian integrals there is an operation $f\mapsto (:f:)$ called
normal ordering,
acting on the integrand, which in a way circumvents these UV divergences
when working with composite insertions.
Let us illustrate this without going into the exact definition of normal
ordering.
First one may show that for linear functions $A_i$ of $\phi$:
$$\langle \prod_{i=1}^n e^{A_i}\rangle =\prod_{i,j} e^{\langle A_i A_j\rangle
/2},$$
whereas the normal ordering operation has the property that
$$\langle \prod_{i=1}^n (:e^{A_i}:)\rangle =\prod_{i\neq j} e^{\langle A_i
A_j\rangle /2}.$$
Now as we saw, in infinite dimensions, the expression $\langle AB\rangle $ is
not
well defined for all combinations $A$ and $B$. Indeed,
suppose the integral is over functions $\phi:\bR^D\rightarrow \bR$, and
let $A_x(\phi):=\phi(x)$. Then we saw that
$$\langle A_x A_y\rangle =\frac{K}{|x-y|^{D-2}}.$$
So $\langle A_x A_x\rangle $ is undefined.
Now $\langle \prod_{i=1}^n \exp(A_{x_i})\rangle $, is undefined because it
involves
$\langle A_{x_i} A_{x_i}\rangle $, but $\langle \prod_{i=1}^n
N(\exp(A_{x_i}))\rangle $ is well
defined if the $x_i$'s are different, because the effect of normal
ordering is not to include the term $i=j$.

Thus, for Gaussian integrals, one knows how to circumvent
ultra violet divergences: By using normal ordering.
Now from a higher point of view there is nothing particularly
special about Gaussian integrals compared to other integrals exept the
fact that they can be explicitly calculated.
So we ask the question: Can we find
an analogue of the normal ordering procedure for non-Gaussian integrals,
so that analogous UV problems can be handled in a similar way?
\subsubsection{Naturality of normal ordering.}
Finally, for geometrical applications, it is important to keep the
naturality of all constructions in mind. Thus, if the weight $S$ is
naturally associated to some geometrical objects, and normal
ordering is naturally associated to weights, then normal ordering
is naturally associated to geometric objects.
\subsection{Motivation for calling $N$ normal ordering.}
\label{motNord}
The map $N$ is related to what is known as normal ordering of
operators. It was initially introduced by Houriet and Kind (1949) to reproduce
Feynman's diagrammatic rules for perturbation theory.
We will here make the link with what is
usually called normal ordering, and show that the so-called canonical
quantization procedure is a way to solve the Schwinger-Dyson equation
for Gaussian integrals.
\subsubsection{Operator normal ordering.}
\begin{rem}
The reader who is familiar with operator normal ordering may note the
following:
Let $J(z)=\oplus_n J_n z^{-n-1}$, with commutation relations
$[J_n,J_m]=n\delta_{n+m,0}$, and $J_{n\geq 0}|0\rangle =0$,
$\langle 0|J_{n<0}=0$. Set $J^+(z):=\oplus_{n\geq 0} z^{-n-1} J_n$,
and $J^-:=J-J^+$.
Then one checks that:
\begin{enumerate}
\item $\langle 0|J^-(z)=0$, $J^+(z)|0\rangle=0$.
\item $[J_+(\zeta),J_+(z)]=0$,
\item $[J_+(\zeta),J(z)]=\frac{1}{(\zeta - z)^2}$ for $|z|<|\zeta|$.
\end{enumerate}
Thus, apart from analytic continuations, we have $[J_+(\zeta),J(z)]
=[J_+(z),J(\zeta)].$
This motivates the following definition, where $W$ is to be thought of
as the algebra of operators, $V$ the ``subspace'' spanned by the $J(z)$'s, and
$\phi,\psi$, say, as $J(z),J(\zeta)$:
\end{rem}
\begin{defn} A Wick algebra will be a combination $(V,W,\pi)$, where
\begin{enumerate}
\item $W$ is an associative, not necessarily commutative algebra with unit, and
$V$ is a subvectorspace of $W$. Thus we have a map $T(V)\rightarrow W$.
Elements of $V$ will be denoted as $\phi,\psi$.
\item $\pi:V\rightarrow W$; $\phi\mapsto \phi^+$; $\phi^-:=\phi-\phi^+$.
\end{enumerate}
such that with $[w_1,w_2]:=w_1 w_2-w_2 w_1$:
\begin{enumerate}
\item $[\phi^+,\psi^+]=[\phi^-,\psi^-]=0$,
\item $[\phi^+,\psi^-]\in K\leq W$ ($K$=scalars times unit),
\item $[\phi^+,\psi]=[\psi^+,\phi]$.
\end{enumerate}
We set $[.\trr .]:V\otimes W\rightarrow W;\;[\phi\trr w]:=[\phi^+,w]$.
An integral for such an algebra is a map:
$\langle 0|.|0\rangle :W\rightarrow K$, such that
$\langle 0| \phi^- w|0\rangle=\langle 0| w\phi^+|0\rangle =0$.
\end{defn}
\begin{defn} For a Wick algebra, we define a map $T(V)\rightarrow W;
t\mapsto (:t:)$, inductively by:
$$(:1:):=1;\;(:s\phi:):=\phi^-(:s:)+(:s:)\phi^+.$$
\end{defn}
\begin{thm}
For $s,t\in T(V)$, and $\phi,\psi\in V$:
\begin{enumerate}
\item $[.\trr.]$ is a derivation on the right,
and $[\phi\trr \psi]=[\phi,\psi^-]=[\phi\trr \psi^-]$.
\item $(:\phi_1..\phi_n:)=(:\phi_{\sigma(1)}..\phi_{\sigma(n)}:)$.
\item $(:\phi s:)=\phi (:s:)-(:[\phi\trr s]:)$.
\label{normprop1}
\item $(:\phi s:)=\phi (:s:)-[\phi\trr (:s:)]$.
\label{normprop2}
\item $\langle 0|\phi_1..\phi_n|0\rangle
=\langle 0| \phi_{\sigma(1)}..\phi_{\sigma(n)} |0\rangle$.
\end{enumerate}
\end{thm}
\begin{pr}
\begin{enumerate}
\item We have
$$[\phi\trr w_1 w_2]=[\phi^+, w_1 w_2]=[\phi^+ ,w_1]w_2+
w_1[\phi^+,w_2]=[\phi\trr w_1]w_2+w_1[\phi\trr w_2],$$
and further, $[\phi\trr \psi]=[\phi^+,\psi]=[\phi^+,\psi^-]=[\phi,\psi^-]$.
\item It suffices to prove that $(:s\phi t:)= (:st \phi:)$. We first prove
this for $|t|=1$, i.e. we prove $(:s\phi\psi:)=(:s\psi\phi:)$. Indeed:
$$LHS=\psi^- (:s\phi:)+(:s\phi:)\psi^+$$
$$=\psi^- \phi^- (:s:) + \psi^- (:s:) \phi^+ + \phi^- (:s:) \psi^+ +
(:s:)\phi^+ \psi^+$$
$$=\phi^-\psi^- (:s:) + \phi^- (:s:) \psi^+ + \psi^- (:s:) \phi^+ + (:s:)
\psi^+ \phi^+=RHS$$
Next we proceed by induction on $n:=|t|$.
So let us assume the identity to hold up to $n=|t|$.
Then:
$$(:s\phi t\psi:)=\psi^- (:s\phi t:)+(:s\phi t:)\psi^+$$
$$=\psi^- (:st\phi:)+(:st\phi:)\psi^+=(:st\phi\psi:)=(:st\psi\phi:).$$
\item Induction on $|s|$. $(:\phi 1:)=\phi=\phi (:1:)-(:[\phi\trr 1]:)$.
Next assume the identity holds for $s$. We will prove it for $s\psi$:
$$(:\phi s \psi:)=\psi^- (:\phi s:)+(:\phi s:)\psi^+$$
$$=\psi^- (\phi (:s:)-(:[\phi\trr s]:))+(\phi (:s:) - (:[\phi \trr
s]:))\psi^+$$
$$=\psi^- \phi (:s:) + \phi (:s:) \psi^+ - (:[\phi\trr s]\psi:)$$
$$=\phi \psi^- (:s:) + \phi (:s:) \psi^+ - [\phi,\psi^-](:s:)-(:[\phi\trr
s]\psi:)$$
$$=\phi (:s\psi:)-(:[\phi\trr s\psi]:).$$
\item Induction on $|s|$. $(:\phi 1:)=\phi=\phi (:1:)-[\phi\trr (:1:)]$. Next
assume the identity holds for $s$. We will prove it for $s\psi$.
$$(:\phi s \psi:)=\psi^- (:\phi s:)+(:\phi s:)\psi^+$$
$$=\psi^- (\phi (:s:)-[\phi\trr (:s:)])+(\phi (:s:) - [\phi\trr (:s:)])\psi^+$$
$$=\phi(\psi^- (:s:)+(:s:)\psi^+)-[\phi,\psi^-] (:s:)$$
$$-\psi^-[\phi\trr (:s:)]-[\phi\trr (:s:)]\psi^+$$
$$=\phi (:s\psi:) -[\phi\trr \psi] (:s:)-\psi^- [\phi\trr (:s:)]-[\phi\trr
(:s:)]\psi^+$$
$$=\phi (:s\psi:)-[\phi\trr \psi^- (:s:) +(:s:) \psi^+]
=\phi (:s\psi:)-[\phi\trr (:s\psi:)].$$
\item It suffices to prove that
\begin{enumerate}
\item $\langle 0| \phi s|0\rangle = \langle 0|s\phi|0\rangle$.
\item $\langle 0| \phi\psi s|0\rangle=\langle 0| \psi\phi s |0\rangle$
\end{enumerate}
Indeed, for the first formula, since $[\phi^+,\psi]=[\psi,\phi^-]$, we have
$[\phi^+,s]=[s,\phi^-]$ for $s\in T(V)$, so that:
$\langle 0|\phi s|0\rangle =\langle 0|\phi^+ s|0\rangle =\langle
0|[\phi^+,s]|0\rangle
=\langle 0|[s,\phi^-]|0\rangle =\langle 0|s\phi^-|0\rangle =\langle
0|s\phi|0\rangle .$
For the second formula:
$$\langle 0|\phi\psi s|0\rangle =\langle 0|[\phi\trr\psi s]|0\rangle =\langle
0|[\phi\trr \psi]s|0\rangle $$
$$+\langle 0|\psi[\phi\trr s]|0\rangle =\langle 0|[\psi\trr \phi]s|0\rangle
+\langle 0|[\psi\trr[\phi\trr s]]|0\rangle $$
$$=\langle 0|[\psi\trr \phi]s|0\rangle +\langle 0|[\phi\trr[\psi\trr
s]]|0\rangle =\langle 0|\psi\phi s|0\rangle .$$
\end{enumerate}
\end{pr}
\begin{rem}
The point of the above theorem is the fact that either of
the properties \ref{normprop1} or \ref{normprop2}
of $(::)$ together with $(:1:)=1$
are defining properties, using only $[.\trr.]$.
This version of the definition of normal ordering is very similar to
the definition of $N$, which was $N(1):=1$ and
$$N(S_{i_0}..S_{i_n}):=S_{i_0} N(S_{i_1}..S_{i_n})-\partial_{i_0}
N(S_{i_1}..S_{i_n}).$$
We see that these two coincide if $S_i\in V$ and $[S_i^+,S_j^-]=g_{ij}$.
If this is so we say that
the Wick algebra is a canonical quantization of the Gaussian action
$S=g_{ij}x^i x^j/2$.
We already know that $N$ is related to the solution of the Schwinger-Dyson
equation. We will now proceed to make the link between the
integrals for a Wick algebra and the solutions of the Schwinger-Dyson
equation:
\end{rem}
\begin{thm}
Let $\langle 0|.|0\rangle$ be an
integral for a Wick algebra $(V,W,\pi)$. Let $Z$ be the zero
projection $T(V)\rightarrow K$. Then
$\langle 0|(:s:)|0\rangle=Z(s).\langle 0|1|0\rangle$.
\end{thm}
\begin{pr}
For $s=1$ this reads $\langle0|1|0\rangle=\langle0|1|0\rangle$.
Further,
$$\langle 0|(:s\psi:)|0\rangle=\langle 0|\psi^-(:s:)+(:s:)\psi^+|0\rangle=0,$$
by definition of an integral.
\end{pr}
\begin{rem}
What we have obtained is the following:
\begin{enumerate}
\item
$N$ is determined using $[\partial_i S\trr\partial_j S]:=\partial_i \partial_j
S$,
by $N(s\psi)=\psi N(s) -[\psi\trr N(s)]$, and any solution $I$ of the
Schwinger-Dyson equation satisfies
$IN(s)=Z(s).I(1)$.
\item
$(::)$ is determined using $[\phi\trr \psi]=[\phi^+,\psi]$,
by $(:s\psi:)=\psi (:s:) -[\psi\trr (:s:)]$, and any
integral $\langle 0|.|0\rangle$ satisfies
$\langle 0|(:s:)|0\rangle=Z(s).\langle 0|1|0\rangle$.
\item If a Wick algebra is a canonical quantization of $S=\frac{1}{2} g_{ij}
x^i
x^j$, then $\partial_i S\in V$ and $[S_i^+,S_j^-]=g_{ij}$.
\end{enumerate}
This amounts to the following:
Canonical quantization is a way to solve the Schwinger-Dyson equation for
Gaussian integrals. In that case normal ordering $(::)$ is the analogue of
the map $N$ that we introduced in the context of not necessarily Gaussian
integrals, and this motivates our calling $N$ normal ordering.
\end{rem}
\subsection{Motivation for the need of a generalized Wick calculus.}
The calculation of commutation relations of normal ordered products like
$$L_n=\frac{1}{2(k+c^\vee)}\sum_{j\in\bZ}g_{ab} (:J_{-j}^a J_{j+n}^b :)$$
using mode expansions is cumbersome.
Now on the one hand, the special case where the $J_n^a$'s
are an abelian Kac-Moody algebra
is much easier because it can be calculated
using Wick's calculus for Gaussian field theories.
But on the other hand on the level of mode
expansion calculations, there is not that much difference between the
free and the non-free calculation. Thus we are led to try and formulate
generalized Wick rules of computation which can be applied to
the non-Abelian Kac-Moody case. This is not a new idea in itself:
In the field of two-dimensional conformal field theory, it
is known how to extend the free Wick calculus to the non-Gaussian conformal
case. See for example \cite[section 5]{sasaki},\cite[appendix A]{bais}.
I have been looking for an extension of the Wick rules that do
not depend on dimensionality or symmetry. Indeed, I found that replacing
for example the free property
$$[a\trr(:a_1 a_2:)]=(:[a\trr a_1]a_2:)+(:a_1 [a\trr a_2]:)$$
by
$$[a\trr(:a_1 a_2:)]=(:[a\trr a_1]a_2:)+(:a_1 [a\trr a_2]:)
+\frac{1}{2}(:[[a_1\trr a]\trr a_2]:)+
\frac{1}{2} (:[[a_2 \trr a]\trr a_1]:)$$
gives the right result for the Kac-Moody calculation. One of the aims
was to give a basis to such a generalized Wick rule, and to find out
more general rules for higher order polynomials. As it turned out
this rule could be derived within the setting of subgaussian contractions.

Next, the operator methods used in higher dimensions lead to
``operators of which the commutation relations do not
satisfy the Jacobi identity''. Although a number of scientists
seem to find this exciting, I would say it is unacceptable.
It seems that the theory of left extensions
provides a good alternative to the operator language.
%
%
%
%
%
\newpage
\section{Review of the Gaussian case.}
\label{gaussiansection}
\subsection{Finite and infinite dimensional Gaussian Schwinger-Dyson
equations.}
\subsubsection{The finite dimensional case.}
The most elementary case
of a weight $S$ where the solution is unique by combinatorics is that of
Gaussian weights, so let us quickly review it:
$S(x)=g(x,x)/2=g_{ij} x^i x^j/2$, where
$g$ is some symmetric bilinear form, not necessarily positive.
In that case, since $\partial_i S=g_{ij} x^j$,
the Schwinger-Dyson equation reduces to $I(g_{ij} x^j s)=I(\partial_i s)$.
If there is a matrix $g^{ij}$
such that $g_{ij} g^{jk}=\delta_i^k$, then $I(x^j s)=g^{jk}I(\partial_k s)$, so
that:
$$I(x^{i_0}x^{i_1}...x^{i_n})=\sum_{j=1}^n g^{i_0 i_j}
I(x^{i_1}..\hat x^{i_j}..x^{i_n}).$$
This is a recurrence relation which determines $I$ completely for polynomials
up to $I(1)$. So we see that
for Gaussian integrals, the Schwinger-Dyson equation is a defining property.
Note by the way that since we have not
required $g$ to be positive, we have in a sense defined the
expression
$$\frac{\int e^{x^2/2} p(x)dx}{\int e^{x^2/2} dx},$$
for polynomial $p$, namely as $\langle p\rangle $ where $\langle .\rangle $ is
the unique
normalized solution to the Schwinger-Dyson
equation with weight $S(x)=-x^2/2$. Note however that this solution is
not positive, since $\langle x^2\rangle =-1$.
Thus we see that combinatorical integration can be an extension of usual
integration. The price we pay is that we lose the normalization constant,
because the Schwinger-Dyson equation will never tell us that
$\int e^{-x^2} dx=\sqrt{\pi}$; We may also lose positivity, and the
solution may not even be unique up to normalization, as we will see in
section \ref{one_var_section}.
\subsubsection{Introduction to Gaussian functional integration.}
\label{gauss_fun_int}
We have now come to the point where we can explain the formula
that we gave at the start of this work:
The idea is that
the Schwinger-Dyson equation is a well defined infinite dimensional
equation, which we may try to solve as well. So let us look at an infinite
dimensional Gaussian Schwinger-Dyson equation.
We will compute the combinatorical integral over the space of differentiable
functions
$\phi:\bR^D\rightarrow \bR$, with weight $S(\phi):=
\int_{\bR^D} \partial_i\phi \partial^i \phi/2$, replacing the vectorfields
$\partial_i$ that
we used before by the generalized vectorfields $\delta_x:=\frac{\delta}{
\delta \phi (x)}$. So we look for a functional $\langle .\rangle $ that
satisfies
$\langle \delta_x(S)f\rangle =\langle \delta_x f\rangle $, i.e. with $\Delta$
being the Laplace operator:
$$\langle (-\Delta\phi)(x) f\rangle =\langle \delta_x f\rangle .$$
Now we recall from the finite dimensional case that we had to find the inverse
matrix $g^{ij}$ such that $g^{ij}g_{jk}=\delta_k^i$
to get to the solution. So let us do the same thing for $\Delta$.
\footnote{If no boundary conditions are imposed, $\Delta$ is not invertible.
This matter will be discussed in greater generality in the section on
integration over quotient manifolds: The point is that the weight $S$ is
invariant under $\phi\mapsto\phi+const.$, and what we are talking about
here is the integral over the quotient $\{\phi\}/\sim$ of functions modulo
constants.
Now $\phi\mapsto \phi(x)$ is not a function on that quotient
and this is reflected in the ambiguity in $f_D$ as defined by
$\Delta(f_D)=\delta$ (like the addition of a constant): It makes
$\langle \phi(x)\phi(y)\rangle $ ill defined, but $\langle \partial_i
\phi(x)\partial_j \phi(y)\rangle $
is well defined, which is good since $\phi\mapsto \partial_i\phi(x)$
is a map on the quotient. What is also well defined on the quotient is
a product $\prod_{k=1}^n e^{ip_k \phi(x_k)}$ if $\sum_k p_k =0$.}
The following should always be read distributionally in $x,y,z$.
In dimension
$D$ we have $\Delta(f_D(x))=\delta(x)$, where $\delta(x)$ is the Dirac
delta function, and
$$f_1(x):={1\over 2}|x|;\;f_2(x):={1\over 2\pi}\ln|x|,$$
$$f_{D\geq 3}(x):={1\over (2-D)Vol(S^{D-1})|x|^{D-2}},$$
Therefore,
$$\langle \phi(x)f\rangle =\int \delta(x-z) \langle \phi(z) f\rangle dz=\int
\Delta(f_D(x-z))\langle \phi(z)f\rangle dz$$
$$=\int f_D(x-z)\langle (\Delta \phi)(z) f\rangle dz=\int f_D(x-z)\langle
-\delta_z f\rangle dz.$$
In particular
$$\langle \phi(x)\phi(y)\rangle =\int f_D(x-z)\langle -\delta(z-y)\rangle
dz=-f_D(x-y).$$
This is the formula that we promised to explain.
\subsection{Gaussian contraction algebras.}
\begin{thm} Gaussian contraction algebras satisfy the following properties:
\label{gauss_wick_th}
\begin{enumerate}
\item They are Abelian: $\forall_{X,Y\in L}\;[X,Y]=0$.
\item $[X\trr Y]=\langle XY\rangle $.
\item $[X\trr N(s)]=N([X\trr s])$.
\item $N(e^X)=e^{X-\langle XX\rangle /2}.$ (As formal power series.)
\item $\langle e^X\rangle =e^{\langle XX\rangle /2}.$
\item $\langle N(e^X)\phi\rangle =\langle e^X \phi\rangle /\langle e^X\rangle
.$
\item $N(e^X) N(e^Y)=N(e^{\langle XY\rangle +X+Y})$.
\item (Wick, \cite[theorem 2]{wick}
\footnote{Wick's theorem expresses products of normal ordered expressions
as normal ordered expressions. One recovers it from the above formulation
by replacing $X$ by $\lambda X$ and taking derivatives
with respect to $\lambda$ at $\lambda=0$.}).
 $N(e^{X_1})..N(e^{X_n})=N(e^{\sum_i X_i+\sum_{i<j} \langle X_i X_j\rangle })$.
\item $\langle \prod_{i=1}^n N(e^{X_i})\rangle =
\prod_{i<j} e^{\langle X_i X_j\rangle }.$
\item For $n\geq 1$: $\langle Y^{2n-1}\rangle =0$, and $\langle Y^{2n}\rangle
=(2n-1)!! \langle Y^2\rangle ^n$.
\end{enumerate}
\end{thm}
\begin{pr}
\begin{enumerate}
\item $[X\trr Y]-[Y\trr X]=[X,Y]\in Sym^0\cap Sym^1=\{0\}$.
\item Since $[X\trr Y]$ is a number, it is equal to $\langle [X\trr Y]\rangle
$, which
equals $\langle XY\rangle $ by the Schwinger-Dyson equation.
\item Recall that for Abelian contraction algebras, we have
$N(Xs)=XN(s)-[X\trr N(s)]$, since by definition \ref{sym_def}:
Abelian $\Rightarrow r(X,s)=0$, so that by theorem
\ref{uec_ident} point \ref{main_uec_ident}: $N(Xs)=\nu(X,s)$.
We now proceed by induction on $|s|$; $[X\trr N(1)]=[X\trr 1]=0=N([X\trr 1]).$
Assume true up to $|t|$. Then:
$$[X\trr N(Yt)]-N([X\trr Yt])$$
$$=[X\trr YN(t)]-[X\trr[Y\trr N(t)]]-N([X\trr Y]t)-N(Y[X\trr t])$$
$$=[X\trr Y]N(t)-[Y\trr[X\trr N(t)]]-[X\trr Y] N(t) -Y N([X\trr t])$$
$$+Y[X\trr N(t)]\;\;\;+[Y\trr N([X\trr t])]=0.$$
\item Define $a(\lambda):=N(e^{\lambda X})$, and
$b(\lambda):=e^{\lambda X-\lambda^2\langle XX\rangle /2}$. Then $a(0)=1=b(0)$,
so it
suffices to prove that both $a$ and $b$ satisfy the differential equation
$\partial_\lambda f(\lambda)=Xf(\lambda)-[X\trr f(\lambda)]$. Indeed:
$$\partial_\lambda a(\lambda)=N(Xe^{\lambda X})=
XN(e^{\lambda X})-[X\trr N(e^{\lambda X})]=Xa(\lambda)-[X\trr a(\lambda)],$$
and
$$\partial_\lambda b(\lambda)=b(\lambda)(X-\lambda\langle XX\rangle )
=Xb(\lambda)-b(\lambda)[X\trr \lambda X-\lambda^2\langle XX\rangle /2]$$
$$=Xb(\lambda)-[X\trr b(\lambda)].$$
\item $1=\langle N(e^X)\rangle =\langle e^X\rangle /e^{\langle XX\rangle /2}.$
\item $\langle N(e^X)\phi\rangle =\langle e^{X-\langle XX\rangle /2}\phi\rangle
$.
\item
$$LHS=e^{X-\langle XX\rangle /2}e^{Y-\langle YY\rangle /2}=e^{X+Y-\langle
(X+Y)^2\rangle /2+\langle XY\rangle }=RHS.$$
\item By induction on $n$ from the previous formula.
\item By taking $N$ of the previous formula since $\langle N(e^{\sum_i
X_i})\rangle =1$.
\item $\langle 1\rangle =1$, $\langle Y\rangle =0$, and
$$\langle Y^{n+2}\rangle =\langle [Y\trr Y^{n+1}]\rangle =(n+1)\langle
Y^n[Y\trr Y]\rangle =(n+1)\langle Y^n\rangle \langle Y^2\rangle ,$$ so
$\langle Y^{2n+1}\rangle =0$, and $\langle Y^{2n+2}\rangle =(2n+1)\langle
Y^{2n}\rangle \langle Y^2\rangle $, so that
$\langle Y^4\rangle =3 \langle Y^2\rangle ^2$, $\langle Y^6\rangle =5.3.\langle
Y^2\rangle ^3$, etc.
\end{enumerate}
\end{pr}
%
%
%

%
%
%
%
%
%
%
%
%
%
\newpage
\section{Examples of easy weights: Polynomials in one variable.}
\label{one_var_section}
This section is meant to give the reader a feeling of what type of
solutions the Schwinger-Dyson equation can have if that solution is
not unique. None of the statements here are very deep since we
will concentrate on polynomial weights $S(x)$ of one real variable only.
What seemed worth mentioning is the fact that among the possible solutions
of the Schwinger-Dyson equation, there are a number of preferred ones
(see def \ref{phasedef}) which we call ``equilibria'', because
these solutions have some properties in common with equilibria:
For example, using
the weight $S(x)=-x^2/2+x^4/4$ one finds equilibria with $\langle x\rangle =\pm
1$, which
are the values where $S$ has a minimum. We don't claim that
these two things are really the same,
but the similarities seemed suggestive enough to use the word equilibrium.

Polynomial weights are particulary illustrative, because both the
solution of the Schwinger-Dyson equation and the positivity condition can
be automated by computer algebra.

\subsection{Algorithmic solution and positivity of the Schwinger-Dyson
equation.}
\subsubsection{An algorithm for the solution of the Schwinger-Dyson equation.}
Let $S(x)$ be a polynomial of degree $D+1$, and let $s:=\partial S$. The
Schwinger-Dyson equation for polynomial integrands $p$ reads
$\langle sp\rangle =\langle \partial p\rangle $.
One way to transform this equation into a recurrence relation is to write
$s$ as $s(x)=\sigma x^D+ \rho(x)$, so that:
$$\langle x^D p\rangle =\sigma^{-1}\langle \partial p-\rho p\rangle ,$$
This is a recurrence relation, which will fix $\langle .\rangle $ once
$\langle x\rangle ,\langle x^2\rangle ,...,\langle x^{D-1}\rangle $ are known.
The Schwinger-Dyson equation does not
determine these values, so that the (normalized) solutions of the
Schwinger-Dyson equation form a $(D-1)-$parameter family. Equivalently, the
Schwinger-Dyson equation in differential form, see footnote \ref{diffsd},
is of order $D$, so we get $D$ integration constants parametrizing the
solutions, of which we substract one for normalization.

A maple procedure that will determine the expectation value $\langle p\rangle $
for
$s=\partial S$ in terms of $y_i:=\langle x^i\rangle $ for low $i$ is the
following:

\begin{verbatim}
Ex:=proc(p:algebraic,s:algebraic,x:name,y:name)
	local r,q,i:
	r:=rem(p,s,x,'q'):
	if q=0 then
		subs({seq(x^i=y.i,i=1..degree(r,x))},r):
	else Ex(p,s,x,y):=simplify(Ex(diff(q,x),s,x,y)+Ex(r,s,x,y))
	fi:
end:
\end{verbatim}

The point is to write $p$ as $p=sq+r$, where the degree of $r$ is lower than
that of $s$. In that case $sq$ can be replaced by $\partial q$, and
the expectation value of $r$ is determined by replacing $x^i$ by $y_i$.

\subsubsection{An algorithm to generate the positivity conditions.}
We will generate a countable number of conditions which will ensure that
for all polynomials $p> 0$, we have $\langle p\rangle  > 0$. To that end we
inductively
define the following bilinear symmetric forms on polynomials, depending
on $\langle .\rangle $:
$$g_0(p,q):=\langle pq\rangle ;
g_{n+1}(p,q):=g_n(1,1)g_n(xp,xq)-g_n(xp,1)g_n(xq,1).$$
\begin{thm} Let $\langle .\rangle $ be a linear form on polynomials. Then we
have the following equivalence:
$$\{\forall_{p>0}\;\langle p\rangle >0\}\Leftrightarrow
\{\forall_n\;g_n(1,1)>0\}.$$
\end{thm}
\begin{pr}
First, every positive real polynomial is a sum of squares
of nonzero polynomials:
Indeed, such a polynomial does not have any real roots, and its complex
roots come in conjugate pairs, so that it can be written as
$\prod_i (x-a_i)(x-\bar a_i)$. Now $(x-a_i)(x-\bar a_i)$ is itself a
positive polynomial, so it suffices to prove that it can be written as
a sum of nonzero squares. Indeed, any second order polynomial is of the form
$(x-b)^2+c$, and this is positive iff $c>0$, so that $c=d^2$.

Thus, to prove positivity of $\langle .\rangle $, it suffices to prove that
$\forall_{p\neq 0}\;\langle pp\rangle > 0$, i.e. $\forall_{p\neq 0}\;g_0(p,p)>
0$.
Let us now prove the following statement:
$$g_n(1,1)>0\Rightarrow [\{\forall_{p:|p|=k+1,p\neq 0}\;g_n(p,p)>
0\}\Leftrightarrow
\{\forall_{p:|p|=k,p\neq 0}\;g_{n+1}(p,p)> 0\}].$$
Indeed, assume that $g_n(1,1)>0$, then we have to prove that $[A\Leftrightarrow
B]$:
$$A\Leftrightarrow \forall_{p\neq 0:|p|=k,b\in \bR}\;0<  g_n(xp+b,xp+b)
=b^2 g_n(1,1) +2b g_n(xp,1) + g_n(xp,xp).$$
The last statement is equivalent to the discriminant being smaller than zero:
$$0> (2 g_n(xp,1))^2-4g_n(xp,xp)g_n(1,1)=-4 g_{n+1}(p,p)
\Leftrightarrow g_{n+1}(p,p)> 0\Leftrightarrow B.$$
With this result, we can prove the theorem:

$(\Rightarrow)$. First assume that $\langle p\rangle $ is positive. Then
$g_0(1,1)>0$.
Therefore, using the above implication, we have
$\forall_{p\neq 0}\;g_0(p,p)>0\Leftrightarrow \forall_{p\neq 0} \;g_1(p,p)>0$,
so that $g_1(1,1)>0$. This in turn allows us to use the implication again, so
we get $g_2(p,p)>0$, etc.

$(\Leftarrow)$. Now assume that $\forall_n g_n(1,1)>0$. Then we have
$\forall_n$:
$$\{\forall_{p:|p|=k+1,p\neq 0}\;g_n(p,p)> 0\}\Leftrightarrow
\{\forall_{p:|p|=k,p\neq 0}\;g_{n+1}(p,p)> 0\}.$$
Thus we may deduce that
$$
\forall_k \; g_k(1,1)> 0\Leftrightarrow
\forall_{k,p\neq 0:|p|=0}\; g_k(p,p)>0 \Leftrightarrow$$
$$\forall_{k,p\neq 0:|p|=k}\; g_0(p,p)> 0 \Leftrightarrow
\forall_{p\neq 0}\; g_0(p,p)>0.$$
\end{pr}

\begin{rem} For example, we have
$$g_0(1,1)=\langle 1\rangle (:=1),$$
$$g_1(1,1)=\langle x^2\rangle -\langle x\rangle ^2,$$
$$g_2(1,1)=(\langle x^2\rangle -\langle x\rangle ^2)(\langle x^4\rangle
-\langle x^2\rangle ^2)-(\langle x^3\rangle -\langle x^2\rangle \langle
x\rangle )^2.$$
\end{rem}

\begin{defn}
\label{phasedef}
By an equilibrium for the weight $S$, we mean a solution $\langle .\rangle $ of
the
Schwinger-Dyson equation for $S$, which satisfies $\forall_n g_n(1,1)\geq 0$.
\end{defn}

\begin{rem}
Note that $g_n(1,1)$ is allowed to be zero. In particular, an equilibrium need
not be
a positive solution of the Schwinger-Dyson equation.

The $g_n$'s can be computed in Maple using the following procedure, which
computes $g_n(p,q)$ for the derived weight $s=\partial S$:

\begin{verbatim}
DEx:=proc(n:nonnegint,p,q,s:algebraic,x:name,y:name)
	if n=0 then Ex(p*q,s,x,y)
	else DEx(n,p,q,s,x,y):=
		DEx(n-1,1,1,s,x,y)*DEx(n-1,x*p,x*q,s,x,y)
		-DEx(n-1,x*p,1,s,x,y)*DEx(n-1,x*q,1,s,x,y)
	fi:
end:
\end{verbatim}
\end{rem}
\begin{defn}
By a null-equilibrium, we mean an equilibrium such that\\
$\exists_N\forall_{n\geq N} g_n(1,1)=0$.
\end{defn}
\begin{thm} (The following are useful to prove that an equilibrium is a
null-equilibrium:)
\label{nullphaseth}
\begin{enumerate}
\item $g_n(1,1)=0\Rightarrow \{g_{n+1}(1,1)=0 \Leftrightarrow g_n(x,1)=0\}$.
\item $\{g_n(1,1)=g_n(x,1)=0\}\Rightarrow g_{n+1}(p,1)=0$.
\end{enumerate}
\end{thm}
\begin{pr}
\begin{enumerate}
\item
$$g_{n+1}(1,1)=g_n(1,1)g_n(x,x)-g_n(x,1)g_n(x,1)=-g_n(x,1)^2.$$
\item
$$g_{n+1}(p,1)=g_n(1,1)g_n(xp,x)-g_n(xp,1)g_n(x,1)=0.$$
\end{enumerate}
\end{pr}
\subsection{Third order and fourth order weights.}

\begin{thm} Using the above procedures for calculations, one can deduce:
\begin{enumerate}
\item $S(x)={1\over 2}x^2+{1\over 3}x^3$ has no equilibria.
\item $S(x)=-{1\over 2}x^2+{1\over 4} x^4$ has multiple equilibria, of which we
list a few:\\\\
\begin{tabular}{|l|l|l|l|l|l|l|l|}
\hline
Name of eq.& $\langle x\rangle $ & $\langle x^2\rangle $ & $\langle x^3\rangle
$ & $\langle x^4\rangle $ & $g_1(1,1)$ & $g_2$ & $g_3$\\
\hline
$1_{r}$ & $1$ & $1$ & $1$ & $2$ & $0$ & $0$ & \\
\hline
$1_{l}$ & $-1$ & $1$ & $-1$ & $2$ & $0$ & $0$ & \\
\hline
$1_{mid}$ & $0$ & $0$ & $0$ & $1$ & $0$ & $0$ & \\
\hline
$2_{mid}$ & $0$ & ${1\over 2}(1+\sqrt{5})$ & $0$ & ${1\over 2}(3+\sqrt{5})$ &
${1\over 2}(1+\sqrt{5})$ & $0$ & $0$ \\
\hline
.. & .. & .. & .. & .. & .. & .. & ..\\
\hline
$\infty$ & $0$ & $\int x^2 e^{-S} dx.$ & $0$ & $\int x^4 e^{-S}dx$ & $>0$ &
$>0$ & \\
\hline
\end{tabular}\\\\
\item There is no equilibrium for $S(x)=-{1\over 2}x^2 + {1\over 4} x^4$ with
$\langle x\rangle ={1\over 2}$.
\end{enumerate}
\end{thm}

\begin{pr}
\begin{enumerate}
\item As an illustration of the use of maple, one may type:
\begin{verbatim}
s:=x+x^2;
ic1:=a;
for i from 1 to 3 do
	expect.i:= simplify(Ex(x^i,s,x,ic));
	eq.i:=simplify(DEx(i,1,1,s,x,ic));
	evalf(solve(eq.i,a));
od;
\end{verbatim}
This will tell us that the first two positivity equations for
the undetermined integration constant $\langle x\rangle =a$ read:
$$-a^2-a\geq 0;\;\;-2 a^3-3a^2-a-1\geq 0.$$
These conditions are equivalent to approximately $a\in [-1,0]$, and $a\leq
-1.4$,
which is impossible, so that this weight has no equilibria.
\item  For this weight there are two integration constants, $a:=\langle
x\rangle $ and $b:=\langle x^2\rangle $.
In terms of these one finds
$$g_1(1,1)=b-a^2,$$
$$g_2(1,1)=b+b^2-b^3-2a^2+a^2 b.$$
Intersecting the zero's of $g_1$ and $g_2$ in the $(a,b)$-plane gives the
points $(-1,1),(0,0),(1,1)$.
Therefore, by theorem \ref{nullphaseth}, $g_{n\geq 1}(1,1)=0$ at those points,
so that
these are null equilibria.
Other points may be found by intersecting
$g_n$ with $g_{n+1}$.
Further, there is obviously a positive equilibrium,
corresponding to usual integration:
$\langle f\rangle :=\int e^{-S} f dx.$
\item Assume $\langle x\rangle ={1\over 2}$, and set $b:=\langle x^2\rangle $.
Then the first three
conditions read:
$$g_1=b-{1\over 4}\geq 0,$$
$$g_2={5\over 4} b + b^2-b^3 -{1\over 2} \geq 0,$$
$$g_3={11\over 4} b^3 - {107\over 16} b^2 + {35\over 16} b + {19\over 4} b^4
-3 b^5 -{3\over 16}\geq 0.$$
These are incompatible. (By just plotting the graphs with high enough
resolution.)
\end{enumerate}
\end{pr}
%
%
%
%
%
\newpage
\section{Non-polynomial contractions.}
\label{nonpolsection}
\subsubsection{More general relations.} Let us drop the topic of
uniqueness of solutions of the Schwinger-Dyson equation for a moment,
and consider for example the case in
which $\partial^2 S$ cannot be written as a function of $\partial S$;
This is the case for example for $S(x)=x^3$.
Let us first recall
$$\langle \partial_{i_1}(S)...\partial_{i_n}(S)\rangle =
\sum_{k=2}^n \langle \partial_{i_2}(S)..\partial_{i_1}\partial_{i_k}(S)..
\partial_{i_n}(S)\rangle .$$
Now if $\partial^2 S$ cannot be written in terms of first derivatives, then
we have an additional independent equation
$$\langle \partial_{i_1}(S)...\partial_{i_n}(S)\partial_l\partial_m(S)\rangle =
\sum_{k=2}^n \langle \partial_{i_2}(S)..\partial_{i_1}\partial_{i_k}(S)..
\partial_{i_n}(S)\partial_l \partial_m (S)\rangle $$
$$+\langle
\partial_{i_2}(S)...\partial_{i_n}(S)\partial_{i_1}\partial_l\partial_m(S)\rangle ,$$
and analogously for higher powers in $\partial^2 S$.
If $\partial^3 S$ can be written as a polynomial in lower degree derivatives,
then there are no more equations to be considered, otherwise we may go on.
So, recalling that $\langle .\rangle $ was defined on the algebra $\cS$
generated by all derivatives of $S$, we see that
that the only things that are relevant
are the algebraic relations that exist between the various derivatives
of $S$. Contraction algebras are objects in which the concept of a
weight $S$ has been repaced by relations.

The universal contraction algebra is the algebra in which there are no
relations at all. It is just the algebra of formal combinations like
$$(\partial_i \partial_j)(\partial_k \partial_l \partial_m),
\;or\;[\partial_i\trr\partial_j][\partial_k\trr[\partial_l\trr\partial_m]] $$
which when given a weight $S$ are mapped to
$(\partial_i \partial_j S)(\partial_k \partial_l \partial_m S);$
The kernel of this map determines relations in the universal contraction
algebra, and so what we mean by a contraction algebra is the
universal contraction algebra subject to a number of relations. So
a contraction algebra is an abstraction of a weight. The introduction
of contraction algebras serves a number of purposes:
\begin{enumerate}
\item They save space, because $XYZ$ is shorter than $(XS)(YS)(ZS)$.
\item Some constructions can be made even for the universal contraction
algebra, and are therefore weight-independent.
\item For applications to the infinite dimensional case, it is useful to
be able to manipulate the expression
$(\frac
{\delta}{\delta \phi(x)}
\frac
{\delta}{\delta \phi(x)}
)$ in the
universal algebra, even if its actual evaluation
$(\frac
{\delta}{\delta \phi(x)}
\frac
{\delta}{\delta \phi(x)}
S)$
is undefined.
\item It may be easier to specify relations than to explicitly give a
weight that actually satisfies these relations.
\end{enumerate}
\subsubsection{The noncommutative case, a number of definitions.}
\begin{rem} Non-commuting vectorfields lead to a number of complications:
\begin{enumerate}
\item We have to take all commutators into account.
\item Like in remark \ref{mot_cont_def}, it is better to talk about
divergences $\nabla$ than about weights $S$.
\item
We will also want to consider
 combinations of the form $(X_1 X_2)(Y_1 Y_2 Y_3)$,
which again are to be thought of as $(X_1 X_2 S)(Y_1 Y_2 Y_3 S)$, or
rather when given a divergence as
$(-X_1\nabla(X_2))(-Y_1 Y_2 \nabla(Y_3))$.
Since in any case we will have $XY(S)-YX(S)=[X,Y](S)$, we lose no
information when dividing out these expressions by the relation $XY-YX=[X,Y]$.
This will also be true for any divergence, because they
always satisfy $\nabla([X,Y])=X(\nabla(Y))-Y(\nabla(X))$, see theorem
\ref{vol_div_th}.
In other words, the expressions like $(X_1 X_2)$ and $(Y_1 Y_2 Y_3)$ between
the brackets are in the
universal enveloping algebra of a Lie algebra, and since we want
to multiply these expressions in turn, we end up defining the
universal contraction algebra as $Sym(UEA(L))$, the
symmetric algebra of the universal enveloping algebra of a Lie algebra $L$.
Finally, if $\tilde I$ satisfies the Schwinger-Dyson equation, we may as well
consider its pullback to $Sym(UEA(L))$:
$$I((X_1 X_2)(Y_1 Y_2 Y_3)):=\tilde I((X_1 X_2 S)(Y_1 Y_2 Y_3S)).$$
For $I$ the Schwinger-Dyson equation now reads $I(Xs)=I([X\trr s])$,
using the notation introduced below:
\end{enumerate}
\end{rem}
\begin{defn} Let $UEA(L)$ denote the universal enveloping algebra of $L$.
We define $UEC(L):=Sym(UEA^\prime(L))$.
\footnote{We write $UEA^\prime$ instead of just $UEA$ to indicate that
we do not include multiples of the unit in $UEA$,
which would lead to confusions
when combined with the unit in $Sym$. Note that when given a weight
$S$ the unit in $UEA$ would
get mapped to $S$, not to $1$. We will not keep writing that prime though.}
We denote the left-multiplicative action of an element $X\in L$ on $UEA(L)$ by
$[X\trr.]$, and idem for the induced action by derivations on $UEC(L)$.
This being so, we will not use the round brackets $(.)$ in $UEC$ any more,
and prefer to write
$[Y\trr Z] Z^2 [X\trr[Y\trr X]] \in UEC(L)$ instead of $(YZ)(Z)(Z)(XYX)$.
$[X\trr Y]$ is referred to as the contraction of $X$ and $Y$.

An ideal in $UEC(L)$ is a subspace $I\leq UEC(L)$ such that $I.UEC(L)\subset I$
and
$[L\trr I]\subset I$. A contraction algebra is the quotient $UEC(L)/I$
by some ideal. The intersection of a collection of ideals is again an ideal.
A contraction algebra can be specified by giving relations, i.e.
a subset $R\subset UEC(L)$; In that case it is understood that the quotient is
taken by the smallest ideal containing $R$.
\end{defn}
\begin{defn} Equivalently: A contraction algebra is a combination
$(L,A,\nabla)$, where $L$ is a Lie algebra, $A$ is an associative commutative
algebra with unit, $L$ is represented on $A$ by derivations, and
$\nabla:L\rightarrow A$, such that:
\begin{enumerate}
\item $\nabla([X,Y])=X(\nabla(Y))-Y(\nabla(X))$,
\item The algebra homomorphism $\nabla:UEC(L)\rightarrow A$ defined by
$$[X_1\trr[..[X_{n-1}\trr X_n]..]\mapsto X_1..X_{n-1}(-\nabla(X_n))$$
is surjective.
\end{enumerate}
The relation with the previous definition being $I:=Ker(\nabla)\leq UEC(L)$,
and $A:=UEC(L)/I$.
\end{defn}
\begin{rem}
In the main body of the text we have restricted our attention to
polynomial contractions.
In terms of the above defintion, this means that $A=Sym(L)$, and
that $\nabla$ is just the inclusion $L\rightarrow Sym(L)$.
Relations of polynomial type defined in \ref{pol_contr_def}
induce a map $UEC(L)\rightarrow Sym(L)$.
\end{rem}
%
%
%
%
%
\newpage
\section{More complicated Schwinger-Dyson equations.}
\label{geometric_section}
\subsection{The Schwinger-Dyson equation for differential forms.}
In this section we will be concerned with generalizing the weight
$e^{-S} dx^1 .. dx^n$ to any volumeform. Since we have the infinite dimensional
case in mind, we will avoid talking about
volumeforms, and instead talk about divergences $\nabla$.
This is enough for our purpose since the Schwinger-Dyson equation
only depends on $\nabla$, not on the whole $\mu$.
More generally, we will look for a formulation of the Schwinger-Dyson
equation for ``integration of forms'' of finite codegree,
applicable in infinite dimensions. Therefore we will first concentrate
our efforts on defining forms of finite codegree in possibly infinite
dimensions.
Only after that will we look for the notion of integration of these objects.
We will also consider integrals with prescribed boundary terms.
\begin{defn}
By an infinitesimal calculus $(A,L)$, we mean an associative symmetric
algebra $A$ with unit, a Lie algebra $L$, a representation $(f,X)\mapsto fX$
of $A$ on $L$ and a representation $(X,f)\mapsto Xf$ of $L$ on $A$ by
derivations, such that: $(gX)f=g(Xf)$ and $[X,fY]=f[X,Y]+(Xf)Y$.
By $\Omega(L,A)$ we mean antisymmetric $A$-linear forms on $L$ with
values in $A$.
\end{defn}
\begin{rem}
To every manifold is associated an infinitesimal calculus, by letting
$A$ be the real functions, and $L$ the vectorfields.
\end{rem}
\begin{defn}
A volume manifold is a combination $(M,\mu)$, where $M$ is a manifold and
$\mu$ is a volume-form, i.e. a differential form of maximal degree
such that $\mu_m$ is nonzero in every point $m\in M$.
In addition to the infinitesimal calculus
$(A,L)$ associated to $M$ alone, we may define
$\nabla:L\rightarrow A$ by the defining property
$\nabla(X)\mu:=L_X\mu.$
\end{defn}
\begin{thm} $\nabla$ above satisfies the following properties:
\label{vol_div_th}
\begin{enumerate}
\item It is closed: $\nabla([X,Y])=X(\nabla(Y))-Y(\nabla(X))$.
\item It is local: $\nabla(fX)=X(f)+f\nabla(X).$
\item $\nabla$ fixes $\mu$ up to multiplication by a locally constant function.
\end{enumerate}
\end{thm}
\begin{pr}
\begin{enumerate}
\item We have
$$\nabla([X,Y])\mu=L_{[X,Y]}\mu=L_X L_Y\mu-L_Y L_X \mu=
L_X(\nabla(Y)\mu)-L_Y(\nabla(X)\mu)$$
$$=X(\nabla(Y))\mu+\nabla(Y)\nabla(X)\mu-Y(\nabla(X))\mu-\nabla(X)\nabla(Y)\mu
$$
$$=(X(\nabla(Y))-Y(\nabla(X)))\mu.$$
\item
$$\nabla(fX)\mu=(i_{fX}d+di_{fX})\mu=d(fi_X\mu)=df\wedge i_X\mu+fdi_X\mu$$
$$=-i_X(df\wedge \mu)+(i_Xdf)\wedge \mu + fL_X\mu=(X(f)+f\nabla(X))\mu.$$
\item Finally, let $\nu$ be another volume form giving the same $\nabla$.
Since volume forms are proportional, there is a function $f$ such that
$\nu=f\mu$. Thus we have
$$X(f)\mu=L_X(f\mu)-fL_X\mu=L_X\nu-f\nabla(X)\mu=\nabla(X)\nu-\nabla(X)\nu=0.$$
So $X(f)=0$, i.e. $f$ is locally constant.
\end{enumerate}
\end{pr}
\begin{rem}
We might roughly state the above as follows: If we are not interested in
a particular normalization of the integral, then all the information
contained in $(M,\mu)$ is in $(A,L,\nabla)$.
\end{rem}
\begin{defn} This motivates the definition of a formal volume manifold:
\begin{enumerate}
\item By a formal volume manifold, we mean a combination $(A,L,\nabla)$, where
$(A,L)$ is an infinitesimal calculus and $\nabla$ is a divergence for
$(A,L)$, meaning that $\nabla:L\rightarrow A$, such that:
\begin{enumerate}
\item $\nabla([X,Y])=X(\nabla(Y))-Y(\nabla(X))$.
\item $\nabla(fX)=X(f)+f\nabla(X).$
\end{enumerate}
\item Next for any infinitesimal calculus, and $k\in \bZ$, we set
$$\bar \cI^{-k}(L,A):=\bigoplus_{n\in \bN} \Omega^{n-k}(L,A) \otimes_A
{\bigwedge}^{n}_A(L);\;\;\bar \cI:=\bigoplus_{k\in \bZ} \bar \cI^k.$$
\end{enumerate}
\end{defn}
\begin{rem}
Note that the
combination $(L,A:=Sym(UEA(L)))$ that we saw before
has properties similar to formal volume
manifolds if we let $\nabla$ just be the inclusion. It is not exactly
a formal volume manifold though, since $A$ does not act on $L$.
Next, just as there is a theory of integration over volume manifolds,
we will now look for a theory of integration over formal volume manifolds.
We will start by formalizing the integrands first,
and use $\bar \cI^{-k}$ to model
the notion of a differential form of codegree $k$, in view of the map
below:
\end{rem}
\begin{defn}
Indeed, suppose that $(A,L)$ comes from $(M,\mu)$, then we define
$$\bar G:\bar \cI^{-k} \rightarrow \Omega^{|M|-k}(M);
\omega\otimes X_1...X_n\mapsto \omega\wedge i_{X_1}...i_{X_n}\mu.$$
\end{defn}
\begin{rem}
The Cartan calculus of Lie derivation $L_X$, interior product $i_X$, and
exterior derivation $d$, (see appendix \ref{nat_sup_section}),
present in $(M,\mu)$ can be largely transported to
$\bar \cI(A,L)$, as we will see in the next theorem. This motivates the
following definition:
\end{rem}
\begin{defn}
Let $(L,A)$ be an infinitesimal calculus.
Then $Car(L)$ is naturally represented on $\bar \cI(A,L)$, by the
supertensorproduct
representation of those on $\Omega(L,A)$ and $\bigwedge(L)$ seperately.
(See appendix \ref{nat_sup_section}).
Denote this tensor representation by $\tilde i_X,\tilde L_X,\tilde d$.
Given a divergence $\nabla$ for $(A,L)$, we define another
map $Car(L)\rightarrow End(\bar \cI)$; (It is not a representation):
\begin{enumerate}
\item $i_X:=\tilde i_X$.
\item $L_X:=\tilde L_X+M_{\nabla(X)}$, where $M_f$ denotes multiplication with
$f$.
\item $d:=\tilde d+ \delta$, where
$$\delta(\omega X_1...X_n):=(-1)^{|\omega|}\sum_{i=1}^n (-1)^{i+1}
\nabla(X_i)\omega X_{[1,n]
\backslash i}.$$
\end{enumerate}
\end{defn}
\begin{rem}
In general the operators $L_X,i_X,d$ above do not form a representation of the
Cartan algebra. They do however modulo elements of the form
$i_{X_1}..i_{X_n}\phi$,
where $\phi$ is an element of positive degree, as we will see in a moment.
Since $\bar G$ annihilates these elements
anyway, we lose nothing of interest when taking the quotient
by the subspace $\cO$ of these "overflow" elements:
\end{rem}
\begin{defn} Let $(A,L)$ be an infinitesimal calculus. Then we define:
\begin{enumerate}
\item The (graded) subspace of overflow forms: $\cO\leq \bar \cI(A,L)$,as
the space generated by elements of the form $i_{X_1}..i_{X_n}\phi$, where
$|\phi|>0$ and $n\geq 0$.
\item Next, we set $\cI(A,L):=\bar \cI(A,L) /\cO$, with induced grading.
\item Finally, if $(A,L)$ comes from $(M,\mu)$, then we define the map
$G:\cI(A,L)\rightarrow \Omega(M)$ to be the one induced by $\bar G$.
\end{enumerate}
\end{defn}
\begin{cor} Thus, we have a number of identities in $\cI(A,L)$, e.g.:
\begin{enumerate}
\item $\cI^{>0}(A,L)=0$: This is the case $i_{X_1}...i_{X_n}\phi=0$ with $n=0$.
\item $X(g)f\otimes 1=(dg)f\otimes X$: $0=i_X(fdg\otimes 1)=fX(g)\otimes 1 -
fdg\otimes X$.
\end{enumerate}
\end{cor}
\begin{thm} If $(L,A,\nabla)$ comes from $(M,\mu)$, then:
\begin{enumerate}
\item $\bar G i_X=i_X\bar G$.
\item $\bar G L_X=L_X\bar G$.
\item $\bar Gd=d\bar G$.
\item For any $(L,A,\nabla)$, the operators $L_X,i_X,d$ descend to a
representation of the Cartan
algebra on $\cI(A,L)$.
\end{enumerate}
\end{thm}
\begin{pr}
\begin{enumerate}
\item
$$i_X \bar G(\omega X_1...X_n)=i_X(\omega i_{X_1}...i_{X_n}\mu)$$
$$=(i_X\omega) i_{X_1}..i_{X_n}\mu+(-1)^{|\omega|}
\omega i_Xi_{X_1}..i_{X_n}\mu$$
$$=\bar G((i_X\omega)X_1...X_n
+(-1)^{|\omega|}
\omega XX_1..X_n)
=\bar G i_X(\omega X_1...X_n).$$
\item
$$L_X\bar G(\omega X_1..X_n)=L_X(\omega i_{X_1}...i_{X_n}\mu)$$
$$=(L_X \omega)i_{X_1}...i_{X_n}\mu+
\omega \sum_{j=1}^n i_{X_1}..[L_X,i_{X_j}]..i_{X_n}\mu
+\omega i_{X_1}..i_{X_n} \nabla(X)\mu$$
$$=\bar G(\tilde L_X(\omega X_1..X_n)+\nabla(X)\omega X_1..X_n)
=\bar G L_X(\omega X_1..X_n).$$
\item
$$d\bar G(\omega X_1..X_n)=d(\omega i_{X_1}..i_{X_n}\mu)$$
$$=(d\omega)i_{X_1}..i_{X_n}\mu+
(-1)^{|\omega|} \sum_{j=1}^n (-1)^{j+1} \omega
i_{X_1}..([d,i_{X_j}]=L_{X_j})..i_{X_n}\mu$$
$$=(d\omega)i_{X_1}..i_{X_n}\mu
+(-1)^{|\omega|}
\sum_{j=1}^n(-1)^{j+1}
\nabla(X_j)\omega i_{X_1}...i_{X_n} \mu$$
$$+(-1)^{|\omega|} \sum_{j=1}^n (-1)^{j+1} \omega i_{X_1}..\hat i_{X_j}
\sum_{k>j} i_{X_{j+1}}..([L_{X_j},i_{X_k}]=i_{[X_j,X_k]})..i_{X_n}\mu$$
$$=\bar G((d\omega)X_1..X_n+\delta(\omega X_1..X_n)
+(-1)^{|\omega|}\omega \sum_{j<k}(-1)^{j+k+1}[X_j,X_k]X_{[1,n]\backslash
ij}).$$
\item
First, using the fact that $\tilde i_X,\tilde L_X,\tilde d$ form a
representation,
we see that a number of commutators are already correct in $\bar \cI$:
\begin{enumerate}
\item $[i_X,i_Y]=[\tilde i_X,\tilde i_Y]=0$.
\item $[L_X,i_Y]=[\tilde L_X,\tilde i_Y]+[M_{\nabla(X)},\tilde i_Y]
=\tilde i_{[X,Y]}+0=i_{[X,Y]}.$
\item $[d,i_Y]-L_Y=[\tilde d+\delta,\tilde i_Y]-\tilde L_Y -M_{\nabla(Y)}
=[\delta,i_Y]-M_{\nabla(Y)}=0:$
$$([\delta,i_Y]-M_{\nabla(Y)})(\omega X_1..X_n)
=(\delta i_Y+i_Y\delta)(\omega X_1..X_n)-\nabla(Y)\omega X_1..X_n$$
$$=\delta ((i_Y\omega)X_1..X_n+(-1)^{|\omega|}\omega YX_1..X_n)
-\nabla(Y)\omega X_1..X_n$$
$$+i_Y(-1)^{|\omega|}\sum_{j=1}^n (-1)^{j+1}
\nabla(X_j)\omega X_{[1,n]\backslash j}$$
$$=(-1)^{|\omega|-1}
\sum_{j=1}^n (-1)^{j+1}
\nabla(X_j)(i_Y\omega) X_{[1,n]\backslash j}$$
$$+\nabla(Y)\omega X_1..X_n
-\sum_{j=1}^n (-1)^{j+1}\nabla(X_j)\omega Y X_{[1,n]\backslash j}
-\nabla(Y)\omega X_1..X_n$$
$$+\sum_{j=1}^n (-1)^{j+1}
\{(-1)^{|\omega|} \nabla(X_j)(i_Y\omega)X_{[1,n]\backslash j}
+\nabla(X_j) \omega Y X_{[1,n]\backslash j}\}=0.$$
\item $$[L_X,L_Y]=[\tilde L_X+M_{\nabla(X)},\tilde L_Y+M_{\nabla(Y)}]$$
$$=[\tilde L_X,\tilde L_Y]+[\tilde L_X,M_{\nabla(Y)}]-[\tilde
L_Y,M_\nabla(X)]$$
$$=\tilde L_{[X,Y]}+M_{X(\nabla(Y))}-M_{Y(\nabla(X))}
=\tilde L_{[X,Y]}+M_{\nabla([X,Y])}
=L_{[X,Y]}.$$
\end{enumerate}
Applying the operators $L_X,i_X,d$ to the expression $i_{X_1}..i_{X_n}\phi$,
and using the above commutation relations proves that $\cO$ is invariant,
so that the operators descend to $\cI(A,L)=\bar \cI(A,L)/\cO$.
So it remains to prove that the following commutators are zero modulo $\cO$:
$R_X:=[L_X,d];\;\Delta:=[d,d]$. First, by lemma \ref{i_Xlemma},
$[i_X,R_Y]=0$, and $[i_X,\Delta]=2R_X$.
Using this,
we will now prove by induction on $n$ that $\Delta$ and $R_X$ are
zero mod $\cO$ on $\omega X_1..X_n$.
Indeed, consider the case $n=0$ first:
\begin{enumerate}
\item
$$R_X(\omega\otimes 1)=(L_X d-dL_X)(\omega\otimes 1)$$
$$=L_X(d\omega\otimes 1)-d(L_X\omega\otimes 1
+
\nabla(X)\omega\otimes 1)$$
$$=L_Xd\omega\otimes 1+\nabla(X)d\omega\otimes 1$$
$$-dL_X\omega\otimes 1-d(\nabla(X))\wedge\omega\otimes 1
-\nabla(X)d\omega\otimes 1\in \cO.$$
\item
$$\Delta(\omega\otimes 1)=2d^2(\omega\otimes 1)=2(d^2\omega\otimes 1)=0.$$
\end{enumerate}
So the statement is true for $n=0$. Assume it has been proved for
$\omega X_1..X_n$, and let $A$ be any linear combination of the
operators $R_X$ and
$\Delta$. We have to prove that $A(\omega Y X_1..X_n)\in \cO$. Using that
$[A,i_Y]=A^\prime$, where $A^\prime$ is also of that form, we have:
$$(-1)^{|\omega|}A(\omega Y X_1..X_n)=A(
i_Y(\omega X_1..X_n)
-
(i_Y \omega)X_1..X_n)$$
$$= [A,i_Y](\omega X_1..X_n)
\pm i_Y A(\omega X_1..X_n)
-A((i_Y\omega)X_1..X_n)$$
$$= A^\prime(\omega X_1..X_n)
\pm i_Y A(\omega X_1..X_n)
-A((i_Y\omega)X_1..X_n)\in \cO,$$
by induction.
\end{enumerate}
\end{pr}
\subsubsection{The Schwinger-Dyson equation for forms.}
\begin{rem}
Now that we have a reasonable grasp of the integrands, we will proceed
to define formal integrals.
\end{rem}
$
$
\begin{defn} Let $(A,L)$ be an infinitesimal calculus, then
by an integral we mean a map $I:\cI(A,L)\rightarrow \bR$.
Such an integral is said to be of codimension $k$ if it is zero on the
homogeneous subspaces $\cI^{l\neq -k}$.
Further, given a divergence $\nabla$, we define $\partial I:=I\circ d$.
In that case we define the Schwinger-Dyson equation (without boundary)
for $I$ to be $\partial I=0$. We will see in a moment that if
$I$ is of codimension zero, then this is the usual Schwinger-Dyson
equation.
\end{defn}
\begin{cor} Let $(A,L,\nabla)$ come from $(M,\mu)$. Let $N\subset M$ be
a submanifold. Define $I_N:\cI\rightarrow K$ by
$$I_N(\phi):=\int_N G(\phi).$$
\label{I_verses_manifold}
Then:
\begin{enumerate}
\item $codim(I_N)=codim(N)$.
\item $\partial I_N=I_{\partial N}.$
\end{enumerate}
\end{cor}
\begin{rem} This ends our discussion of the integration of forms.
Our final aim in this section is to show that the formulation of the
Schwinger-Dyson equation for some functional $I$ in the geometric language
reduces to the usual one in codegree zero, and to make a statement
concerning the equation Schwinger-Dyson equation with fixed nonzero
boundary term $\tilde J$ (Compare remark \ref{boundary_rem}):
\end{rem}
\begin{thm} Let $I$ be of codimension zero. Then the following are equivalent:
\label{boundary_schdys}
\begin{enumerate}
\item $\partial I=\tilde J$.
\item $I(X(f)+\nabla(X))=\tilde J(f\otimes X)$.
\end{enumerate}
\end{thm}
\begin{pr} First, for any $I$ we have the identity
$$\partial I(f\otimes X)=I(di_X(f\otimes 1))=I(di_X+i_Xd)(f\otimes 1)
=IL_X(f\otimes 1)=I(X(f)+\nabla(X)),$$
which proves the implication $(1)\Rightarrow (2)$. On the other hand,
starting from $(2)$, it proves that $\partial I(f\otimes X)=\tilde J(f\otimes
X)$,
so it remains to prove that $\partial I(\omega\otimes X_0..X_n)=
\tilde J(\omega\otimes X_0..X_n)$, for $|\omega|=n$,which we prove by
induction.
We just saw that this is true for $n=0$, and for $n+1$, we note that
if $|\omega|=n+1$, then $\omega YX_0..X_n=(-1)^{|\omega|} (i_Y\omega)
X_0..X_n$,
since $i_Y(\omega X_0..X_n)$ is an overflow form.
\end{pr}
\begin{rem}
For $\tilde J=0$ this is the usual Schwinger-Dyson equation: Indeed, taking
$\mu:=e^{-S}\mu_{Leb}$, we have:$\nabla(\partial_i)=-\partial_i S$, so that
$0=I((\partial_i f-f\partial_i(S))\otimes 1)$.
The equation for more general vectorfields follows from
$\nabla(fX)=X(f)+f\nabla(X)$.
\end{rem}
\pagebreak[4]
\
\subsection{The Schwinger-Dyson equation on quotient manifolds.}
\nopagebreak[4]
\subsubsection{Integration over quotient manifolds.}
\begin{rem} We will first note some properties of integration
over quotient manifolds, and after that make the algebraic abstraction
for the Schwinger-Dyson equation. The content of this section is
a compact reformulation of formulas from
\cite{fadeev}, \cite{becchi} and \cite{batalin}.
\end{rem}
\begin{defn} Let $G$ be a connected Lie group with Lie algebra $\gog$.
By a $G$-structure for a volume manifold $(M,\mu)$ we will understand
an action of $G$ on $M$, and an element $Q_\mu\in Z^1(\gog)$
such that:
\begin{enumerate}
\item $\pi:M\rightarrow M/G$ is a principal $G$-bundle.
\item $\forall_{A\in\gog}\;L_A\mu=Q_\mu(A)\mu$.
\end{enumerate}
We set $Q_\gog(X):=Tr(Ad(X))$. We define the background charge
as $Q_{back}:=Q_\gog+Q_\mu$.
\end{defn}
\begin{thm}(``Fadeev-Popov'',\cite{fadeev}).
\label{fadpopth}
Let $G$ be a connected Lie group, and
$(M,\mu)$ a $G$-volume manifold with $\partial M=\emptyset$.
Let $\phi:M\rightarrow
\bR^{|G|}$, such that the zero set $Z(\phi)$ is an $n$-sheeted cover of $M/G$.
Let $F\in L^2(\bR^{|G|})$ such that $\int F=1$.
Let $\omega$ be a top-form on $M/G$, let $f:M\rightarrow \bR$,
let $\{T_i\}$ be a basis for $\gog$, and
let $v:=T_1..T_{|G|}\in \bigwedge^{max}(\gog)$.
Then:
\begin{enumerate}
\item $\int_{M/G}\omega={\pm 1\over n} \int_M \pi^* \omega F(\phi)
d\phi^1\wedge...\wedge d\phi^{|G|}.$
\item $fi_v\mu$ is a basic form
\footnote{A form $\alpha$ on $M$ is called basic
iff it can be written as $\pi^* \omega$
where $\omega$ is a form on the base $M/G$. If $G$ is connected this condition
is equivalent to $\forall_{A\in\gog}\; i_A \alpha=L_A\alpha=0$:
Indeed this implies that for any vertical vectorfield $X=f^iT_i$ on $M$ we have
$L_X\alpha= di_X\alpha+i_Xd\alpha= df^i\wedge
i_{T_i}\alpha+f^iL_{T_i}\alpha=0$,
so that we can take $\omega:=\sigma^*\alpha$, because it is then independent of
the section $\sigma$. To check that $\alpha=\pi^*\omega=\pi^*\sigma^*\alpha$,
we note that both sides are equal on the horizontal space spanned by $\sigma$,
and since $i_A\alpha=0$ this is enough.
For more details, see H. Cartan.
}
iff $\forall_{A\in\gog}L_A(f)=-Q_{back}(A)f$.
In that case:
\item $\int_{M/G} \pi_*(f i_v \mu)={\pm 1\over n} \int_M f
F(\phi) det_{ij}(T_i(\phi^j))\mu.$
\item Let $e^{-H}$ be the Fourier transform of $F$. Then there is a
constant $K$ such that
$$\int_{M/G} \pi_*(f i_v \mu)=K \int db \int_{Ber} dcd\bar c\int_M
f e^{-H(b)+ib_j\phi^j+i\bar c_i T_j(\phi^i)c^j}\mu.$$
\label{fadpopint}
\end{enumerate}
\end{thm}
\begin{pr}
\begin{enumerate}
\item Since $Z$ is an $n$-sheeted cover, and since $d\pi^*\omega=
\pi^*d\omega=0$, we have:
\footnote{The Thom class of a vectorbundle $E\rightarrow M$
is a form $\Theta$ on $E$ such that for $d\alpha=0$,
$\int_{Z(\phi)}\alpha=\int_M \alpha
\wedge \phi^*\Theta$, where $\phi$ is a section of $E$.
If $E=M\times V$ then the Thom class is given as a topform
$Fdx^1..dx^n$ on $V$, normalized to $\int_V Fdx=1$. The idea is that
if we take $F$ to be the delta function then this will reproduce exactly the
zero locus of $\phi$, but since $d\alpha=0$, we can replace $\Theta$ by any
other fast decreasing representative, for example a Gaussian on $V$.}
$$\int_{M/G} \omega = {1\over n} \int_{Z(\phi)} \pi^* \omega =
{1\over n} \int_M \pi^* \omega\wedge \phi^*(Thom(M\times \bR^{|G|}))$$
$$={\pm 1\over n} \int_M \pi^*\omega F(\phi)d\phi^1..d\phi^{|G|}.$$
\item Since $v$ is of highest degree, we have $i_Ai_v\mu=0$. Further,
\footnote{If $M:V\rightarrow V$ is linear, then $M$ acts on $\bigwedge^{max}$
by derivations. The latter action is $Tr_V(M)$ times the identity which one
verifies by making $M$ act on some fixed element $e_1...e_{max}$. Therefore,
$[A,v]= Tr(Ad(A)).v= Q_\gog(A).v$.}
$$L_A(fi_v\mu)=(L_A f)i_v\mu+fi_{[A,v]}\mu+fi_v L_A\mu$$
$$=(L_Af+Q_\gog(A)f+Q_\mu(A)f)i_v\mu=(L_Af+Q_{back}(A)f)i_v\mu.$$
Therefore, since $i_v\mu$ is nowhere zero, $L_A(fi_v\mu)=0\Leftrightarrow
L_A(f)=-Q_{back}(A)f$.
\item We now specialize point $(1)$ to the case
$\omega=\pi_*(fi_v\mu)$, i.e. $\pi^*\omega = f i_v\mu$, so that
the left hand side equals:
$${\pm 1\over n} \int_M f i_v\mu F(\phi)d\phi^1..d\phi^{|G|}
={\pm 1\over n}\int_M f F(\phi) i_v (d\phi^1..d\phi^{|G|})\mu=RHS.$$
\item Here we used the Fourier representation of $F$, and
the Berezin representation of $\det$.
The variables $c,\bar c$ are known as Fadeev-Popov ghosts, and the
function $G(\phi)$ as a gauge fixing term.
\end{enumerate}
\end{pr}
\begin{defn}(BRST, \cite{becchi}).
\label{brstcharge}
\begin{enumerate}
\item Define the following derivation on functions
$\cO$ of $m\in M, b,c$, and $\bar c$, i.e. on $\cF(M)=Map(M,\bR)$ tensored with
the polynomial algebra on $b$'s and the exterior algebra on $c,\bar c$'s:
$$Q:=c^j T_j -{1\over 2} f_{ij}^k c^i c^j
{\partial\over\partial c^k}
-b_i
{\partial\over\partial \bar c_i}
.$$
Here $T_i$ is a basis element of $\gog$ acting on $\cF(M)$.
\item A function $\cO$ will be called basic iff $Q\cO=-Q_{back}(T_i)c^i \cO$.
\end{enumerate}
\end{defn}
\begin{thm}(BRST).
\label{brst_thm}
\begin{enumerate}
\item $Q^2=0$.
\item  Assume $H$ is polynomial. Then there is a Berezin-odd function $\Psi$,
i.e. having an odd total number of $c$ and $\bar c$'s,
(depending on $F$ and $\phi$) and a
number $K$ such that
$$\int_{M/G} \pi_*(f i_v \mu)=K \int db \int_{Ber} dcd\bar c\int_M\mu
f e^{Q(\Psi)}.$$
($\Psi$ is known as a gauge fixing fermion).
\end{enumerate}
\end{thm}
\begin{pr}
\begin{enumerate}
\item Since $Q$ is a derivation, and since $Q$ is odd, $Q^2$ is also a
derivation, so that it suffices to prove that $Q^2$ is zero on generators.
Indeed:
\begin{enumerate}
\item
$$Q^2(f)=Q(c^j T_j(f))=Q(c^j)T_j(f)-c^j Q(T_j(f))
=-{1\over 2} f_{lm}^j c^l c^m T_j(f)$$
$$ - c^j c^k T_k T_j(f)
=-{1\over 2} c^l c^m [T_l,T_m](f)-{1\over 2} c^j c^k [T_k,T_j](f)=0.$$
\item $$Q^2(\bar c_i)=Q(-b_i)=0=Q^2(b_i).$$
\item $Q^2(c_i)=0$, because in that case only $-{1\over 2} f_{ij}^k c^i c^j
{\partial\over\partial c^k}$ matters, which is proportional to the Lie algebra
cohomology operator.
\end{enumerate}
\item Since we allow for an arbitrary constant $K$, we may assume that
$H(0)=0$, so that $H$ is a polynomial of degree at least $1$. But all
such polynomials are $Q$-exact since $b_i=Q(-\bar c_i)$. Further,
$$Q(-i \bar c_j \phi^j)=-i Q(\bar c_j)\phi^j+i \bar c_j Q(\phi^j)
=i b_j \phi^j + i \bar c_j c^k T_k(\phi^j),$$
so that we see that there is a $\Psi$ such that
$-H(b)+ib_j\phi^j+i\bar c_i T_j(\phi^i)c^j=Q(\Psi).$
\end{enumerate}
\end{pr}
\subsubsection{Algebraic abstraction.}
\begin{rem}
Just like we replaced the notion of a volume form by the notion of a divergence
in order to find the infinite dimensional version of integration over
volume manifolds, in the same way will we now make algebraic abstractions
hoping to define infinite dimensional integration over quotients.
What we have to avoid now is the use of any element $v\in\bigwedge^{max}(L)$,
because there need not be any maximal degree.
The Fadeev-Popov representation, point \ref{fadpopint} in
theorem \ref{fadpopth}, is more
practicable in the infinite dimensional case, because it only involves
summation over an infinite basis, for example in $T_j(\phi^i)c^j$.
It is outside the scope of this work to consider regularizations of this
summation, so from now on we assume that this kind of summation is possible.
Thus, for example we also assume that $X\mapsto Tr(Ad(X))$ exists.
\end{rem}
\begin{defn}
A $\gog$-symmetric formal volume manifold is defined to be a combination
$(A,L,\nabla,Q_\mu,\gog)$, where
\begin{enumerate}
\item $(A,L,\nabla)$ is a formal volume manifold.
\item $\gog\leq L$ is a sub Lie algebra, called the zero-mode or gauge algebra.
\item $Q_\mu \in Z^1(\gog)$.
\item $\forall_{n\in \gog}\; \nabla(n)=Q_\mu (n)$.
\end{enumerate}
Further, we set $Q_\gog(X):=Tr(Ad(X))$, and $Q_{back}:=Q_\gog+Q_\mu.$
For $\Psi$ a Berezin-odd element, $D\in\{T_j,{\partial\over\partial
c^k},{\partial\over\partial \bar c_i},{\partial\over\partial b_i}\}$,
and with $\nabla({\partial\over\partial c^k}):=\nabla({\partial\over\partial
\bar c_i}):=\nabla({\partial\over\partial b_i}):=0$, the Schwinger-Dyson
equation is defined as:
$$\langle D(\cO)+\nabla(D)\cO+(-1)^{\cO D} \cO DQ(\Psi)\rangle =0,$$
for maps
$\cO\mapsto \langle \cO\rangle $, where $\cO$ is a combination of functions
on $M$ and functions of $c,\bar c,b$.
\end{defn}
\begin{rem} As in the usual case, the above Schwinger-Dyson
equation is motivated by the same property of
$$\cO\mapsto \langle \cO\rangle:=\int dbdcd\bar c\mu \cO e^{Q(\Psi)}.$$
As we have seen using Thom's theorem, the Fadeev-Popov integral does not
depend on the choice of $\phi$ or $F$, since the quotient integral
does not involve these objects. An alternative proof is given in
the following theorem:
\end{rem}
\begin{thm} (Gauge invariance using the Schwinger-Dyson equation.)
\begin{enumerate}
\item $\langle Q\cO\rangle =-\langle Q_{back}(T_i) c^i \cO\rangle $.
\item If $\cO_1$ is basic, then $\langle \cO_1 Q(\cO_2)\rangle =0$.
\item Let $t\mapsto \Psi_t$ be a family of Berezin-odd elements.
Let $\langle .\rangle _\Psi$ be a family of solutions of the corresponding
Schwinger-Dyson equations
satisfying $\partial_t\langle \cO\rangle _{\Psi_t}=\langle \cO Q\partial_t
\Psi_t\rangle _{\Psi_t}$,
and let $\cO$ be basic, then $\partial_t \langle \cO\rangle _{\Psi_t}=0$.
\end{enumerate}
\end{thm}
\begin{pr}
\begin{enumerate}
\item
$$\langle Q\cO\rangle =\langle c^j T_j\cO-{1\over 2} f_{ij}^k c^i c^j {\partial
\cO\over
\partial c^k} - b_i {\partial \cO\over \partial \bar c_i}\rangle $$
$$=\langle T_j(c^j \cO)-{\partial\over\partial c^k}({1\over 2} f_{ij}^k c^i c^j
\cO)
+{\partial\over\partial c^k}({1\over 2}f_{ij}^k c^i c^j)\cO
-{\partial\over\partial \bar c_i}(b_i \cO)\rangle $$
$$=\langle -c^j \cO T_j Q\Psi +{1\over 2} f_{ij}^k c^i c^j \cO (-1)^\cO
{\partial\over\partial c^k}Q\Psi+ b_i \cO (-1)^\cO
{\partial\over\partial \bar c_i}Q\Psi\rangle $$
$$+\langle -c^j \cO \nabla(T_j) + {1\over 2} f_{kj}^k c^j \cO
-{1\over 2} f_{ik}^k c^i\cO\rangle $$
$$=\langle -(-1)^\cO\cO\{
c^j T_j -{1\over 2} f_{ij}^k c^i c^j
{\partial\over\partial c^k}
-b_i
{\partial\over\partial \bar c_i}
\}Q\Psi\rangle$$
$$+\langle -c^j \cO Q_\mu(T_j)-Q_\gog(T_i)c^i \cO\rangle $$
$$=\langle -(-1)^\cO \cO Q^2 \Psi\rangle -\langle
(Q_\mu+Q_\gog)(T_i)c^i\cO\rangle =-\langle Q_{back}(T_i)c^i\cO\rangle .$$
\item
$$\langle \cO_1 Q(\cO_2)\rangle
=(-1)^{\cO_1}(
\langle Q(\cO_1\cO_2)\rangle -\langle (Q\cO_1)\cO_2\rangle) $$
$$=(-1)^{\cO_1}\langle -Q_{back}(T_i)c^i \cO_1 \cO_2\rangle
+(-1)^{\cO_1}\langle Q_{back}(T_i)c^i \cO_1 \cO_2\rangle =0.$$
\item By taking $\cO_2:=\partial_t\Psi_t$ in the previous point.
\end{enumerate}
\end{pr}
\begin{ex} (Maxwell-Feynman). Consider the set $M=\Omega^1(\bR^D)$ of
connections on
a product bundle $\bR^D\times U(1)$ over $\bR^D$. Let $G$ be
$\Omega^0(\bR^D)$, which acts on $M$ by $\varphi:A\mapsto A+d\varphi$.
The weight $S(A):=\int_{\bR^D} {1\over 4} F_{\mu\nu}F^{\mu\nu}dx$, where $F=dA$
is well-defined on the quotient $M/G$. Let us try to compute
$$\int_{M/G} f(A)e^{-S(A)}DA,$$
where $DA$ means that we use the affine structure of $M$ to produce some
preferred vectorfields $\delta/\delta A(x)$ and then use the
Schwinger-Dyson equation.
Since we go to the quotient, we
need a gauge fixing function $\phi$,
for which we take $\phi^x(A):=\partial_\mu A^\mu(x)$,
together with a condition at infinity to ensure that no remaining group
elements with $\Delta\varphi=0$ leave the condition $\phi(A)=0$ invariant.
Then the functions $T_j(\phi^i)$ are $A$-independent:
$$\partial_t \phi^x(A+td\varphi)=\Delta\varphi(x),$$
So that the term $X(T_j(\phi^i))$ does not enter the Schwinger-Dyson equation.
For $F$, let us take a Gaussian bump function:
$$F(\phi):=\exp(-{1\over 2}\int_{\bR^D}  \phi^x\phi^x dx).$$
What we see then, is that the quotient calculation is reduced to
the calculation of $\int_{M} e^{-S_1(A)}f(A)DA$, where
$$S_1(A):=\int_{\bR^D} {1\over 4} F_{\mu\nu}F^{\mu\nu}
+{1\over 2} (\partial_\mu A^\mu)^2.$$
This is a Gaussian weight and we can now even integrate
the function $A\mapsto A^\mu(x)$, even if it is not $G$-invariant.
Since ${\delta S_1\over \delta A^\mu(x)}=-\Delta A_\mu(x)$, we see that
$\langle A_\mu(x)A_\nu(y) \rangle=
-g_{\mu\nu} f_D(x-y)$, where the functions $f_D$ were
defined in section \ref{gauss_fun_int}. From this we can also read off the
contractions for $G$-invariant objects like $F_{\mu\nu}(x)$,
or $I_L(A):=\oint_L A$:
$$\langle I_L I_M\rangle =-\oint_L dx^\mu\oint_M dy^\nu g_{\mu\nu} f_D(x-y).$$
\end{ex}
\begin{rem} It was noted by Batalin and Vilkovisky \cite{batalin} that
the BRST expression
$$Q:=c^j T_j -{1\over 2} f_{ij}^k c^i c^j
{\partial\over\partial c^k}
-b_i
{\partial\over\partial \bar c_i}
$$
used in theorem
\ref{brst_thm} can be generalized by replacing it
by a more general polynomial in the symbols
$f(m),c,\bar c, b, f_i^*:=T_i, c_i^*:=
{\partial\over \partial c^i},
\bar c^{i*}:=
{\partial\over \partial \bar c_i},
b^{i*}:={\partial\over \partial b_i}$.
Just like in the BRST case where $Q$ occurs in the integral only through
$Q(\Psi)$, one is led to:
\begin{enumerate}
\item Assign Berezin parities to the starred variables
as if they are evaluated on the odd element $\Psi$:
$$Parity(\phi_\alpha^*):=Parity({\partial \Psi\over \partial \phi^\alpha})
=Parity(\phi^\alpha)+1.$$
\item Consider the BRST integral, but now with the more general $Q$.
In that case $Q(\Psi)$ means replacing $\phi_\alpha^*$ by
${\partial \Psi\over \partial \phi^\alpha}$ in $Q$.
\item Find a condition on $Q$ such that the end-result is independent of
$\Psi$.
\end{enumerate}
The fact that the $Q$ of BRST satisfies $Q^2=0$ can now be reformulated as
the fact that $\{Q,Q\}=0$, where $\{.,.\}$ is the Schouten bracket.
For more general $Q$'s there is also a condition that will guarantee
independence of $\Psi$, known as the BV master equation
\cite[formula 16]{batalin},
but it does not read $\{Q,Q\}=0$. Instead, there is an additional
term in the equation, which is zero in the special BRST-case.
\end{rem}
%
%
%
%
%
\newpage
\section{Reference material on some natural super Lie algebras.}
\label{nat_sup_section}
\begin{defn} Let $(G,+)$ be an abelian group, and let $<.,.>:G\times G
\rightarrow \bZ_2$ be bilinear symmetric.
\begin{enumerate}
\item A $G$-graded vectorspace or $G$-vectorspace
is a vectorspace $V$ together with a direct sum decomposition
$V=\oplus_{g\in G} V_g$.  For homogeneous elements $a,b,..\in V$, we will
denote the degree by $|a|,|b|,..\in G$. The number $(-1)^{<|a|,|b|>}$ will
be denoted simply by $(-1)^{ab}$. Thus, for example, we have
$(-1)^{a(b+c)}=(-1)^{ab+ac}$, whereas $(-1)^{abc}$ is undefined.
\item A $G$-algebra of degree $w\in G$
is an algebra of which the underlying vectorspace is
$G$-graded, and such that the composition is from $A_g\otimes A_h$
to $A_{g+h+w}$. By setting $A_g^\prime:=A_{g-w}$, we get a $G$-algebra
of degree $0$. Thus, if one considers only one algebraic structure at a time,
one may restrict to degree zero, which we will do from now on.
\item A linear map $M:A\rightarrow B$ is said to be of degree $|M|\in G$ iff it
maps
$A_h$ to $B_{h+|M|}$. We write $(-1)^{M.}$ for $(-1)^{<|M|,.>}$.
\item The above notion of $G$-gradation does not depend on $<.,.>$. We will now
introduce some concepts that do depend on $<.,.>$, and this situation is
usually referred to as ``super''. We will stop mentioning the dependence
on $G$: Thus, algebra means $G$-graded algebra.
In most cases, we have $G=\bZ$ and $<g,h>:=gh$ mod $2\bZ$.
\item $M$ as above is called a derivation w.r.t. $<.,.>$ or a superderivation
iff $M(ab)=M(a)b+(-1)^{Ma}aM(b)$.
\item An algebra is called symmetric
(w.r.t. to $<.,.>$) or supersymmetric, iff $ab=(-1)^{ab}ba$.
\item It is antisymmetric iff $ab=-(-1)^{ab}ba$.
\item An algebra is called associative iff $a(bc)=(ab)c$.
\item We will now also drop the word ``super'': It will be understood
whenever a definition depends on $<.,.>$.
\item An algebra is called pre-Lie \cite{gerstenhaber}
iff $a(bc)-(-1)^{ba}b(ac)=
(ab)c-(-1)^{ab}(ba)c$. Pre-Lie compositions will usually be
denoted as $[a\trr b]$, in view of their connection with subgaussian
contraction algebras.
\item An algebra is called Jacobi iff $\sum_{Cycl(a,b,c)} (-1)^{ac} a(bc)=0$.
\item An algebra is called deriving iff $a(bc)=a(bc)+(-1)^{ab}b(ac)$.
\item An algebra is called Lie iff it is antisymmetric and Jacobi,
or equivalently antisymmetric and deriving. A Lie composition
will be denoted by $[.,.]$ or $\{.,.\}$.
\item An algebra with two compositions is called Poisson iff the first
is symmetric associative with unit, the second, denoted by $\{.,.\}$ is Lie,
and the two are compatible in the sense that
$\{a,bc\}=\{a,b\}c+(-1)^{bc}\{a,c\}b$
\item A Batalin-Vilkovisky algebra $A$ is a Poisson algebra of which the
associative product is of degree $0$ and the bracket of degree $-1$,
together with a differential $\partial:A\rightarrow A$, i.e. $\partial^2=0$,
of degree $-1$, such that:
$$\{a,b\}=a(\partial b)+(-1)^{|a|}(\partial a)b-(-1)^{|a|}\partial (ab).$$
See Getzler, \cite{getzler}.
\end{enumerate}
\begin{rem}
\label{sup_lazy_rem}
Proofs of identities that hold in the usual case where $<.,.>=0$ need in
general not be repeated in the supercase. Indeed, if the ungraded identity
concerns polynomial expressions in a number of variables $a_1,...,a_n\in A$,
say
$$[a_1,[a_2,a_3]]=[[a_1,a_2],a_3]+[a_2,[a_1,a_3]],$$
then the above definitions are such that one gets the corresponding
super identity by multiplying each seperate term with $\pm 1$, according
to the permutation that the variables have undergone on paper w.r.t.,
say, the order $a_1a_2a_3$; so in the above case this gives the
expression
$$[a_1,[a_2,a_3]]=[[a_1,a_2],a_3]+[a_2,[a_1,a_3]](-1)^{a_1 a_2}.$$
One may thus consider the first formula to be shorthand notation for the
second, and in this way, one can consider the proofs of ungraded identities
to be shorthand notation for the supercase, since in every expression in
the proof one may add the corresponding signs. For example this reasoning
applies to the proof that the semi-direct product defined below produces
a new super Lie algebra as claimed: First prove this fact in the ungraded
case, and then graded proof is produced by adding minus signs according to
the permutation rule.
\end{rem}
We list a number of functors between these categories.
\begin{enumerate}
\item Direct sums and tensorproducts of $G$-spaces are $G$-graded.
\item The tensoralgebra of $V$ is defined as $T(V):=\oplus_{n} V^{\otimes n}$.
\item The symmetric algebra is defined as $Sym(V):=T(V)/I$, where $I$ is
the two-sided ideal generated by elements of the form $ab-(-1)^{ab}ba$.
It is an associative $\bZ\oplus G$ algebra with unit containing $V$,
such that $v\in V$ has degree $(1,|v|)$ in $Sym(V)$. It is symmetric w.r.t. the
form
$(n\oplus g,m\oplus h):=<g,h>$.
\item The exterior algebra is defined as $\bigwedge(V):=T(V)/J$, where
$J$ is the two-sided ideal generated by elements of the form $ab+(-1)^{ab}ba$.
It is a $\bZ\oplus G$-graded algebra, symmetric w.r.t.
the form $(n\oplus g,m\oplus h):=nm+<g,h>$.
\item Direct sums of $G$-algebras are again $G$-algebras.
\item The tensorproduct of two associative algebras is understood to have
the following associative multiplication (which depends on $<.,.>$):
$(a\otimes b)(c\otimes d):=(-1)^{bc}(ac\otimes bd)$.
\item The tensor product of two Poisson algebras is again Poisson
using the above associative multiplication and bracket
$$\{a_1\otimes b_1,a_2\otimes b_2\}:=(-1)^{a_2 b_1}\{a_1,a_2\}\otimes b_1 b_2+
(-1)^{a_2 b_1}a_1 a_2\otimes \{b_1,b_2\}.$$
\item Further, we define Associative$\otimes$Lie to be Lie as follows:
$[a\otimes X,b\otimes Y]:=(-1)^{bX}ab\otimes[X,Y]$.
\item The semi-direct product of two Lie algebras $A$,$B$
w.r.t. $\pi:A\rightarrow Der(B)$ (degree preserving)
is defined as
the vectorspace $A\oplus B$, with composition $[a_1\oplus b_1, a_2\oplus b_2]
:=[a_1,a_2]\oplus[b_1,b_2]+\pi(a_1)(b_2)-(-1)^{b_1 a_2}\pi(a_2)(b_1).$
\item Associative$\rightarrow$Pre Lie; $[a\trr b]:=ab$.
\item Pre-Lie $\rightarrow$ Lie; $[a,b]:=[a\trr b]-(-1)^{ab}[b\trr a]$. In
particular,
the inner endomorphisms $I_a(b):=[a\trr b]$ form a representation of the Lie
algebra, since by definition of pre Lie, we have $[I_a,I_b]=I_{[a,b]}$.
\item Associative $\rightarrow$ Lie; $A\mapsto Der(A)$. (The derivations are
closed
under the commutator bracket.)
\item Lie$\rightarrow$Poisson; $L\mapsto Sym(L)$, with obvious associative
product, and symmetric Schouten bracket $\{X_1..X_n,Y_1..Y_m\}:=\sum_{i,j}
X_{[1,n]\backslash i}
[X_i,Y_j] Y_{[1,m]\backslash j}$. (With signs added in supercase.)
\item Lie$\rightarrow$BV; $L\mapsto \bigwedge(L)$, with the antisymmetric
Schouten
bracket:
$$[X_1..X_n,Y_1..Y_m]:=\sum_{i,j}(-1)^{i+j}[X_i,Y_j]
X_{[1,n]\backslash i}Y_{[1,m]\backslash j},$$ and differential
$$\partial(X_1..X_n):=\sum_{i<j} (-1)^{i+j+1} [X_i,X_j]X_{[1,n]\backslash
ij}.$$
\item Whenever $d: G\rightarrow \bZ$ is a group homomorphism, we can extend
a $G$-Lie algebra with a ``counting'' element $N_d$ of $G$-degree $0$,
by setting
$[N_d,N_d]:=0$ and $[N_d,a]:=-[a,N_d]:=d(|a|)a$.
\end{enumerate}
\end{defn}
\begin{defn}
Let $L$ be a Lie algebra.
To it we associate $Car(L)$, the Cartan
algebra of $L$, which is is a $\bZ$-Lie algebra
generated by
symbols $d,i_X,L_X$, linear in $X\in L$, with degrees and relations as in
the following table
\footnote{More generally, one can make a $\bZ\oplus G$-Lie algebra
from a $G$ algebra, by giving $i_X$, say, degree $(-1,|X|)$. This
is what happens if you want to have a Cartan calculus on a supermanifold.}
:\\\\
\begin{tabular}{|lr|r|r|r|}
\hline
$[.,.]$ & $(deg)$ & $d$ & $i_Y$ & $L_Y$\\
\hline\hline
$d$ & $(1)$ & $0$ & $L_Y$ & $0$\\
\hline
$i_X$ & $(-1)$ & & $0$ & $i_{[X,Y]}$\\
\hline
$L_X$ & $(0)$ & & & $L_{[X,Y]}$\\
\hline
\end{tabular}\\\\
We will see in a minute that this is a Lie algebra.
Further, there is a natural map $K:Car(L)\rightarrow End(\bigwedge(L))$, as
follows:
\begin{enumerate}
\item $d(X_1..X_n):=\sum_{i<j} (-1)^{i+j+1} [X_i,X_j] X_{[1,n]\backslash ij}$,
\item $i_X(X_1..X_n):=XX_1..X_n$.
\item $L_X(X_1..X_n):=\sum_{i=1}^n X_1..[X,X_i]..X_n$.
\end{enumerate}
\end{defn}
\begin{thm} $[K(A),K(B)]=K([A,B])$.
\label{carmod}
\end{thm}
\begin{pr}
We will prove this in the order $ii,id,Li,LL,dd,Ld$:
\begin{enumerate}
\item $[i_X,i_Y](Z_1..Z_n)=(XY+YX)Z_1..Z_n=0$.
\item $[i_X,d](Z_1..Z_n)=Xd(Z_1..Z_n)+d(XZ_1..Z_n)$
$$=X\sum_{i<j} (-1)^{i+j+1}[Z_i,Z_j]Z_{[1,n]\backslash ij}
+\sum_{j=1}^n (-1)^{j+1} [X,Z_j]Z_{[1,n]\backslash j}$$
$$+\sum_{i<j} (-1)^{i+j+1} [Z_i,Z_j]X Z_{[1,n]\backslash ij}
=L_X(Z_1..Z_n).$$
\item $[L_X,i_Y](Z_1..Z_n) =[X,YZ_1..Z_n]-Y[X,Z_1..Z_n]$
$$=[X,Y]Z_1..Z_n=i_{[X,Y]}Z_1..Z_n.$$
\item $X\mapsto L_X$ is the adjoint representation on $\bigwedge(L)$.
\item We prove this together with
\item First define $\Delta:=[d,d]$, and $R_X:=[L_X,d]$.
\begin{lemma} $[i_X,R_Y]=0$, and $[i_X,\Delta]=2R_X$.
\label{i_Xlemma}
\end{lemma}
\begin{pr}
Using the known commutators involving $i_X$, we have:
\begin{enumerate}
\item $[i_X,R_Y]=[i_X,[L_Y,d]]=[[i_X,L_Y],d]+[L_Y,[i_X,d]]=[i_{[X,Y]},d]
+[L_Y,L_X]=L_{[X,Y]}+L_{[Y,X]}=0,$ and
\item $[i_X,\Delta]=[i_X,[d,d]]=[[i_X,d],d]-[d,[i_X,d]]=[L_X,d]-[d,L_X]=2R_X$.
\end{enumerate}
\end{pr}
This allows us to prove that $\Delta(Z_1..Z_n)=R_X(Z_1..Z_n)=0$ by induction on
$n$: The case $n=0$ is clear since $d(1)=L_X(1)=0$. \label{i_Xind}
Assume the statement to be true up to $n$, then:
\begin{enumerate}
\item $\Delta(YZ_1..Z_n)=[\Delta,i_Y](Z_1..Z_n)=2 R_Y(Z_1..Z_n)=0$, and
\item $R_X(YZ_1..Z_n)=[R_X,i_Y](Z_1..Z_n)=0$.
\end{enumerate}
\end{enumerate}
\end{pr}
\begin{thm} $Car(L)$ is a super Lie algebra.
\end{thm}
\begin{pr} One may check this directly, but we will just prove it
by intimidation:
In order to prove that an algebra is a Lie algebra, it suffices to find
a faithful representation. We already have a representation, but it
need not be faithful, since for example if $L$ is Abelian, then $d$ acts
as $0$. To that end, let $V$ be the vectorspace underlying the Lie algebra
$L$, and let $\tilde L$ be the free Lie algebra over $V$.
Then the map $\tilde \rho: Car(\tilde L)\rightarrow End(\bigwedge(\tilde L))$
is injective: One can see this by noting that all the operators have
different degree, and the operator $L_{X-Y}$ is zero iff $X-Y=0$.
So $\tilde \rho$ is injective, and therefore $Car(\tilde L)$
is a super Lie algebra. But there is a surjective morphism
$Car(\tilde L)\rightarrow Car(L)$, so $Car(L)$ is super Lie.
\end{pr}
\begin{thm} Using the above module, we can make some other modules:
\begin{enumerate}
\item $Car(L)$ is represented on $\bigwedge(L)^{dual}$ by
$d\omega:=-\omega\circ d$, $i_X\omega:=\omega\circ i_X$, and
$L_X \omega:=-\omega \circ L_X$.
\item The subspace ${\bigwedge(L)}^{dual,[0,n]}\oplus B^{n+1}(L)\subset
\bigwedge(L)^{dual}$ is a submodule, where $B^{n+1}$ denotes the exact elements
$\alpha$ of degree $n+1$, i.e. such that $\exists_{\beta}\;\alpha=d\beta$.
\end{enumerate}
\end{thm}
\begin{pr}
\begin{enumerate}
\item Denote the operators on $\bigwedge(L)$ by $d,i_X,L_X$, and those on
the dual by $*d,*i_X,*L_X$. Then $*d=-d^T,$ $*i_X=i_X^T$, and $*L_X=-L_X^T$,
where $A^T\omega:=\omega\circ A$. using $[A^T,B^T]=[B,A]^T$, both for
commutators and anti-commutators, we have:
\begin{enumerate}
\item $[*i_X,*i_Y]=[i_X^T,i_Y^T]=[i_Y,i_X]^T=0$.
\item $[*i_X,*d]=[-d,i_X]^T=-L_X^T=*L_X$.
\item $[*L_X,*i_Y]=[i_Y,-L_X]^T=-i_{[Y,X]}^T=*i_{[X,Y]}$.
\item $[*L_X,*L_Y]=[-L_Y,-L_X]^T=L_{[Y,X]}^T=*L_{[X,Y]}$.
\item $[*d,*d]=[-d,-d]^T=0$.
\item $[*d,*L_X]=[-L_X,-d]^T=0$.
\end{enumerate}
\item It is closed under $i_X$ because $i_X$ lowers degree and annihilates
zero degree. It is closed under $L_X$, because $L_X$ preserves degree and
commutes with $d$. It is closed under $d$, because it increases degree by $1$,
maps ${\bigwedge}^n(L)$ to $B^{n+1}$, and maps $B^{n+1}$ to zero, since
$d^2=0$.
\end{enumerate}
\end{pr}
\begin{defn} Associated to any module $V$ of $Car(L)$ is associated a new Lie
algebra $Car(L,V)$, namely the semi-direct product of $Car(L)$ and $V$, where
$V$ is
seen as an Abelian algebra. Thus, the extension looks as follows, with $v\in
V$:\\\\
\begin{tabular}{|lr|r|r|r|r|}
\hline
$[.,.]$ & $(deg)$ & $d$ & $i_Y$ & $L_Y$ & $M_v$\\
\hline\hline
$d$ & $(1)$ & $0$ & $L_Y$ & $0$ & $M_{dv}$\\
\hline
$i_X$ & $(-1)$ & & $0$ & $i_{[X,Y]}$ & $M_{i_X v}$\\
\hline
$L_X$ & $(0)$ & & & $L_{[X,Y]}$ & $M_{L_X v}$\\
\hline
$M_w$ & $(|w|)$ & & & & $0$\\
\hline
\end{tabular}\\\\
We have already constructed a number of natural modules, so we get a number
of larger super Lie algebras associated to $L$, for example
$Car(L,\bigwedge(L)^{dual})$, or
$$Car^{(n)}(L):=Car(L,{\bigwedge}^{\leq n}(L)^{dual}\oplus B^{n+1}).$$
Of special interest is an algebra which we choose to call the $BRST$-algebra
of $L$, which is $Car^{(1)}(L)$, extended with a counting element $N$:
\end{defn}
\begin{thm} Explicitly, $BRST(L)$ is isomorphic to an algebra generated by
symbols $N,d,i_X,L_X,c_\alpha,e_\alpha,1$,
where $X\in L$ and $\alpha\in L^{dual}$, with the following composition:\\\\
\begin{tabular}{|lr||r|r|r|r|r|r|r|}
\hline
$[.,.]$ & $(deg)$ & $N$ & $d$ & $i_Y$ & $L_Y$ & $c_\beta$ & $e_\beta$ & $1$\\
\hline\hline
$N$ & $(0)$ & $0$ & $d$ & $-i_Y$ & $0$ & $c_\beta$ & $2e_\beta$ & $0$ \\
\hline
$d$ & $(+1)$ & & $0$ & $L_Y$ & $0$ & $e_\beta$ & $0$ & $0$ \\
\hline
$i_X$ & $(-1)$ & & & $0$ & $i_{[X,Y]}$ & $\beta(X)1$ & $c_{[X,\beta]}$ & $0$ \\
\hline
$L_X$ & $(0)$ & & & & $L_{[X,Y]}$ & $c_{[X,\beta]}$ & $e_{[X,\beta]}$ & $0$ \\
\hline
$c_\alpha$ & $(+1)$ & & & & & $0$ & $0$ & $0$\\
\hline
$e_\alpha$ & $(+2)$ & & & & & & $0$ & $0$ \\
\hline
$1$ & $(0)$ & & & & & & & $0$ \\
\hline
\end{tabular}\\\\
Here $[X,\beta]$ denotes the coadjoint action: $[X,\beta](Y):=-\beta([X,Y])$.
\end{thm}
\begin{pr}
Define $1:=M_1$, and for $\alpha\in L^{dual}$, set
$c_\alpha:=M_\alpha$, and $e_\alpha:=M_{d\alpha}$. Then, since we already know
the commutation relations of $Car(L)$ itself, and the that of $N$, it remains
to check that:
\begin{enumerate}
\item $[d,c_\beta]=[d,M_\beta]=M_{d\beta}=e_\beta$.
\item $[d,e_\beta]=[d,M_{d\beta}]=0$.
\item $[i_X,c_\beta]=[i_X,M_\beta]=M_{i_X\beta}=\beta(X)1$.
\item $[i_X,e_\beta]=[i_X,M_{d\beta}]=M_{i_Xd\beta}=c_{i_Xd\beta}$, and
$i_Xd\beta(Y)=d\beta(X,Y)=-\beta([X,Y])=[X,\beta](Y)$.
\item $[L_X,c_\beta]=[L_X,M_\beta]=M_{L_X\beta}=c_{L_X\beta}$, and
$L_X\beta(Y)=-\beta([X,Y])=[X,\beta](Y)$.
\item All commutators among $1,c,$ and $e$ are of the form $[M_1,M_2]$
and therefore zero.
\end{enumerate}
\end{pr}
\begin{cor} A number of subalgebras can be seen:
\begin{enumerate}
\item First of course $Car(L)\subset BRST(L)$.
\item Next $bc(L)\subset BRST(L)$, where $bc(L)$ is the subalgebra generated by
$i_X,c_\alpha,$ and $1$.
\item $Weil(L)\subset UEA(BRST(L))$, where $Weil(L):=\bigwedge(L^{dual})\otimes
Sym(L^{dual})$, as follows: $\alpha_1..\alpha_n\otimes \beta_1..\beta_m
\mapsto c_{\alpha_1}..c_{\alpha_n}e_{\beta_1}..e_{\beta_m}$.
\end{enumerate}
\end{cor}
\begin{thm}
We can extend the action $L\rightarrow End(\bigwedge(L))$
to $BRST(L)\rightarrow End(\bigwedge(L))$ as follows:
\begin{enumerate}
\item $N(X_1..X_n):=-n.X_1..X_n$.
\item $c_\alpha(X_1..X_n):=\sum_{i=1}^n (-1)^{i+1} \alpha(X_i)
X_{[1,n]\backslash i}$.
\item $e_\alpha(X_1..X_n):=\sum_{i<j} (-1)^{i+j+1} \alpha([X_i,X_j])
X_{[1,n]\backslash ij}$.
\item $1(X_1..X_n):=X_1..X_n$.
\end{enumerate}
\end{thm}
\begin{pr}\\
$(1)$ We will prove the commutation relations in the order of the
superscripts indicated below:\\\\
\begin{tabular}{|l||r|r|r|r|r|r|}
\hline
$[.,.]$ & $N$ & $d$ & $i_Y$ & $L_Y$ & $c_\beta$ & $e_\beta$ \\
\hline\hline
$N$ & $0^{16}$ & $d^{17}$ & $-i_Y^{18}$ & $0^{19}$ & $c_\beta^{20}$ &
$2e_\beta^{21}$ \\
\hline
$d$ & & $0^6$ & $L_Y^2$ & $0^5$ & $e_\beta^{11}$ & $0^{12}$  \\
\hline
$i_X$ & & & $0^1$ & $i_{[X,Y]}^3$ & $\beta(X)1^7$ & $c_{[X,\beta]}^9$  \\
\hline
$L_X$ & & & & $L_{[X,Y]}^4$ & $c_{[X,\beta]}^8$ & $e_{[X,\beta]}^{15}$  \\
\hline
$c_\alpha$ & & & & & $0^{10}$ & $0^{13}$ \\
\hline
$e_\alpha$ & & & & & & $0^{14}$  \\
\hline
\end{tabular}\\\\
\begin{enumerate}
\item-(6) were already proved in theorem \ref{carmod}.
\setcounter{enumi}{6}
\item $[i_X,c_\beta](Z_1..Z_n)=Xc_\beta(Z_1..Z_n)+c_\beta(XZ_1..Z_n)
=\sum_{i=1}^n (-1)^{i+1} \beta(Z_i) X Z_{[1,n]\backslash i}$
$$+\beta(X) Z_1..Z_n+\sum_{i=1}^n (-1)^i \beta(Z_i) X Z_{[1,n]\backslash i}
=\beta(X).1(Z_1..Z_n).$$
\item Both $L_X$ and $c_\beta$ are superderivations.
Therefore we only need to prove
the identity on $(Z)$. Indeed:
$$[L_X,c_\beta](Z)=L_X c_\beta(Z)-c_\beta L_X
(Z)=L_X(\beta(Z))-c_\beta([X,Z])$$
$$=-\beta([X,Z])=[X,\beta](Z)=c_{[X,\beta]}(Z).$$
\item $[i_X,e_\beta](Z_1..Z_n)=X e_\beta(Z_1..Z_n)-e_\beta(XZ_1..Z_n)$
$$=X\sum_{i<j} (-1)^{i+j+1} \beta([Z_i,Z_j])Z_{[1,n]\backslash ij}
-\sum_{j=1}^n (-1)^{j+1} \beta([X,Z_j])Z_{[1,n]\backslash j}$$
$$-\sum_{i<j} (-1)^{i+j+1} \beta([Z_i,Z_j]) XZ_{[1,n]\backslash ij}
=\sum_{i=1}^n (-1)^{j+1} [X,\beta](Z_j)Z_{[1,n]\backslash j}$$
$$=c_{[X,\beta]}(Z_1..Z_n).$$
\item All $c_\alpha$'s are superderivations. Thus it suffices to check that
$$[c_\alpha,c_\beta](Z)=c_\alpha c_\beta (Z)+c_\beta c_\alpha(Z)
=\beta(Z)c_\alpha(1)+\alpha(Z)c_\beta(1)=0.$$
\item Let $r_\alpha:=[d,c_\alpha]$. Then $[i_X,r_\alpha-e_\alpha]=0$, since
$$[i_X,r_\alpha]=[i_X,[d,c_\alpha]]=[L_X,c_\alpha]-[d,[i_X,c_\alpha]]$$
$$=c_{[X,\alpha]}-[d,1]\alpha(X)=c_{[X,\alpha]}=[i_X,e_\alpha].$$
Thus to prove that $r_\alpha=e_\alpha$,
using the same reasoning as in lemma \ref{i_Xlemma},
it remains to be shown that
$r_\alpha(1)=e_\alpha(1)$. Indeed, $e_\alpha(1)=0$ by definition, and
$r_\alpha(1)=0$ because $d(1)=0$ and $c_\alpha(1)=0$.
\item $[d,e_\alpha]=[d,[d,c_\alpha]]=[[d,d],c_\alpha]-[d,[d,c_\alpha]]
=-[d,e_\alpha]\;\;\Rightarrow [d,e_\alpha]=0.$
\item This we will prove simultaneously with
\item Set $r_{\alpha\otimes\beta}:=[c_\alpha,e_\beta];\
q_{\alpha\otimes\beta}:=[e_\alpha,e_\beta]$.
\begin{lemma} $[i_X,r_{\alpha\otimes\beta}]=0;\;
[i_X,q_{\alpha\otimes\beta}]
=r_{[X,\alpha]\otimes \beta-[X,\beta]\otimes \alpha}$.
\end{lemma}
\begin{pr}
\begin{enumerate}
\item $[i_X,r_{\alpha\otimes\beta}]=[i_X,[c_\alpha,e_\beta]]
=[[i_X,c_\alpha],e_\beta]-[c_\alpha,[i_X,e_\beta]]$
$$=[\alpha(X)1,e_\beta]-[c_\alpha,c_{[X,\beta]}]=0.$$
\item $[i_X,q_{\alpha\otimes\beta}]=[i_X,[e_\alpha,e_\beta]]
=[[i_X,e_\alpha],e_\beta]+[e_\alpha,[i_X,e_\beta]]$
$$=[c_{[X,\alpha]},e_\beta]+[e_\alpha,c_{[X,\beta]}]
=r_{[X,\alpha]\otimes\beta}-r_{[X,\beta]\otimes\alpha}.$$
\end{enumerate}
\end{pr}
Again, like in lemma \ref{i_Xlemma}, it suffices to prove that
$r_{\alpha\otimes\beta}(1)=
q_{\alpha\otimes\beta}(1)=0$. This is true because $c_\alpha(1)=e_\alpha(1)=0$.
\item $[L_X,e_\beta]=[L_X,[d,c_\beta]]=[d,[L_X,c_\beta]]=[d,c_{[X,\beta]}]
=e_{[X,\beta]}.$
\item to $(21)$: These commutators follow from the fact that if
$A:{\bigwedge}^k\rightarrow {\bigwedge}^{k+n}$, then $[N,A]=-nA$:
$[N,A](Z_1..Z_k)=NA(Z_1..Z_k)-AN(Z_1..Z_k)$
$$=-(n+k)A(Z_1..Z_k)+A(kZ_1..Z_k)=-nA(Z_1..Z_k).$$
\end{enumerate}
\end{pr}
\begin{rem}
Finally we include some often used theorems in the Hamiltonian approach to
BRST symmetry. Note however still the similarity between $H(d)$ below,
and $Q$ in definition \ref{brstcharge}.
\end{rem}
\begin{thm} (Associative BRST construction). If $L$ is finite dimensional with
basis
$\{T_a\}$, and dual basis $\{T^a\}$, then there is a Lie algebra morphism
$H:BRST(L)\rightarrow UEA(L)\otimes UEA(bc(L))$, where the right-hand side is
regarded as an associative algebra, as follows:
\begin{enumerate}
\item $H(N):=-1\otimes i_{T_a}c_{T^a}$.
\item $H(d):=T_a\otimes c_{T^a}-1\otimes {1\over 2}
i_{[T_a,T_b]} c_{T^a} c_{T^b}.$
\item $H(i_X):=1\otimes i_X$.
\item $H(L_X):=X\otimes 1+1\otimes i_{[X,T_a]} c_{T^a}$.
\item $H(c_\alpha):=1\otimes c_\alpha$.
\item $H(e_\alpha):=-{1\over 2}\alpha([T_a,T_b])c_{T^a}c_{T^b}$.
\item $H(1):=1\otimes 1$.
\end{enumerate}
In that case the representation on $\bigwedge(L)$ factorises through this
homomorphism.
\end{thm}
\begin{cor}
Let $L$ be a finite dimensional Lie algebra. If $L$ is represented on $V$, and
$bc(L)$ is represented on $W$, then $BRST(L)$
is naturally represented on $V\otimes W$. Indeed
$$
BRST(L)
\rightarrow
UEA(L) \otimes UEA(bc(L))
\rightarrow
End(V) \otimes UEA(bc(L))$$
$$\rightarrow
End(V) \otimes End(W)
\rightarrow
End(V\otimes W).
$$
\end{cor}
\begin{thm}(Poisson BRST construction.)
Let $L$ be a finite dimensional Lie algebra,
Then there is a super natural Lie homomorphism
$H:BRST(L)\rightarrow Sym(L) \otimes Sym(bc(L))$, as
follows: ($\{T_a\}$ is a basis for $L$, and $\{T^a\}$ its dual basis):
\begin{enumerate}
\item $H(1):=1\otimes 1$
\item $H(i_X):=1\otimes i_X$
\item $H(c_\alpha):=1 \otimes c_\alpha$
\item $H(N):=1 \otimes c_{T^a} i_{T_a}$
\item $H(d):=T_a\otimes c_{T^a} -{1\over 2} c_{T^a} c_{T^b} i_{[T_a,T_b]}$
\item $H(L_X):=X\otimes 1+c_{T^a}i_{[T_a,X]}$
\item $H(e_\alpha):=-{1\over 2} c_{T^a} c_{T^b} \alpha([T_a,T_b])$
\end{enumerate}
\end{thm}
\begin{ex} The above is often seen in the following situation:
\begin{enumerate}
\item Let $\cF$ be a Poisson algebra with $1$,
and let $h:L\rightarrow \cF$ be a Lie homomorphism.
Then we get a Lie homomorphism
$$BRST(L)\rightarrow \cF\otimes Sym(bc(L)).$$
\item Let $\cF$ be the functions on a symplectic manifold $M$. Then
the Poisson algebra $\cF \otimes Sym(bc(L))$ is usually referred to
as the functions on extended phase space. In this language the above
theorem says that if phase space $M$ is endowed with a $L$-hamiltonian, then
extended phase space has a natural $BRST(L)$-hamiltonian.
\end{enumerate}
\end{ex}

\end{document}